\documentclass{emulateapj}
\usepackage{color}
\usepackage{framed}
\usepackage{natbib}

\newcommand{\ms}{\ensuremath{\mathrm{m\,s}^{-1}}}

\newcommand{\mass}{\mathcal{M}}
\newcommand{\radius}{\mathcal{R}}

\newcommand{\msun}{\ensuremath{\mass_\sun}}
\newcommand{\mearth}{\ensuremath{\mass_\earth}}
\newcommand{\msini}{\ensuremath{\mass\sin i}}

\newcommand{\rsun}{\ensuremath{\radius_\sun}}

\newcommand{\logl}{\ensuremath{\log\mathcal{L}}}

\newcommand{\s}[1]{\mathrm{#1}}

\begin{document}

\title{A Six-Planet System Orbiting HD~219134}
\author{Steven S. Vogt\altaffilmark{1}, Jennifer Burt\altaffilmark{1}, Stefano Meschiari\altaffilmark{3}, R. Paul Butler\altaffilmark{2}, Gregory W. Henry\altaffilmark{4}, Songhu Wang\altaffilmark{1,5}, Brad Holden\altaffilmark{1}, Cyril Gapp\altaffilmark{1}, Russell Hanson\altaffilmark{1}, Pamela Arriagada\altaffilmark{2}, Sandy Keiser\altaffilmark{2}, Johanna Teske\altaffilmark{2}\\ Gregory Laughlin\altaffilmark{1}}

 \altaffiltext{1}{UCO/Lick Observatory, Department of Astronomy and Astrophysics, University of California at Santa Cruz,Santa Cruz, CA 95064}
 \altaffiltext{2}{Department of Terrestrial Magnetism, Carnegie Institute of Washington, Washington, DC 20015}
 \altaffiltext{3}{McDonald Observatory, University of Texas at Austin, Austin, TX 78752}
 \altaffiltext{4}{Center of Excellence in Information Systems,
 Tennessee State University, Nashville, TN 37209}
 \altaffiltext{5}{School of Astronomy and Space Science and Key Laboratory of Modern Astronomy and Astrophysics in Ministry of Education, Nanjing University, Nanjing 210093, China}

\begin{abstract}
We present new, high-precision Doppler radial velocity (RV) data sets for the nearby K3V star HD 219134. The data include 175 velocities obtained with the HIRES Spectrograph at the Keck I Telescope, and 101 velocities obtained with the Levy Spectrograph at the Automated Planet Finder Telescope (APF) at Lick Observatory. Our observations reveal six new planetary candidates, with orbital periods of $P=3.1$, $6.8$, $22.8$, $46.7$, $94.2$ and $2247$ days, spanning masses of $\msini = 3.8$, $3.5$, $8.9$, $21.3$, $10.8$ and $108\,\mearth$ respectively. Our analysis indicates that the outermost signal is unlikely to be an artifact induced by stellar activity. In addition, several years of precision photometry with the T10 0.8~m automatic photometric telescope (APT) at Fairborn Observatory demonstrated a lack of brightness variability to a limit of $\sim$0.0002 mag, providing strong support for planetary-reflex motion as the source of the radial velocity variations. The HD 219134 system, with its bright ($V=5.6$) primary provides an excellent opportunity to obtain detailed orbital characterization (and potentially follow-up observations) of a planetary system that resembles many of the multiple-planet systems detected by Kepler, and which are expected to be detected by NASA's forthcoming TESS Mission and by ESA's forthcoming PLATO Mission. 


\end{abstract}

\section{Introduction}

Extrasolar planets are a source of substantial fascination and excitement, and with thousands of examples now known, the statistics of the global distribution are coming into focus. Concentrations of planets -- well-delineated populations in the mass-period diagram -- have been evident for some time, with super-Earths, hot Jupiters and longer-period eccentric giant planets forming groupings that, while distinct, are of effectively still unknown province. 

An apt analogy can be drawn with the gradual discovery of the orbital distributions of asteroids within our own Solar System. Once minor planets had been discovered in significant quantity, clear structures such as the Kirkwood Gaps \citep{Kirkwood1866} became progressively more apparent, although their origin remained mysterious. It is sobering to remark that even after more than two centuries, and the identification of the gaps as arising from resonant dynamics, the formation and evolution of the asteroids remain topics of active research.

In this paper, we report that our set of 276 velocities for HD~219134 (including 138 measurements with 2-hour binning obtained from long-term Keck planet surveys, 37 measurements with 2-hour binning obtained from spectra taken at Keck by the NASA Q01 Program\footnote{NASA program: ``TPF Preparatory Science: Low Mass Short-Period Companions to TPF Target Stars'; P. I.: W. Cochran'}, and 101 measurements with 2-hour binning made with the Automated Planet Finder (APF) telescope) reveal that this star hosts a multi-planet system. 

Indeed, the radial velocities obtained at Keck have, since 2010, strongly suggested that HD~219134 is accompanied by a multiple-planet system, but the orbital architecture at periods $P<100\,{\mathrm d}$ was unclear; the observing cadence at Keck was insufficient to adequately define the orbital parameters of this rather complex multi-planet system. We find that the new APF data, however, with their high velocity precision and improved observing cadence, permit much fuller orbital characterizations for the planetary candidates. Our best model indicates that the star is accompanied by an inner configuration of five low-amplitude planets (having radial velocity half-amplitudes of $K$=1.9 $\ms$, 1.4 $\ms$, 2.3 $\ms$, 4.4 $\ms$, and 1.8 $\ms$, all with orbital periods $P<100$ days). The system also displays a longer-period signal with $P=2247\pm43$ d, and $M\sin(i)=0.34\pm0.02\,M_{\rm jup}$, which is similar to the mass of Saturn. The presence of this outer planet has interesting consequences for current planet formation theories. 

Taken as a whole, HD~219134 presents a planetary system  of substantial scientific interest. Its retinue of multiple super-Earth category planets is highly reminiscent of many of the systems discovered by Kepler, albeit in association with a star that is thousands of times brighter than the median star in the Kepler catalog.

The plan for this paper is as follows: in \S 2, we review what is currently known about HD~219134. In \S 3, we provide a pro-forma review of our Doppler technique, as well as an up-to-date report on the performance and recent results obtained with the APF telescope and Levy spectrograph. In \S 4, we discuss our radial velocity observations for HD~219134, and the 6 planet model that we use to interpret the velocity variations exhibited by the star. In \S 5 we discuss our photometric observations of the system. In \S 6 we very briefly assess the system in the light of current theories regarding planet formation and conclude.

\section{Stellar parameters}

HD 219134 (HR 8832; GJ 892; HIP 114622 ) is located high in the northern sky (RA = +23:13:17, DEC +57:10:06). As a bright ($V=5.57$), nearby ($d=6.55\,$pc) K-type main sequence dwarf of naked-eye visibility, it has long been of interest as a potential planet-bearing star. It was among the original 23 UBC Precise Radial Velocity program stars observed with the CFHT starting in the early 1980s \citep{Walker1995, Walker2012}, and it was an early target of interest at Keck. The first of our 138 Keck velocity measurements dates to JD 2450395 (November, 1996). HD~219134 is currently the 99th nearest known stellar system\footnote{www.recons.org/TOP100.posted.htm}, and as a consequence, any planetary system that it harbors would rank among the ten closest known systems (a plurality of which orbit much dimmer M-dwarf primaries). Among stars known to harbor planets with masses $M_{\rm pl}\sin(i)<10\,M_\s{jup}$, only the Sun, Alpha Cen B, 61 Virginis, and HD 20794 have brighter V magnitudes.

 \begin{figure}
 \plotone{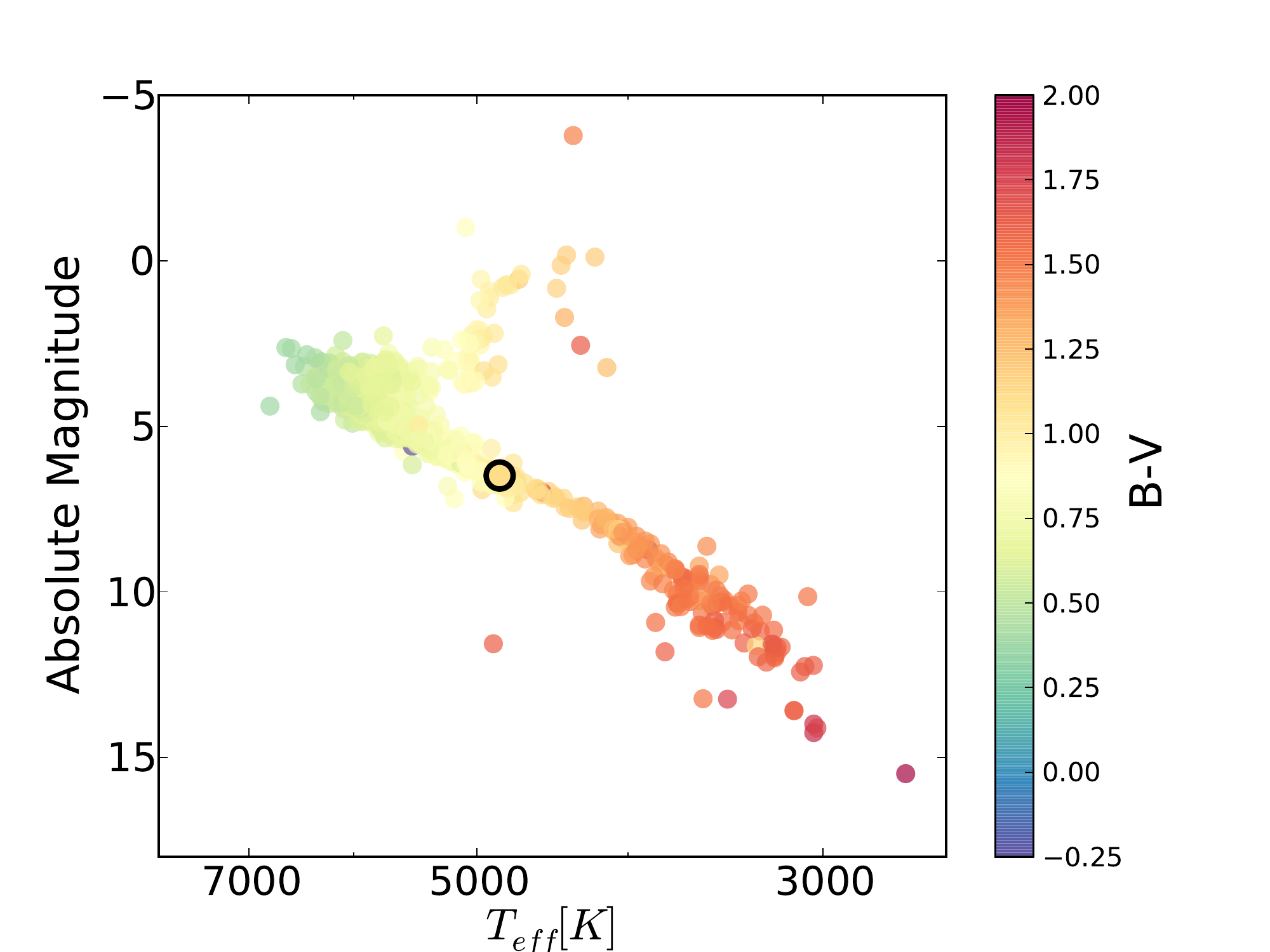}
 \caption{HR diagram with HD~219134's position indicated as a small open circle. Absolute magnitudes, $M$, are estimated from V-band apparent magnitudes and Hipparcos distances using $M=V+5\log_{10}(d/10~{\rm pc})$. All 956 stars in our catalog of radial velocity measurements for which more than 20 Doppler measurements exist are shown, color-coded by their B-V values.}
 \label{fig:HRDiag}
 \end{figure}

Figure \ref{fig:HRDiag} shows HD~219134's position on a color-magnitude diagram containing all of the stars in the current Lick-Carnegie Keck database that have accumulated more than 20 Doppler measurements, and emphasizes the star's entirely ordinary Main Sequence location.
\citet{Takeda2007} derive mass and radius estimates of  $M/M_{\odot}=0.794$ and $R/R_{\odot}=0.77$, along with an age, $\tau= 12.9$ Gyr. \citet{FischerValenti05} measure $v\sin(i)=1.8\,{\rm km s^{-1}}$, which, if we assume equator-on geometry and a radius $R/R_{\odot}=0.77$, implies $P_{\rm rot}\sim20\,$d.

\citet{Tanner2010} used high-contrast imaging of HD~219134 with PALOA/PHARO on the Palomar 200-inch telescope to identify three brown-dwarf companion candidates at $\delta \sim\,$7$\,''$, $\delta \sim\,$10$\,''$, and $\delta \sim\,$11$\,''$ separations. These candidates, however, were determined from archival second-epoch observations to be background stars. \citet{Eggleton2008}, in a multiplicity survey of bright nearby stars, report that HD~219134 has a V=9.4 companion at $\delta \sim\,$106.6$\,''$ separation. If physically bound to the primary, this companion would be a low-mass M-dwarf with a projected separation $d\sim700$ light years, and an orbital period $P\sim20\,$Kyr. Depending on the orbital configuration, such a companion might be capable of exerting non-negligible long-term gravitational perturbations on the HD~219134 system, and so follow-up work to determine the status of the asterism is warranted.

\begin{figure}
\centering
\plotone{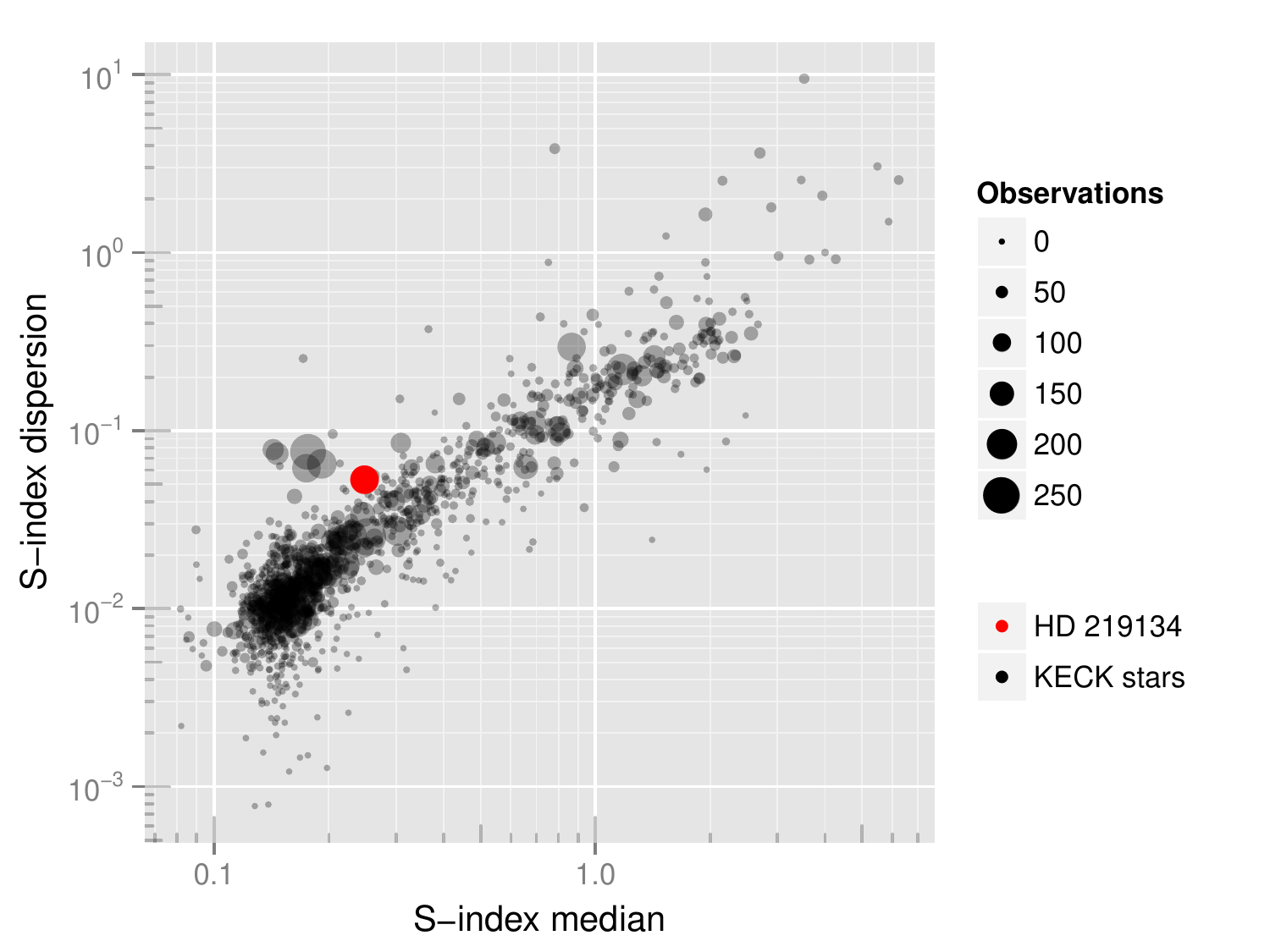}
\caption{\label{fig:sindex} The median $S$-index values and dispersions of the $S$-index measurements  for the stars in the current Keck sample. HD~219134 is shown in red. The size of the points is proportional to the number of observations.}
\end{figure}

 \begin{figure}
 \plotone{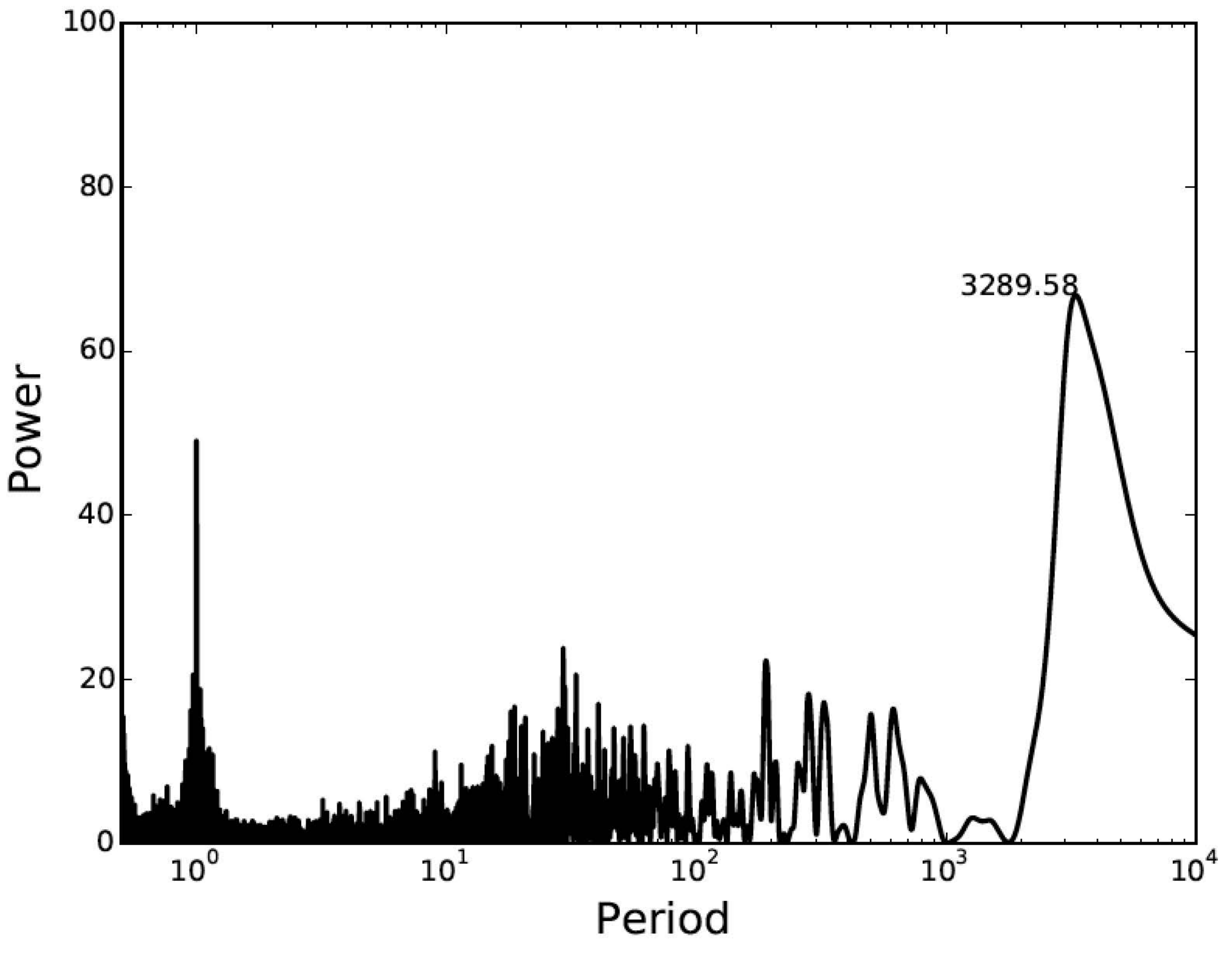}
 \plotone{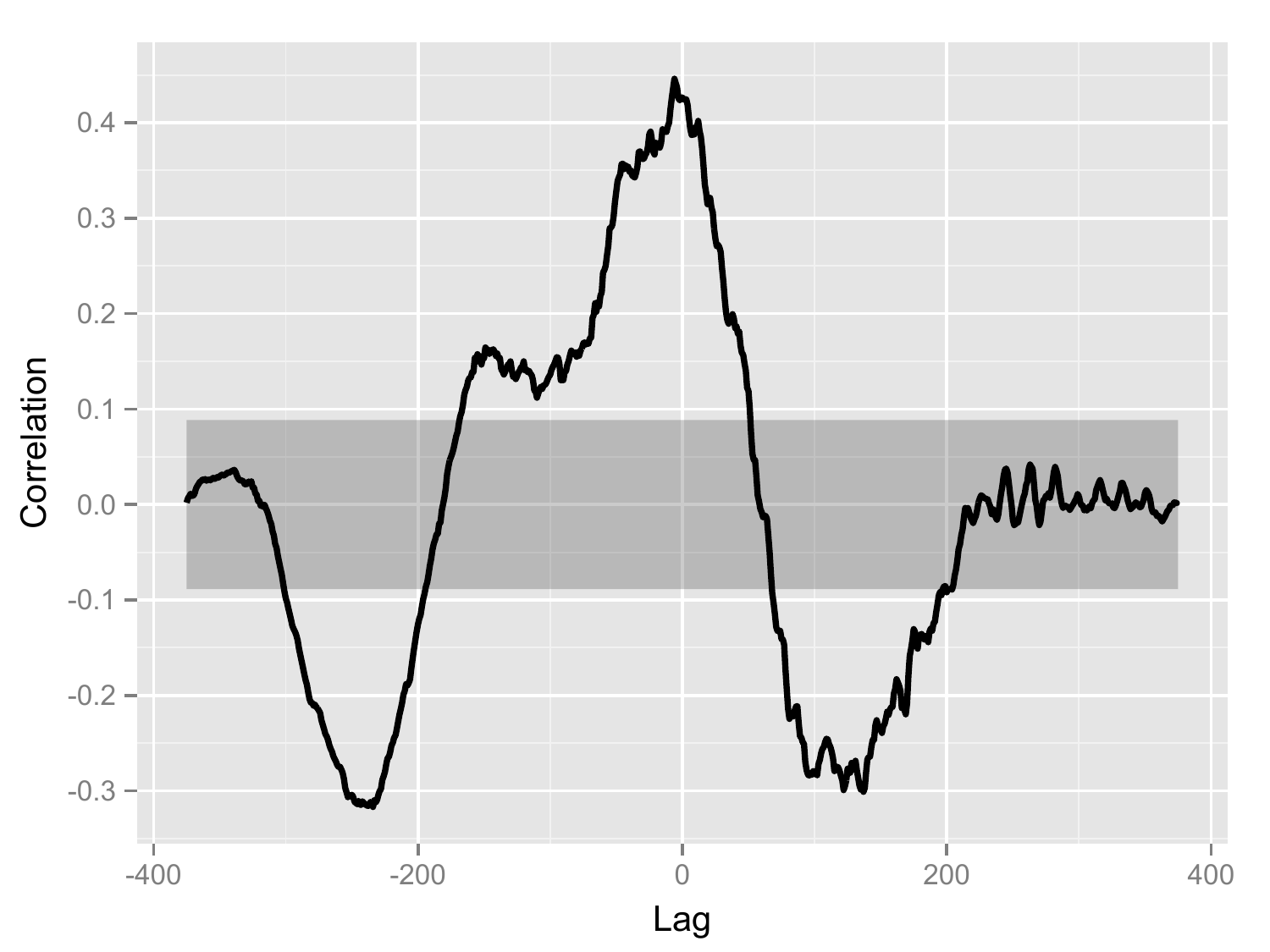}
 \caption{{\it Top Panel} shows a periodogram of the Mt. Wilson $S$-index values associated with our Keck and APF spectra of HD~219134. {\it Bottom Panel} shows a correlation plot for RV data points and their associated $S$-index values. The shaded area marks the 95\% confidence interval for the Pearson correlation coefficient, estimated using sets of white noise data.}
 \label{fig:sindex_info}
 \end{figure}
 
 \begin{figure}
 \plotone{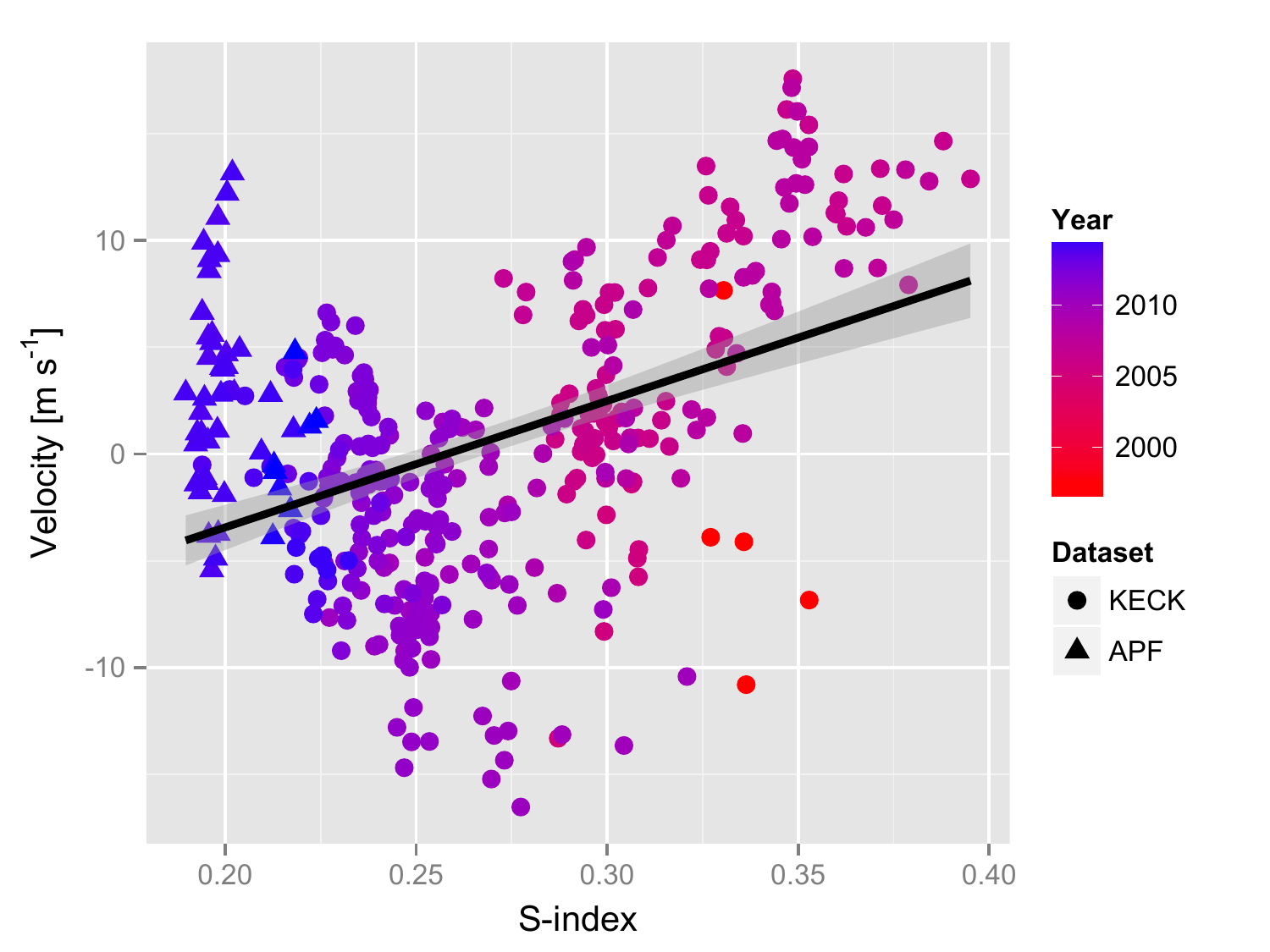}
 \plotone{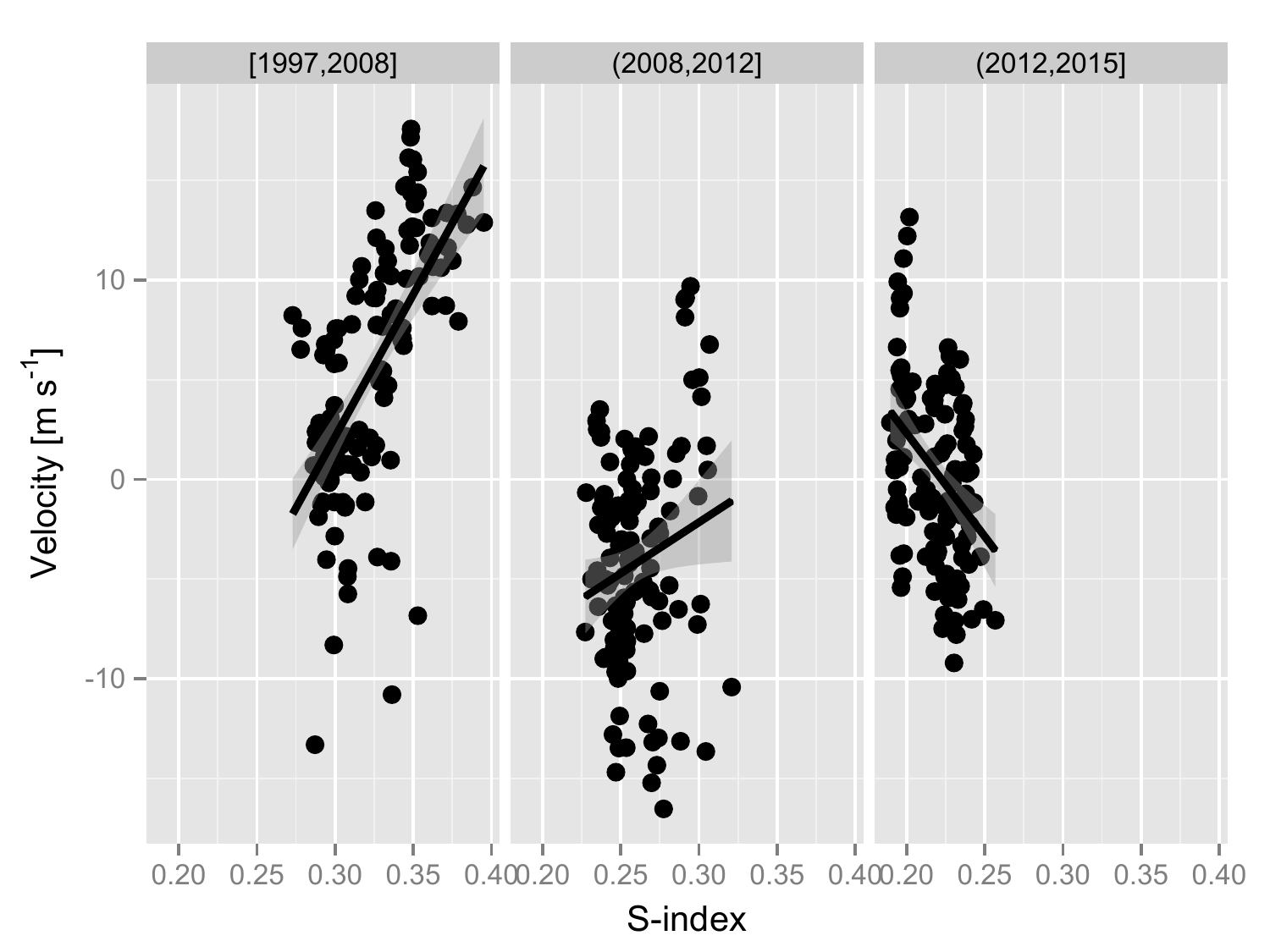}
 \plotone{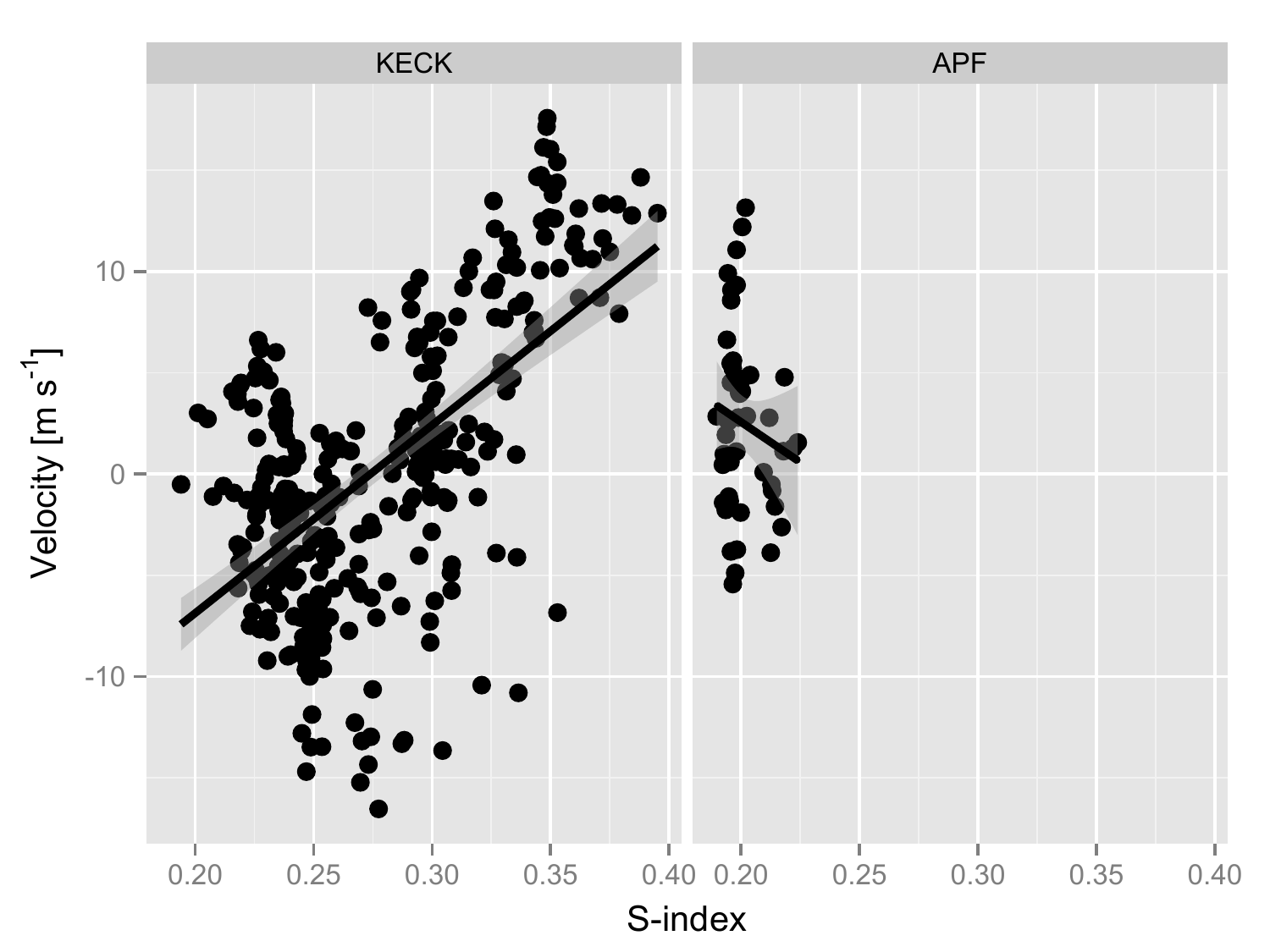} 
 \caption{{\it Top Panel} shows a scatter plot of the radial velocity data for our Keck and APF observations and their associated $S$-index value. Each point is marked according to the time of observation. The best linear fit is shown as a solid line. {\it Middle Panel} shows the same data, faceted in time to emphasize the time-dependent correlation between the radial velocities and the S-index activity indicator. Each panel contains the same number of points. {\it Bottom Panel} shows the same data, faceted by dataset.}
 \label{fig:sindex2_info}
 \end{figure}
 \begin{figure}
 \plotone{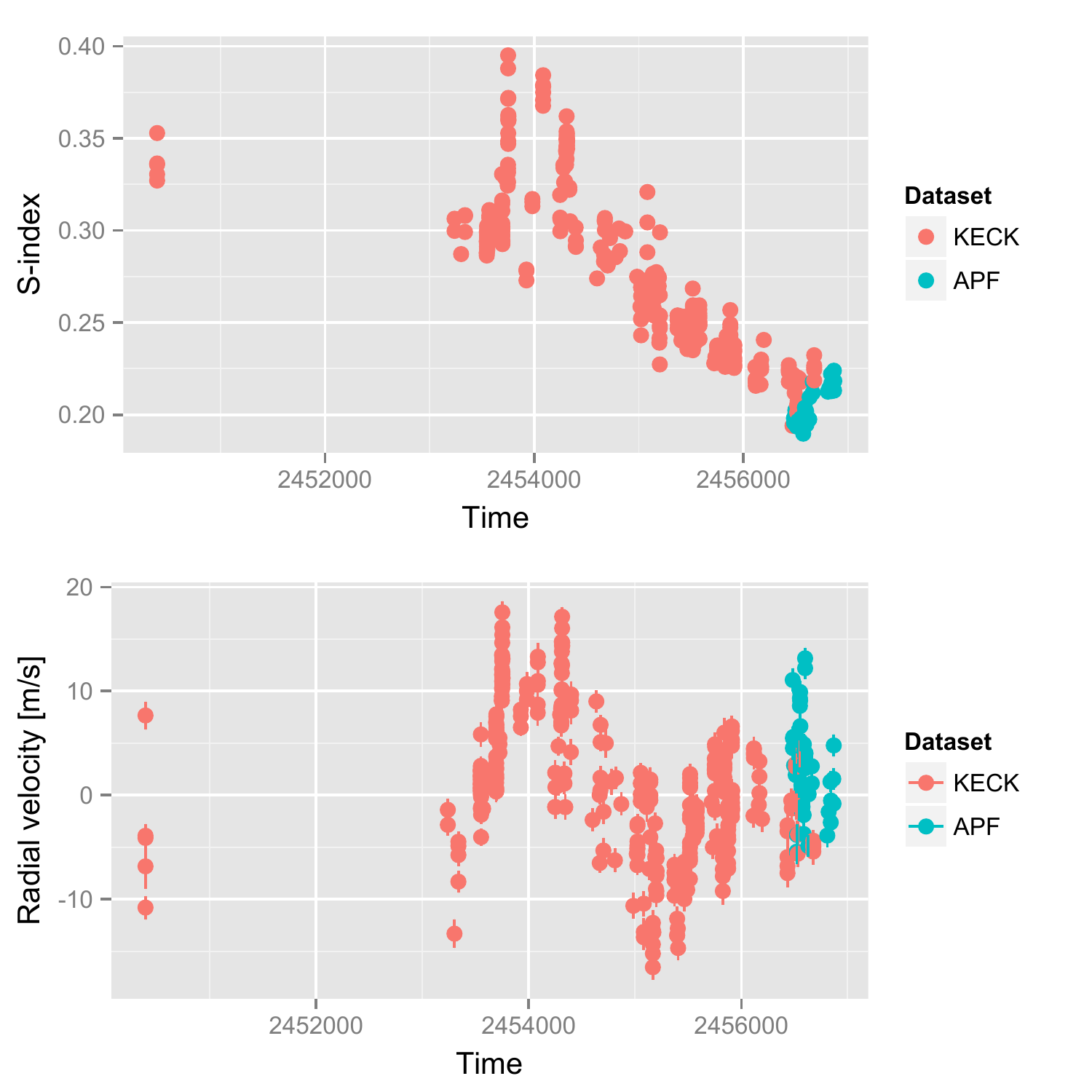}
 \caption{Plot of the S-index values compared to the corresponding RV observation. Velocity measurements and S-index values are shown here for each ${\rm I}_2$ spectrum in our database for HD 219134. For purposes of RV modeling, we use 2-hour binning.\label{fig:sindex_scatter}}
 \end{figure}

In spite of its potentially great age, HD~219134 does show indications of stellar activity. It is listed as a ``Flare Star'' in the SIMBAD Database, and both its median $S$-index value and the standard deviation of its individual $S$-index measurements from our Keck spectra are higher than those of the main locus of stars in our Keck survey (see Figure \ref{fig:sindex}). \citet{Isaacson2010} report a stellar velocity jitter of $1.57\,{\rm m s^{-1}}$ for HD~219134, and this relatively low value is corroborated by the analysis of this paper (Section \ref{sec:analysis}). We do find, however, that the radial velocities for the star are potentially correlated with stellar activity over the decade-plus time baseline of our observations. A periodogram of the $S$-index values (including measurements at all of our Keck epochs, and at all APF epochs for which photoelectron counts in the ${\rm I}_{2}$ region of the spectrum exceed $N=25,000$)  is shown in the top panel of Figure \ref{fig:sindex_info}. There is a significant peak in this periodogram at $P\sim3300$ days. This period is greater than and district from the $P\sim2300$ day periodicity that is present in the Doppler Velocity data for this star. Figure \ref{fig:sindex_info} also shows a correlation plot of the radial velocity observations and their $S$-index values. The peak at zero lag (observation record) indicates that there is correlation between the long-period signal in the radial velocities and the $S$-index values, as manifested by the long period of decline in both time series. It is therefore possible that some of the observed velocity variation can be attributed to surface activity. Caution is always warranted in interpreting long-term RV variations.

The Mt. Wilson $S$-index measures the ratio of flux from 1${\rm \AA}$ bins surrounding the line centers of the Ca II H\& K lines (at 3968.47${\rm \AA}$ and 3933.66${\rm \AA}$), as compared to two broader 25${\rm \AA}$ bandpasses lying 250${\rm \AA}$ to either side of the Ca II H\& K line location (Duncan et al. 1991). In the standard picture, an increase in the $S$-index, whose flux emerges from above the mean photospheric depth of the star, corresponds with an increase in spot activity on the stellar surface. Spots, in turn, suppress convection in their vicinity, which decreases the overall convective blueshift of the star, leading to the expectation of a correlation with the Doppler velocity of the star. A star with a magnetic cycle that modulates the number of spots can therefore present a long-term Doppler trend with an amplitude and periodicity that mimics the Keplerian signal from a distant planet  (Dumusque et al. 2011).

The upper panel of Figure \ref{fig:sindex2_info} charts the velocity measurements taken with the Keck and the APF telescopes (with the median value for each data set subtracted) against the corresponding $S$-index measurements. While the strength of the overall positive correlation is indicated by a linear fit to the data, the color-coded time-ordering of the points indicates that the correlation was much stronger during epochs from 2000 through approximately 2010. Indeed, as is indicated by the middle panel of Figure \ref{fig:sindex2_info}, the correlation has reversed sign from 2012 through present, and a weak negative correlation (having less than $1\, \sigma$ significance) is present in the recent APF observations. Figure \ref{fig:sindex_scatter} permits comparison of the time evolution of the $S$-index values to the corresponding RV observations.

The unusual features in the time development of the RV -- $S$-index correlation imply that considerable caution must be exercised in interpreting the source of the multi-year periodicity that is present in the Doppler velocity time series. Our candidate planetary signal could, for example, be produced by the superposition of the stellar magnetic activity cycle and a giant planet on a Keplerian orbit. Further monitoring, accompanied by detailed analysis, is clearly required.

\citet{FischerValenti05} report a number of additional spectroscopically derived properties for HD~219134. It appears to be somewhat metal-rich in comparison to the Sun, with $[{\rm M/H}]=0.09$, and individual abundances that include $[{\rm Na/H}]=0.13$, $[{\rm Si/H}]=0.02$, $[{\rm Ti/H}]=0.02$, $[{\rm Fe/H}]=0.12$,  and $[{\rm Ni/H}]=0.09$. \citet{Ramirez2013} report a high oxygen abundances of $[{\rm O/H}]=0.23$. We note that this high value seems discrepant in light of the star's other abundance measurements, as well as the value $[{\rm Fe/H}]=0.04$ found by \citep{AllendePrieto2004}, and so must be treated with caution. One could speculate, however, that there might be a connection between the possible high abundance of iron and the apparent ease with which the system formed multiple planets having $P<100$ days \citep{Robinson2006}. \citet{Tanner09} observed HD~219134 at 160$\,\mu m$ using the Spitzer Space Telescope's MIPS spectrometer, and found no excess emission characteristic of a remnant debris disk, in keeping with the large apparent age of the star.

\begin{deluxetable}{rll}
\tablecaption{\label{tab:stellar} Stellar parameters for HD~219134}
\tablehead{{Parameter}&{Value}&{Reference}}
\tablecolumns{3}
\startdata
Spectral type & K3V & \citep{Soubiran2008} \\
$M_V$ & 6.46 & \citep{Soubiran2008} \\
$V$ & 5.57 & \citep{Oja1993} \\
$B-V$ & 0.99 & \cite{Oja1993} \\
Mass (\msun) & $0.794 ^{+0.037} _{-0.022}$ & \citep{Takeda2007} \\
Radius (\rsun) & 0.77 $\pm$ 0.02 & \citep{Takeda2007} \\
Luminosity ($L_\sun$) & 0.31 & This work \\
Distance (pc) & 6.546 $\pm$ 0.012 & \citep{Soubiran2008} \\ 
$S_{hk}$ & 0.25 & This work \\
Age (Gyr) & 12.46 & \citep{Takeda2007} \\
$[\mathrm{Fe/H}]$ & 0.08 & \citep{Soubiran2008} \\
$T_\s{eff}$ ($K$) & 4913 & \citep{Soubiran2008} \\
$\log(g)$ (cm s$^{-2}$) & 4.51 & \citep{Soubiran2008} \\
\enddata
\end{deluxetable}

\section{Radial velocity observations}
The HIRES spectrometer, located at the Keck-I telescope \citep{Vogt94}, and the Automated Planet Finder's Levy spectrometer \citep{Vogt2014a} were employed to obtain the Doppler measurements of HD 219134 that form the basis of this paper.  In accordance with long-established practice,  Doppler shifts at both telescopes are obtained by imprinting an iodine absorption spectrum on the collected starlight prior to its incidence on the spectrograph slit \citep{Butler96}. The forest of added I$_2$ lines generates a stable wavelength calibration and permits the measurement of the spectrometer point spread function (PSF).  For each spectrum so obtained, the $5000~\rm{\AA} \lesssim \lambda \lesssim 6200~\rm{\AA}$ region containing a sufficient density of I$_2$ lines is subdivided into 700 individual segments of width $2\rm{\AA}$, with each segment providing independent measures of the wavelength, the PSF, and the Doppler shift.  Our reported overall stellar velocity from a given spectrum is a weighted mean of the individual velocity measurements. The uncertainty for each velocity is the RMS of the individual segment velocity values about the mean divided by the square root of the number of segments. This ``internal" uncertainty represents primarily errors in the fitting process, which are dominated by Poisson statistics. The velocities are expressed relative to  the  solar  system  barycenter,  but  are not referenced to any absolute fiducial point  As a consequence, the velocity zero-point offset between the measurements at the two telescopes must be treated as a free parameter.

For the data set being considered here, there is an 8-year gap between the first Keck velocity measurement and the second Keck velocity measurement. The Keck HIRES CCD was upgraded during the interval between the two observations. In our reduction pipeline, we first analyze the entire Keck data set using only the spectral chunks that are present in both the pre and post-upgrade detector CCDs, thereby obviating the need for any additional internal velocity offsets.  We then reanalyze the post-fix spectra, using all the spectral chunks, thereby improving the post-fix precision.

The Automated Planet Finder (APF) is a 2.4m telescope at Lick Observatory. Coupled with the high-resolution Levy echelle spectrometer, it was designed to detect planets in the liquid water habitable zone of their host stars. It works at a typical spectral resolution of $R \sim 110,000$ and delivers a peak overall system efficiency (fraction of photons striking the telescope primary that are detected by the CCD) of 15\% \citep{Vogt2014a}. Currently, 80\% of the telescope's time is dedicated to the detection of extrasolar planets, and the APF has scheduling software capable of making decisions on what target to observe based on the ambient atmospheric transparency, atmospheric seeing, and moon phase. This allows the telescope to operate efficiently throughout the year without the need for human supervision.

Since it began acquiring scientific data in Q2 2013, the telescope has contributed to three planetary system discoveries, all with radial velocity measurements having a 1-3 $\ms$ level of precision \citep{Vogt2014b,Burt2014,Fulton2015}. The detection of these planets, along with the candidates described in the current paper, indicates that the APF telescope is well-suited to the discovery of low-mass planets orbiting low-mass stars.

\begin{figure}
\centering
\plotone{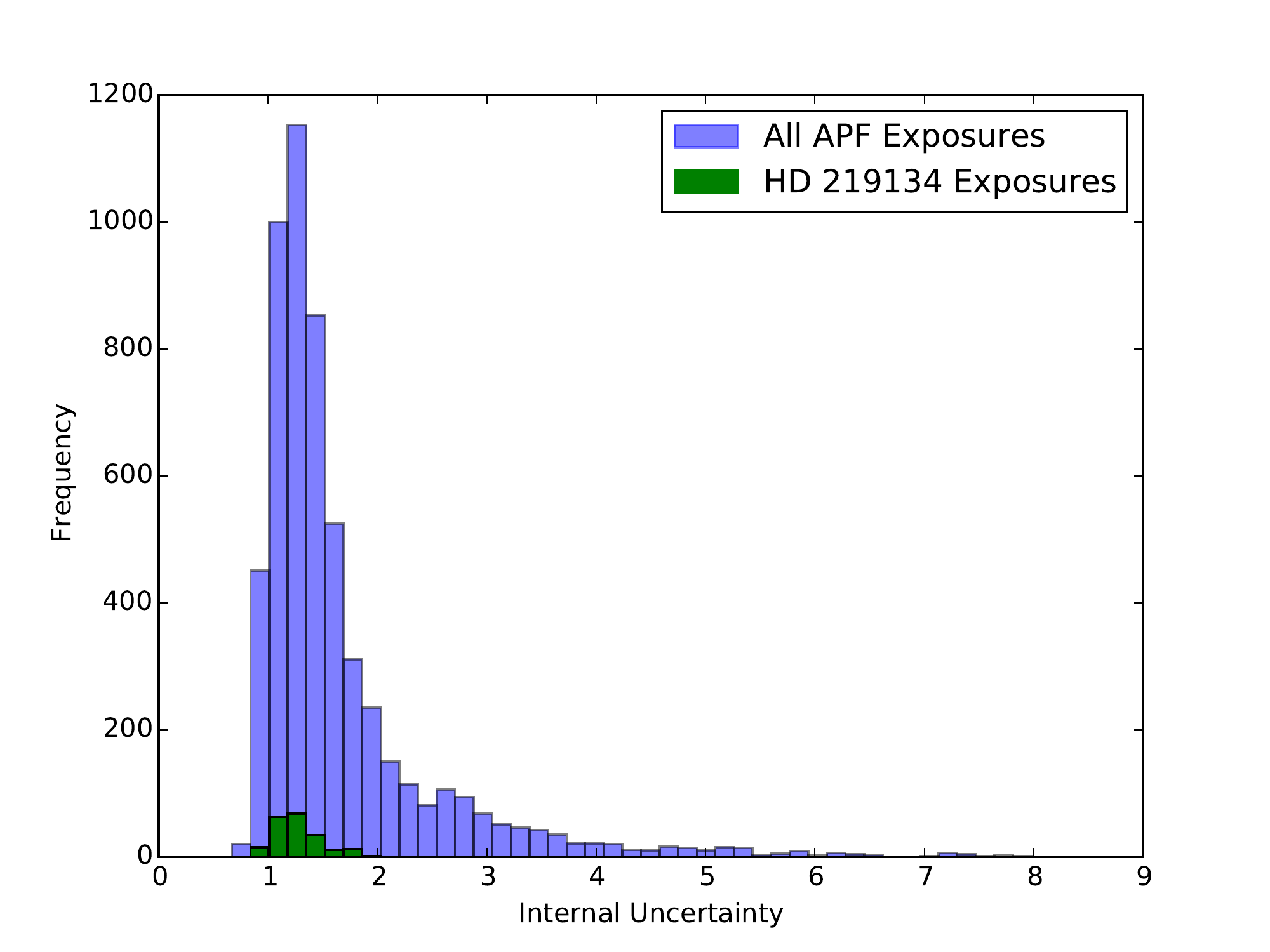}
\caption{\label{fig:AllObs} Histogram showing internal precisions obtained to date with APF while observing bright, nearby Main Sequence Stars. The subset of measurements taken of HD~219134 are shown in green. Observations are unbinned.}
\end{figure}

The APF has consistently achieved internal velocity precision of order $\sigma \lesssim 2\,{\rm m\, s^{-1}}$ on bright (e.g. $V\sim7$) stars. Figure \ref{fig:AllObs} is a histogram showing the internal precisions obtained to date for scientific target stars in our Doppler velocity program. The median internal (photon shot noise) precision is $1.18\,{\rm m\, s^{-1}}$, and $2485$ of the measurements (binned at 2-hour cadence) have $\sigma< 2\,{\rm m\, s^{-1}}$.

\citet{Fulton2015} report use of radial velocity data obtained using the APF telescope to characterize a multiple-planet system orbiting HD~7924. In the course of their analysis, they investigated correlations with environmental parameters such as air pressure and CCD temperature, and find improvements up to a factor of two in the RMS of APF velocities when the data is de-trended against these factors.

In light of these results, we initiated a similar set of experiments. We compared the velocity values for the RV standard stars HD~185144, HD~9407 and HD~10700 against all of the environmental parameters stored for each observation. We detect no notable correlation between the Doppler velocity measurements and any of these factors. When fitting linear trends to the velocities as a function of air pressure, we see the RMS change from 1.84 $\mathrm{m s}^{-1}$ to 1.81 $\mathrm{m\,s}^{-1}$ for HD185144, 2.13 $\mathrm{m\,s}^{-1}$ to 2.11 $\mathrm{m\,s}^{-1}$ for HD~10700 and 2.46 $\mathrm{m\,s}^{-1}$ to 2.33 $\mathrm{m\,s}^{-1}$ for HD~9407. Similarly, when fitting against the CCD temperature we see changes in the RMS of 1.84 $\mathrm{m\,s}^{-1}$ to 1.71 $\mathrm{m\,s}^{-1}$, 2.13 $\mathrm{m\,s}^{-1}$ to 2.13 $\mathrm{m\,s}^{-1}$ and 2.46 $\mathrm{m\,s}^{-1}$ to 2.42 $\mathrm{m\,s}^{-1}$ for HD~185144, HD~10700 and HD~9407, respectively. 

In our view, it is likely that the difference in correlation strengths stems from the different instrument focusing procedures and data reduction pipelines used in the separate analyses of the HD~7924 and HD~219134 spectra. Due to the lack of evident trends, we have elected not to decorrelate our data set against environmental variables.

\begin{table}[ht]
\centering
\begin{tabular}{lcccc}
  \hline
 & Time [JD] & RV [$\ms$] & Uncertainty [$\ms$] & Dataset \\ 
  \hline
1 & 2450395.74 & -4.50 & 0.50 & KECK \\ 
  2 & 2453239.05 & -2.14 & 0.74 & KECK \\ 
  3 & 2453301.77 & -13.31 & 1.37 & KECK \\ 
  4 & 2453338.71 & -5.86 & 0.53 & KECK \\ 
  5 & 2453547.10 & 1.56 & 0.45 & KECK \\ 
  6 & 2453548.09 & 1.39 & 0.48 & KECK \\ 
  7 & 2453549.12 & -0.66 & 0.55 & KECK \\ 
  8 & 2453550.09 & 2.79 & 0.43 & KECK \\ 
  9 & 2453551.09 & 0.81 & 0.38 & KECK \\ 
  10 & 2453552.05 & -1.88 & 0.44 & KECK \\ 
  11 & 2453571.06 & 0.00 & 0.48 & KECK \\ 
  12 & 2453692.77 & 6.20 & 0.71 & KECK \\ 
  13 & 2453693.70 & 1.81 & 0.60 & KECK \\ 
  14 & 2453693.92 & 1.54 & 0.57 & KECK \\ 
  15 & 2453694.70 & 1.48 & 0.58 & KECK \\ 
  16 & 2453694.85 & 1.99 & 0.61 & KECK \\ 
  17 & 2453695.85 & 7.63 & 0.59 & KECK \\ 
  18 & 2453696.71 & 6.48 & 0.55 & KECK \\ 
  19 & 2453713.68 & -1.53 & 0.64 & Q01 \\ 
  20 & 2453713.68 & -1.19 & 0.65 & Q01 \\ 
   \hline
\end{tabular}
\caption{Radial Velocity observations (sample)} 
\label{tab:data}
\end{table}

Table \ref{tab:data} presents the complete set of our RV observations for HD~219134. The RV coverage spans approximately 19 years of monitoring over 276 (two-hour binned) measurements. The median internal uncertainty for our observations is $\sigma_i \approx$ 0.75 \ms, and the peak-to-peak velocity is $\approx$ 31.3 \ms. The velocity scatter around the average RV is $\approx$ 5.7 \ms. Observations marked as ``Keck'' are HIRES velocities from spectra obtained by the Lick-Carnegie Exoplanet team, or, in some cases, from publicly available archived spectra obtained by the California Planet Survey \citep{Howard2010}. Observations marked ``Q01'' are velocities computed using our pipeline from archival HIRES spectra from the 2005 NASA program: ``TPF Preparatory Science: Low Mass Short-Period Companions to TPF Target Stars", to Principal Investigator W. Cochran. Observations marked ``APF'' are from spectra obtained with the APF telescope and Levy Spectrometer.

\section{Keplerian solution}\label{sec:analysis}

\begin{figure}
\centering
\plotone{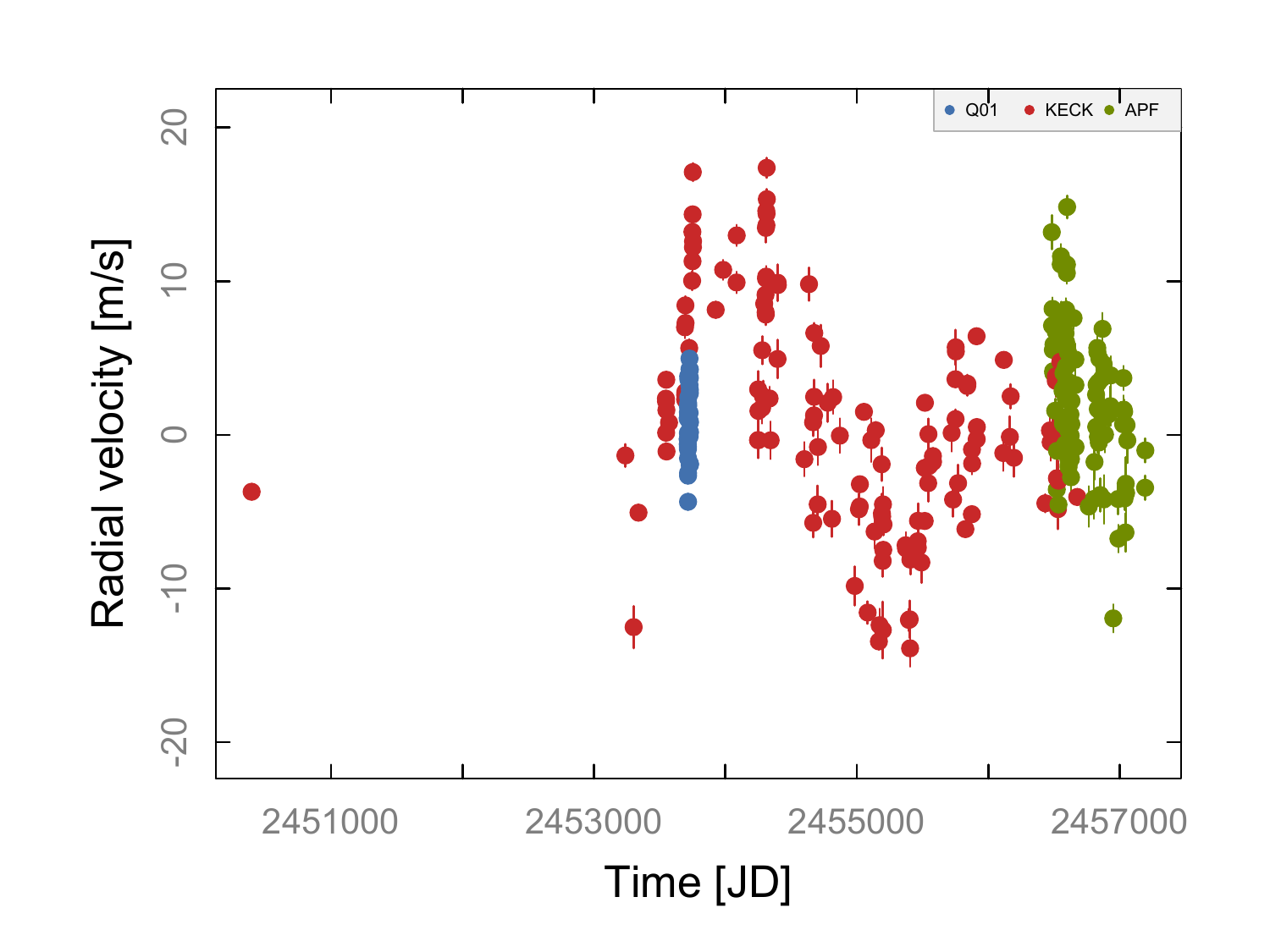}
\caption{\label{fig:data} Radial velocity observations for HD~219134.}
\end{figure}

\begin{figure}
\plotone{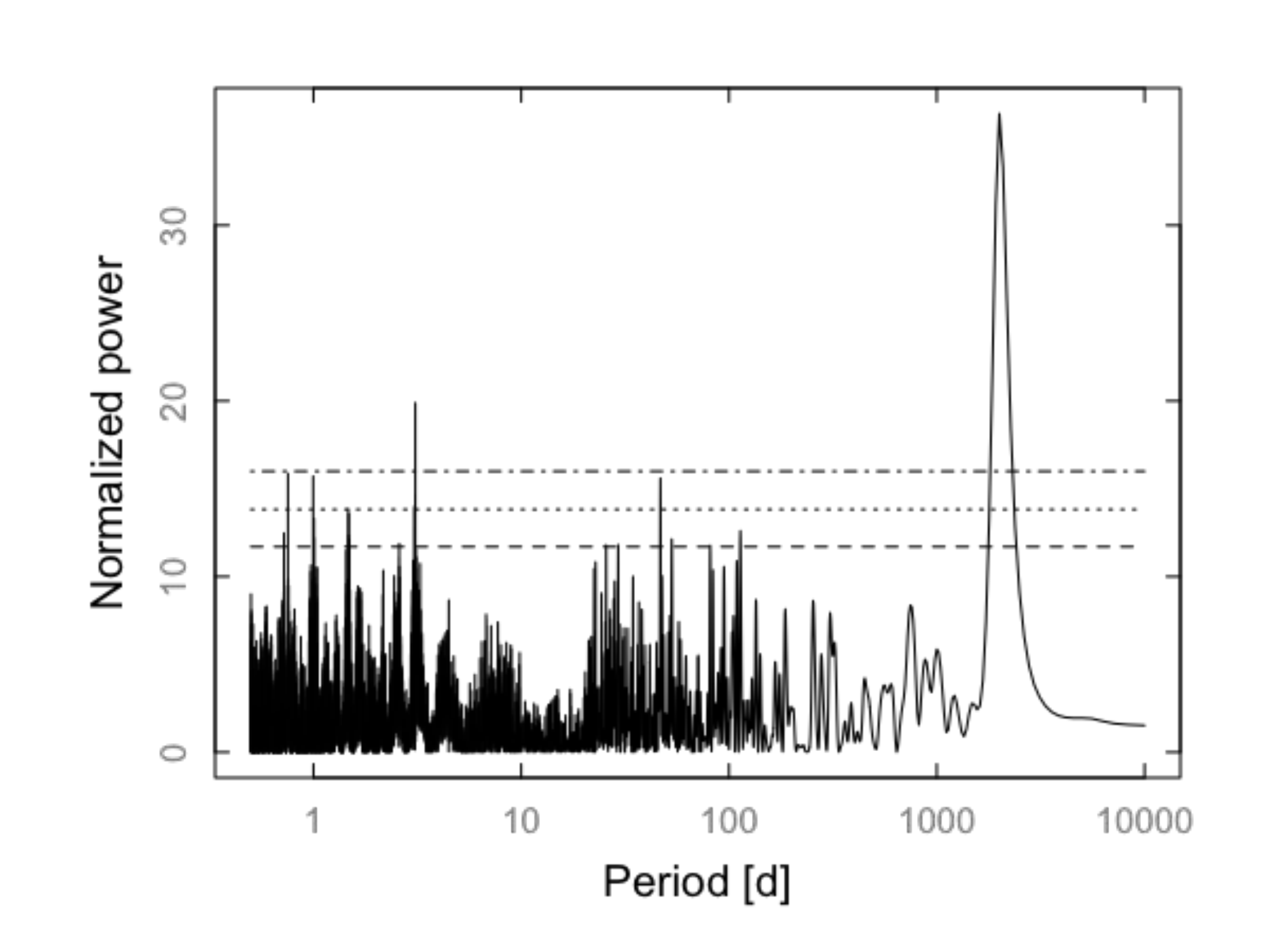}\\
\plotone{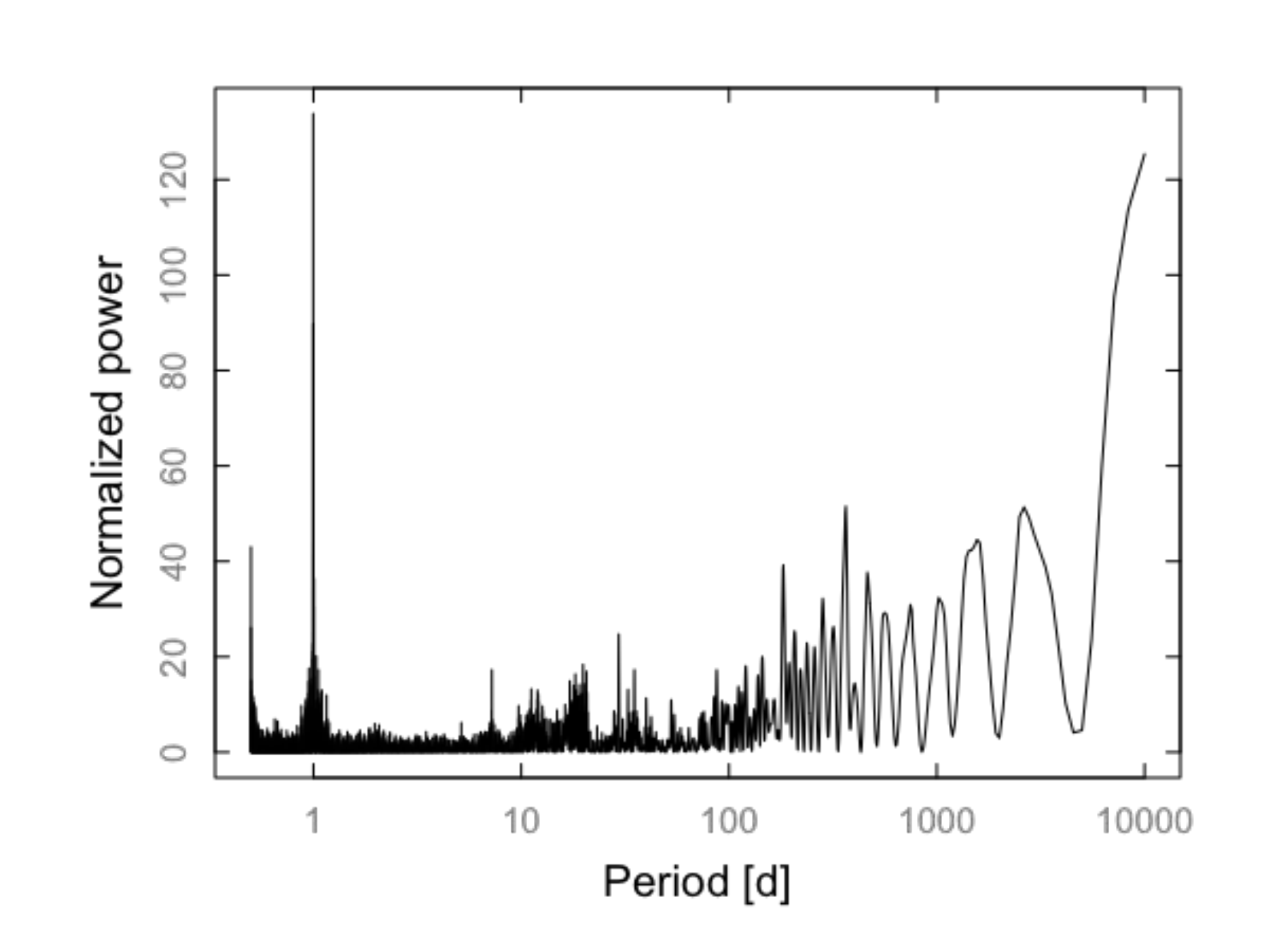}\\
\caption{\label{fig:ls} \textit{Top panel:} Error-weighted Lomb-Scargle periodogram for HD~219134. False-alarm probability levels are shown at the 10\%, 1\% and 0.1\% level. \textit{Bottom panel:} Spectral window function.}
\end{figure}

Figure \ref{fig:data} shows the RV measurements after binning to two-hour increments. We again note that there is a single Doppler measurement at BJD 2450395.74 (November 8th, 1996) followed by a nearly eight-year gap to the next measurement at BJD 2453239.05 (August 21, 2004). The single early epoch point is useful for cementing the lack of any apparent long-term large-scale Doppler velocity trend. The top panel of Figure \ref{fig:ls} shows the error-weighted, normalized Lomb-Scargle periodogram \citep{Zechmeister09}. The three horizontal lines in the plot represent different levels of false alarm probability (FAP; 10\%, 1\% and 0.1\%, from bottom to top, respectively). The FAPs were computed by scrambling the dataset 100,000 times and sampling the periodogram at 100,000 frequencies, in order to determine the probability that the power at each frequency could be exceeded by chance \citep[e.g.][]{Marcy05}. The bottom panel of Figure \ref{fig:ls} shows the spectral window, displaying the usual peaks due to observational cadence, arising from the sidereal and solar days, and from the solar year \citep{Dawson2010}.

We fit the radial velocities with a Keplerian model with a vector of parameters $\bar{\theta}$, consisting of the orbital elements (period, mass, mean anomaly, eccentricity and longitude of pericenter for each planet) and vertical offsets for each dataset (to account for differences in the zero point among datasets). Each radial velocity measurement $v_i$, taken at time $t_i$, is represented as 
\begin{equation}
v_i = V(t_i, \bar{\theta}) + e_i + s_j \, ,
\end{equation}
where $V(t_i, \bar{\theta})$ is the predicted velocity, and $e_i$ is normally distributed with variance $e_i^2$ (fixed, and corresponding to the formal uncertainties quoted by the observer). The term $s_j$ accounts for additional sources of scatter (e.g. underestimated measurement errors, stellar jitter, and other astrophysical sources of RV variation), modeling the residual noise in each $j$-th dataset as normally distributed with variance $s_j^2$ \citep[e.g.][]{Gregory05}. Therefore, scatter from the model is modeled with a Gaussian distribution of variance $e_i^2 + s_j^2$. The best-fit parameters are derived by optimizing the log-likelihood of the model:
\begin{equation}
\logl = -\frac{1}{2} \left[\chi^2 + \sum_{i=1}^{N_{\mathrm{o}}} \log(e_i^2 + s_i^2) + N_{\mathrm{o}} \log(2\pi)\right],
\end{equation}\label{eqn:logl}
where 
\begin{equation}
\chi^2 = \sum_{i=1}^{N_{\mathrm{o}}} {(V_i-v_i)^2}/({e_i^2 + s_i^2})\, .
\end{equation}\label{eqn:chisq}

\begin{figure*}
\centering
\epsscale{0.45}
   \plotone{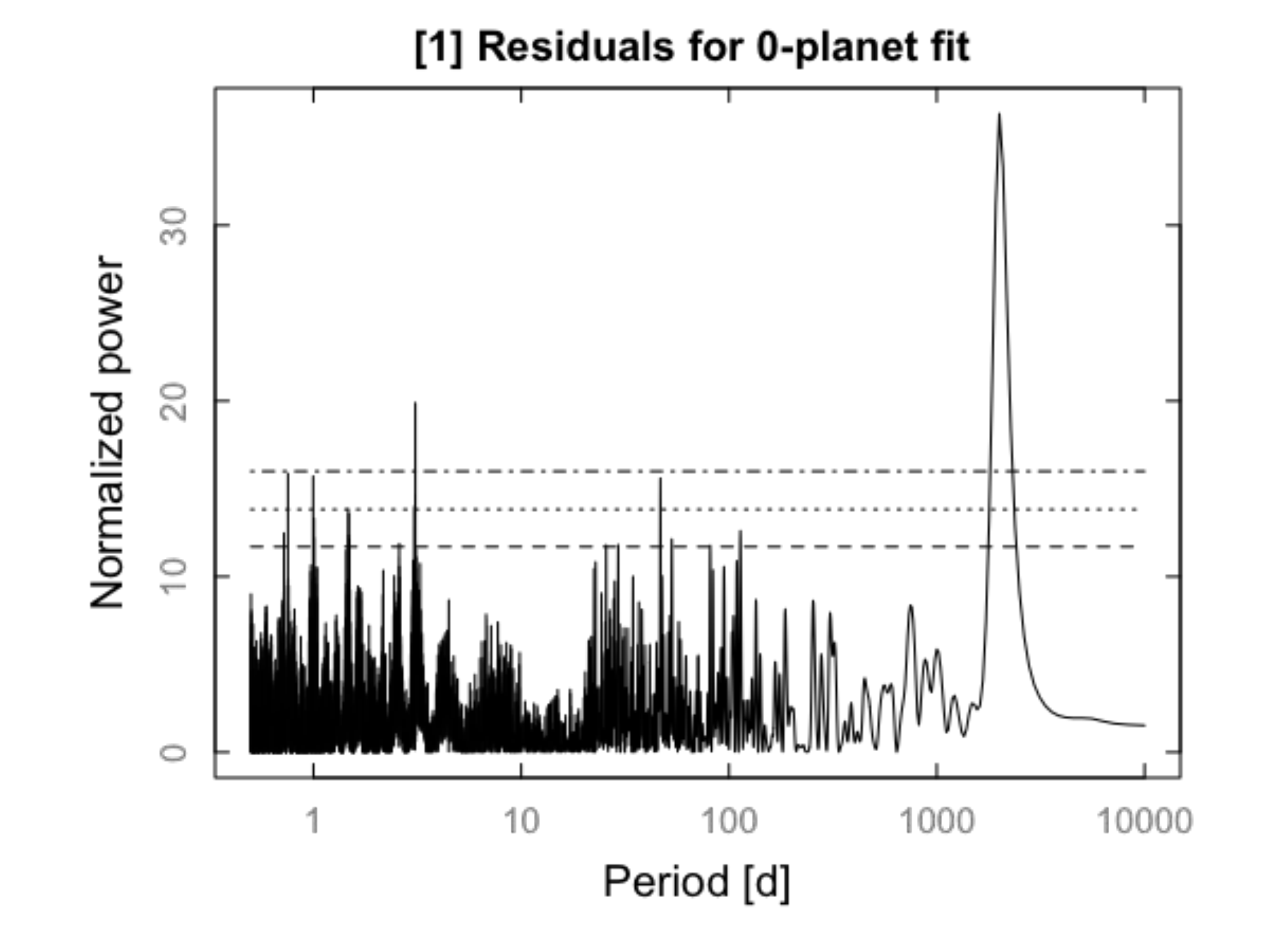}\plotone{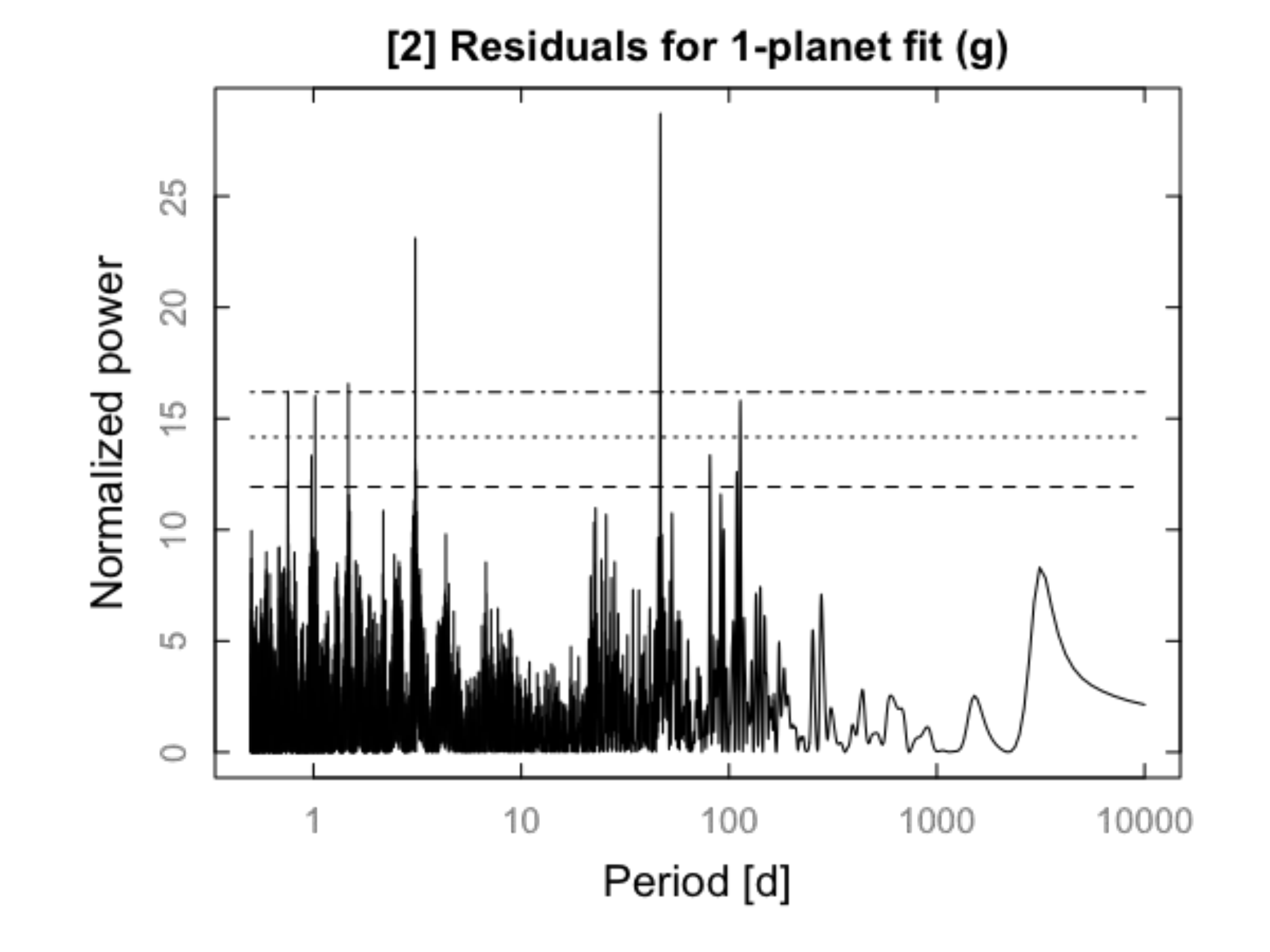}
   \plotone{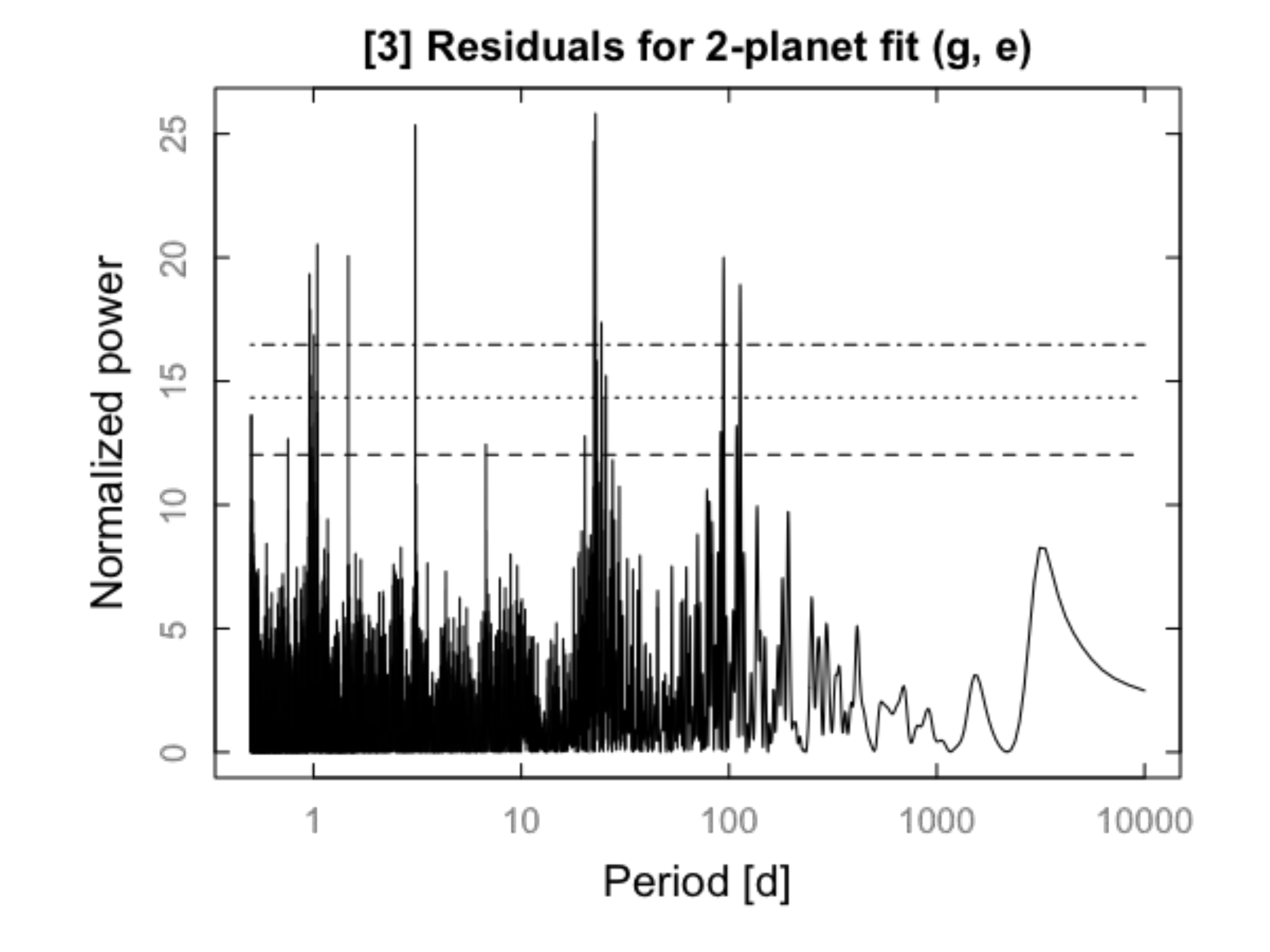}\plotone{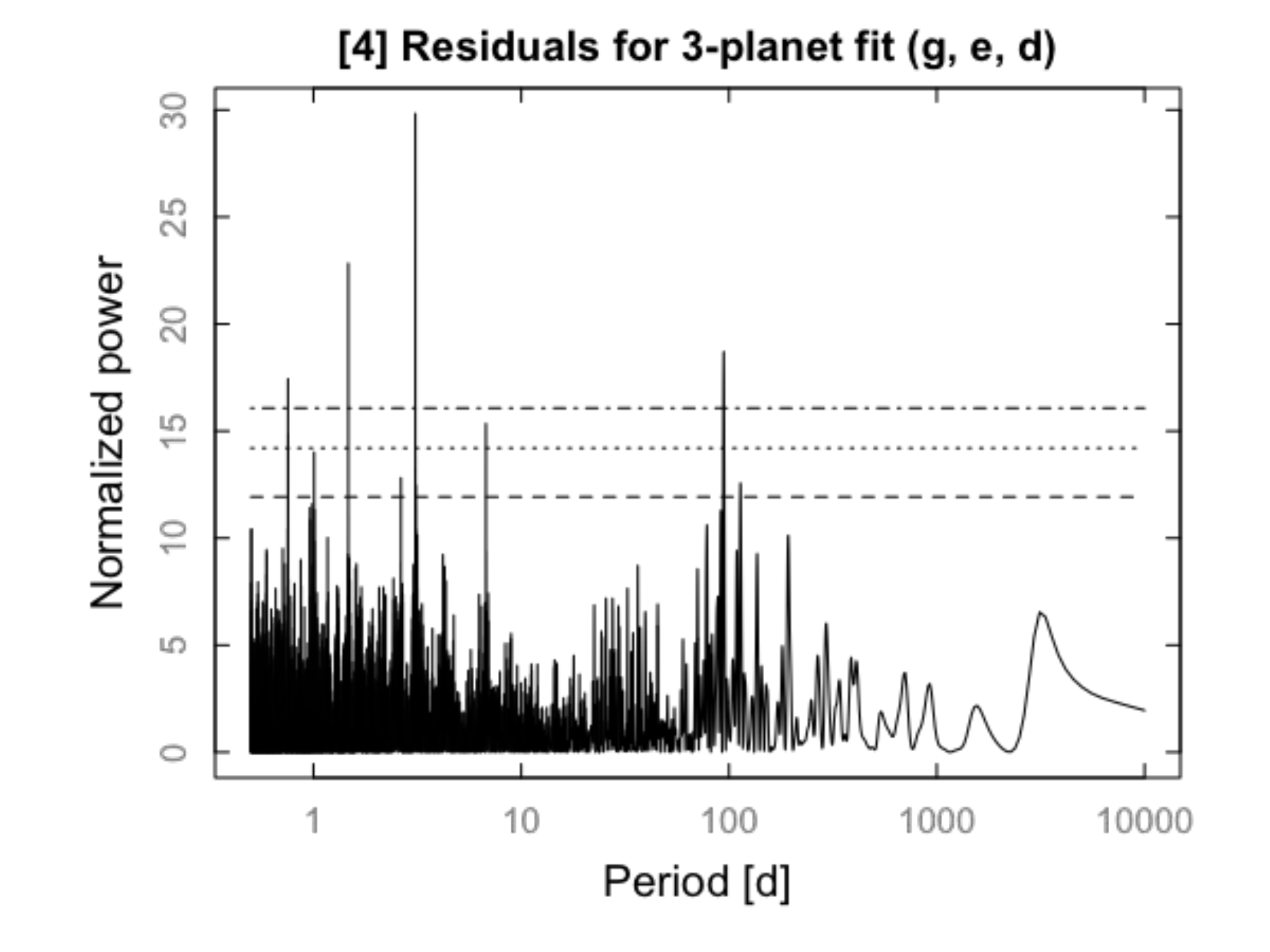}
   \plotone{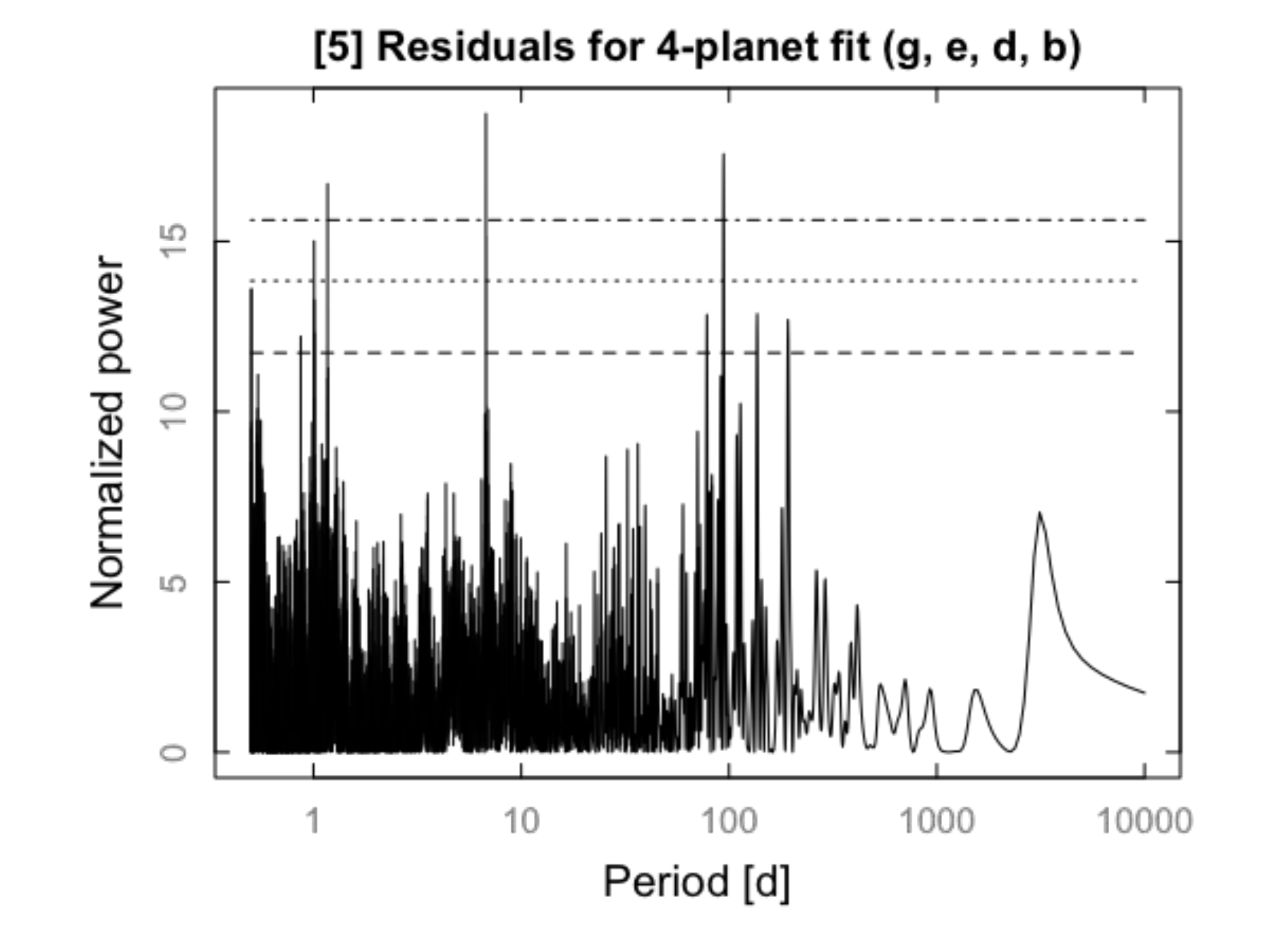}\plotone{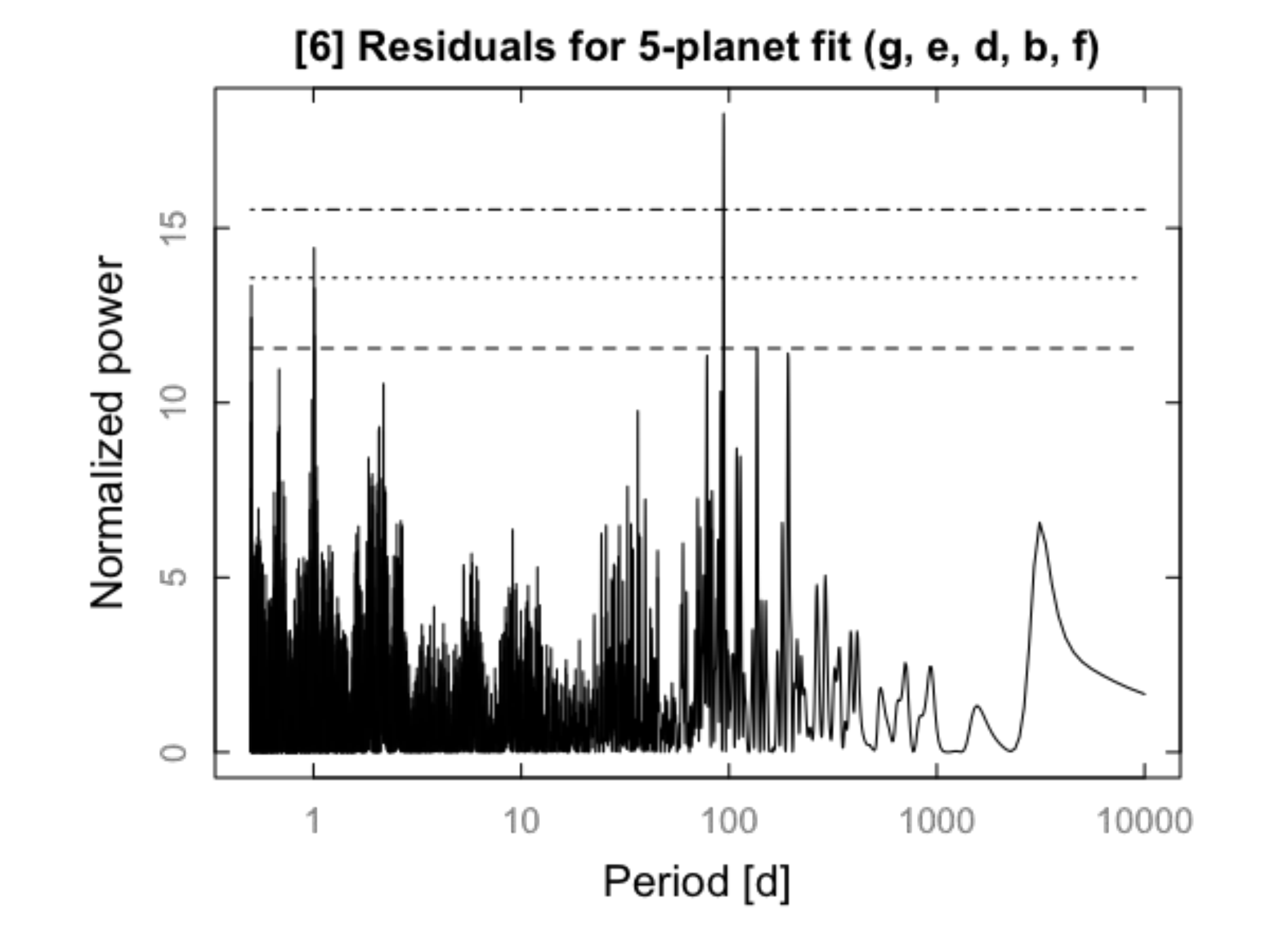}
   \plotone{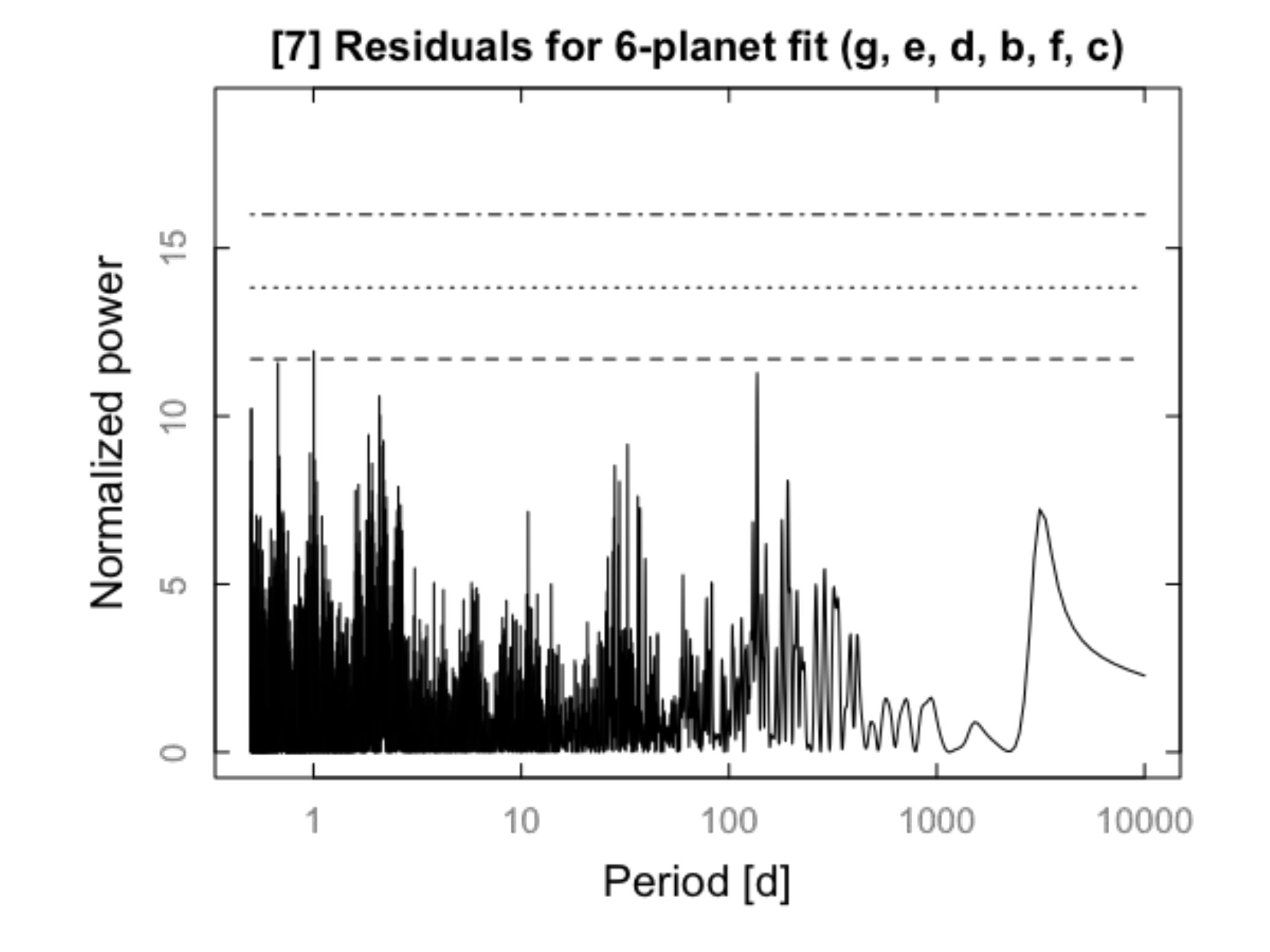}
\caption{\label{fig:periodograms} \textit{Panels (1)-(7):} Lomb-Scargle periodograms computed for each model (0-planet, 1-planet, 2-planet, 3-planet, 4-planet, 5-planet and 6-planet fits).}
\end{figure*}

In order to derive the starting values of the parameters, we fit our data by removing peaks in the Lomb-Scargle periodogram of the residuals. For each $N$-planet model, we compute bootstrapped periodograms and investigate the strongest peaks with FAP $< 10^{-3}$. Figure \ref{fig:periodograms} shows the periodogram of the residuals for each stage of the fit construction. 

\begin{table*}[ht]
\centering\footnotesize
\begin{tabular}{rllllll}
  \hline
 & HD~219134b & HD~219134c & HD~219134d & HD~219134e & HD~219134f & HD~219134g \\ 
  \hline
$P$ [days] & 3.0931 [0.0001] & 6.7635 [0.0006] & 22.805 [0.005] & 46.71 [0.01] & 94.2 [0.2] & 2247 [43] \\ 
  $\mass \sin(i)$ [$\mass_\s{jup}$] & 0.012 [0.001] & 0.011 [0.002] & 0.028 [0.003] & 0.067 [0.004] & 0.034 [0.004] & 0.34 [0.02] \\ 
  $M$ [deg] & 57 [20] & 78 [27] & 263 [20] & 277 [11] & 107 [35] & 209 [56] \\ 
  $e$  & 0 & 0 & 0 & 0 & 0 & 0.06 [0.04] \\ 
  $\omega$ [deg] & 0 & 0 & 0 & 0 & 0 & 215 [50] \\ 
  $K$ [$\mathrm{m s}^{-1}$] & 1.9 [0.2] & 1.4 [0.2] & 2.3 [0.2] & 4.4 [0.2] & 1.8 [0.2] & 6.1 [0.3] \\ 
  $a$ [AU] & 0.0384740 [$8\times 10^{-7}$] & 0.064816 [$4\times 10^{-6}$] & 0.14574 [$2\times 10^{-5}$] & 0.23508 [$4\times 10^{-5}$] & 0.3753 [0.0004] & 3.11 [0.04] \\ 
  $T_\s{peri}$ [JD] & 2449999.5 [0.2] & 2449998.5 [0.5] & 2449983 [1] & 2449964 [1] & 2449972 [9] & 2448725 [356] \\ 
\hline
  Q01 $s_\s{noise}$ [$\mathrm{m s}^{-1}$] & 1.1 [0.2] &  &  &  &  &  \\ 
  KECK $s_\s{noise}$ [$\mathrm{m s}^{-1}$] & 2.5 [0.2] &  &  &  &  &  \\ 
  APF $s_\s{noise}$ [$\mathrm{m s}^{-1}$] & 1.8 [0.2] &  &  &  &  &  \\ 
  Q01 $\Delta v$ [$\mathrm{m s}^{-1}$] & -0.9 [0.6] &  &  &  &  &  \\ 
  KECK $\Delta v$ [$\mathrm{m s}^{-1}$] & -0.8 [0.2] &  &  &  &  &  \\ 
  APF $\Delta v$ [$\mathrm{m s}^{-1}$] & -2.3 [0.6] &  &  &  &  &  \\ 
\hline
  $\mass_\s{\star}$ [$\mass_\odot$] & 0.794 &  &  &  &  &  \\ 
  $\chi^2$  & 280.407 &  &  &  &  &  \\ 
  $-\log\mathcal{L}$  & 593.311 &  &  &  &  &  \\ 
  RMS [$\mathrm{m s}^{-1}$] & 2.223 &  &  &  &  &  \\ 
  $\sigma_\s{\star, Jitter}$ [$\mathrm{m s}^{-1}$] & 2.038 &  &  &  &  &  \\ 
  Epoch [JD] & 2450000 &  &  &  &  &  \\ 
  Data points  & 276 &  &  &  &  &  \\ 
   \hline
\end{tabular}
\caption{Best-fit 6-Keplerian Model for HD 219134} 
\end{table*}

\begin{figure*}
\centering
\plotone{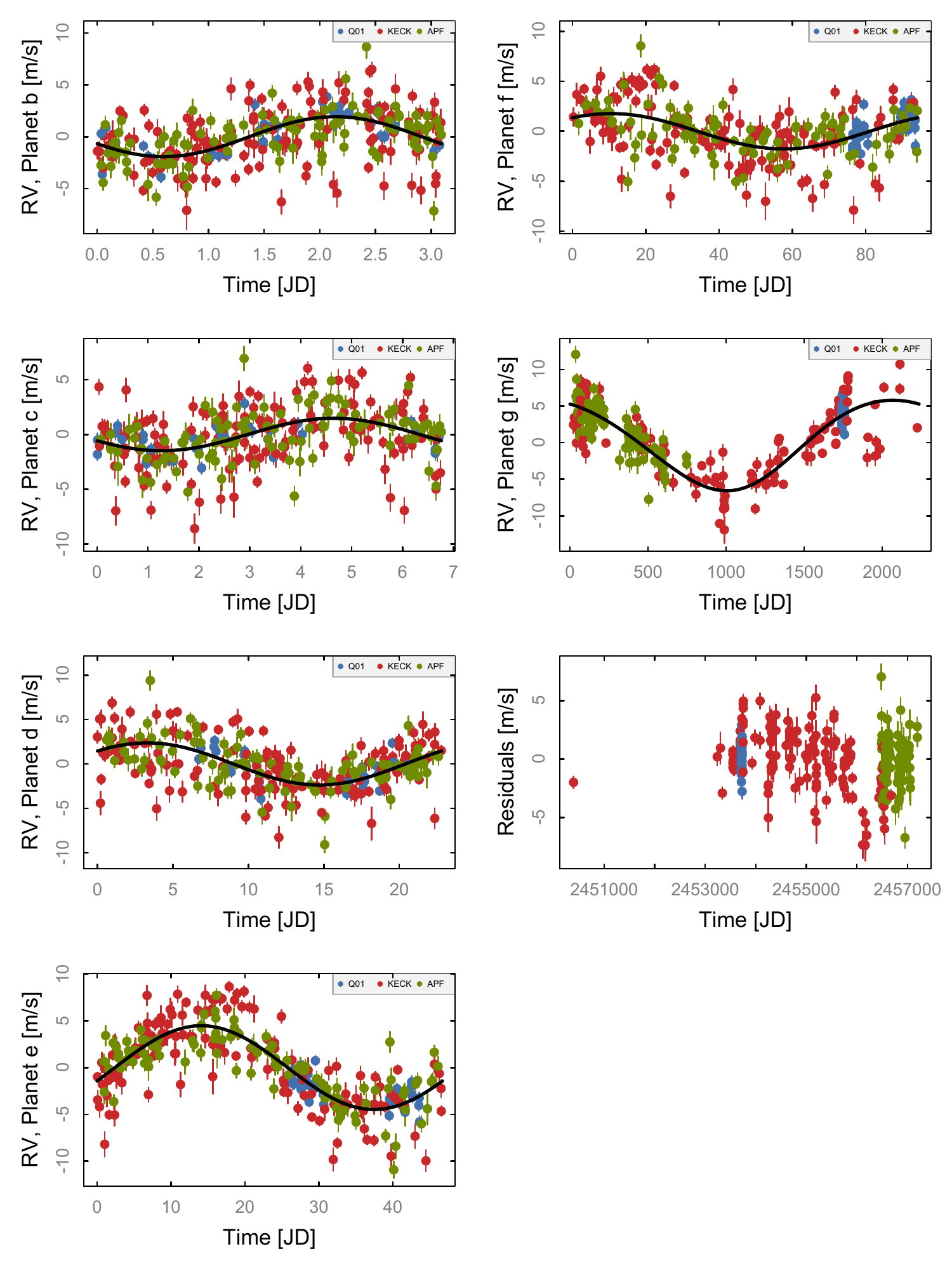}\\
\caption{\label{fig:fit} Best-fit Keplerian model.}
 \label{tab:elements}
\end{figure*}

\begin{figure*}
\centering
\epsscale{0.2}
\begin{framed}
\plotone{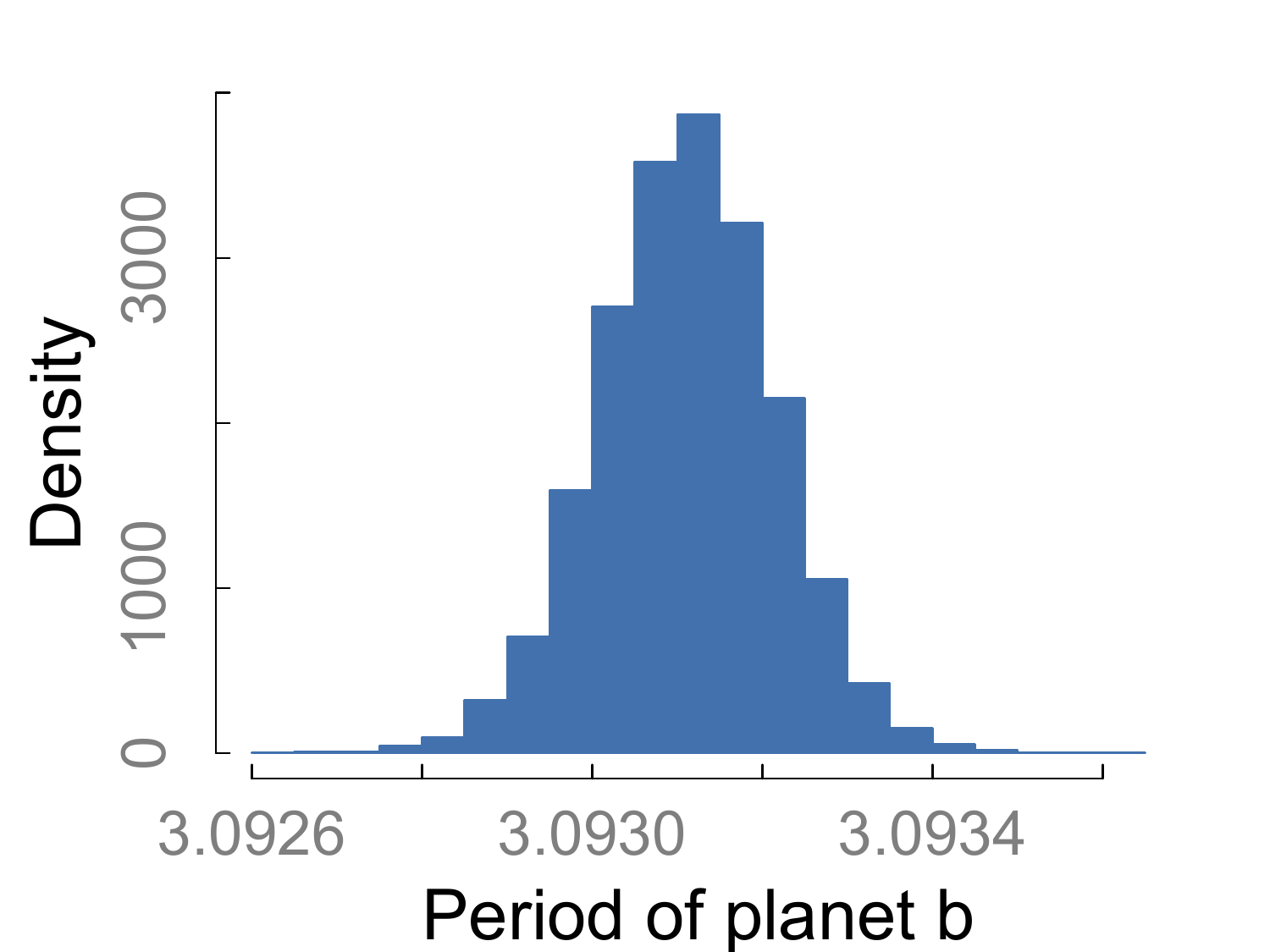}
\plotone{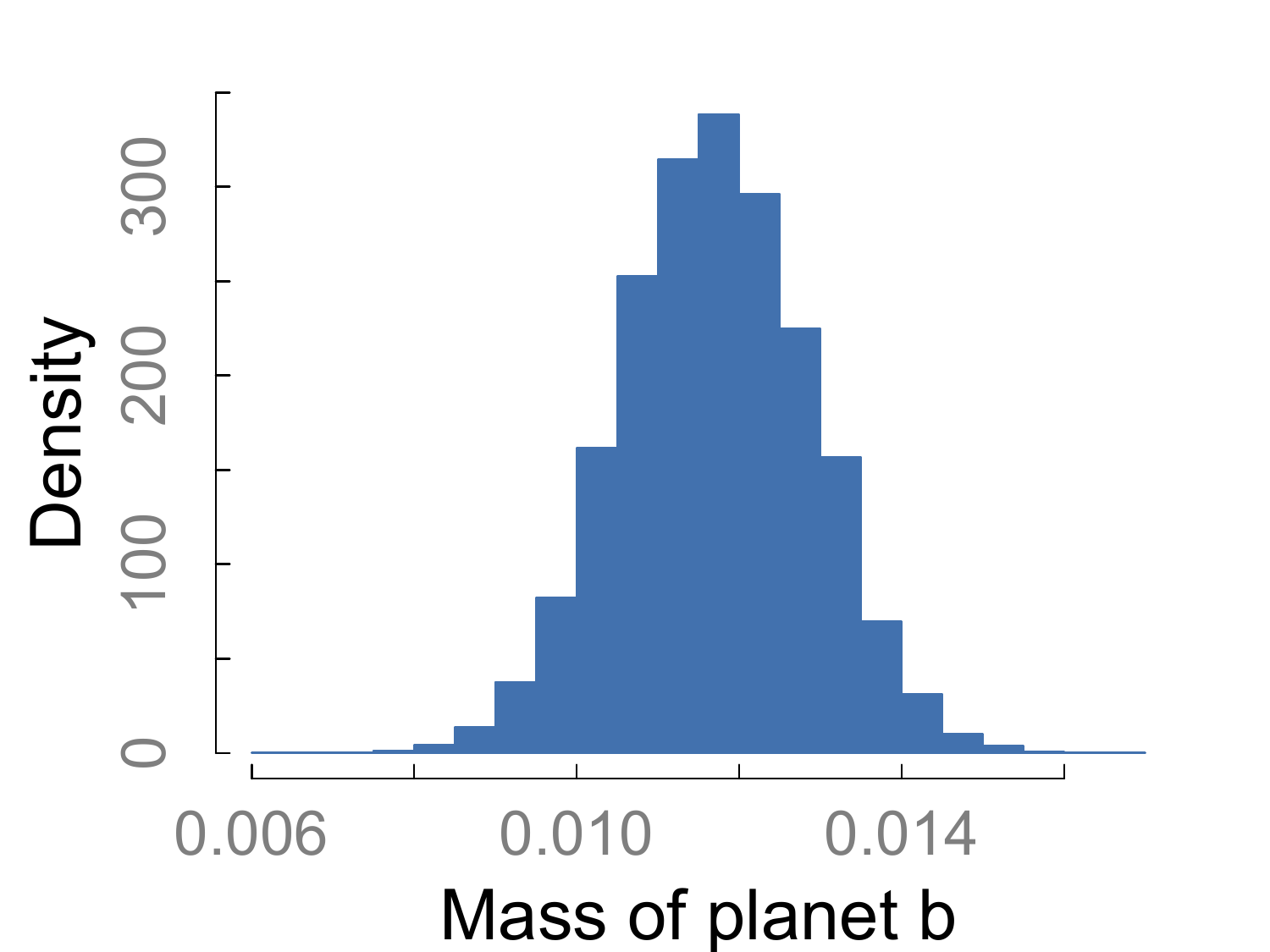}
\plotone{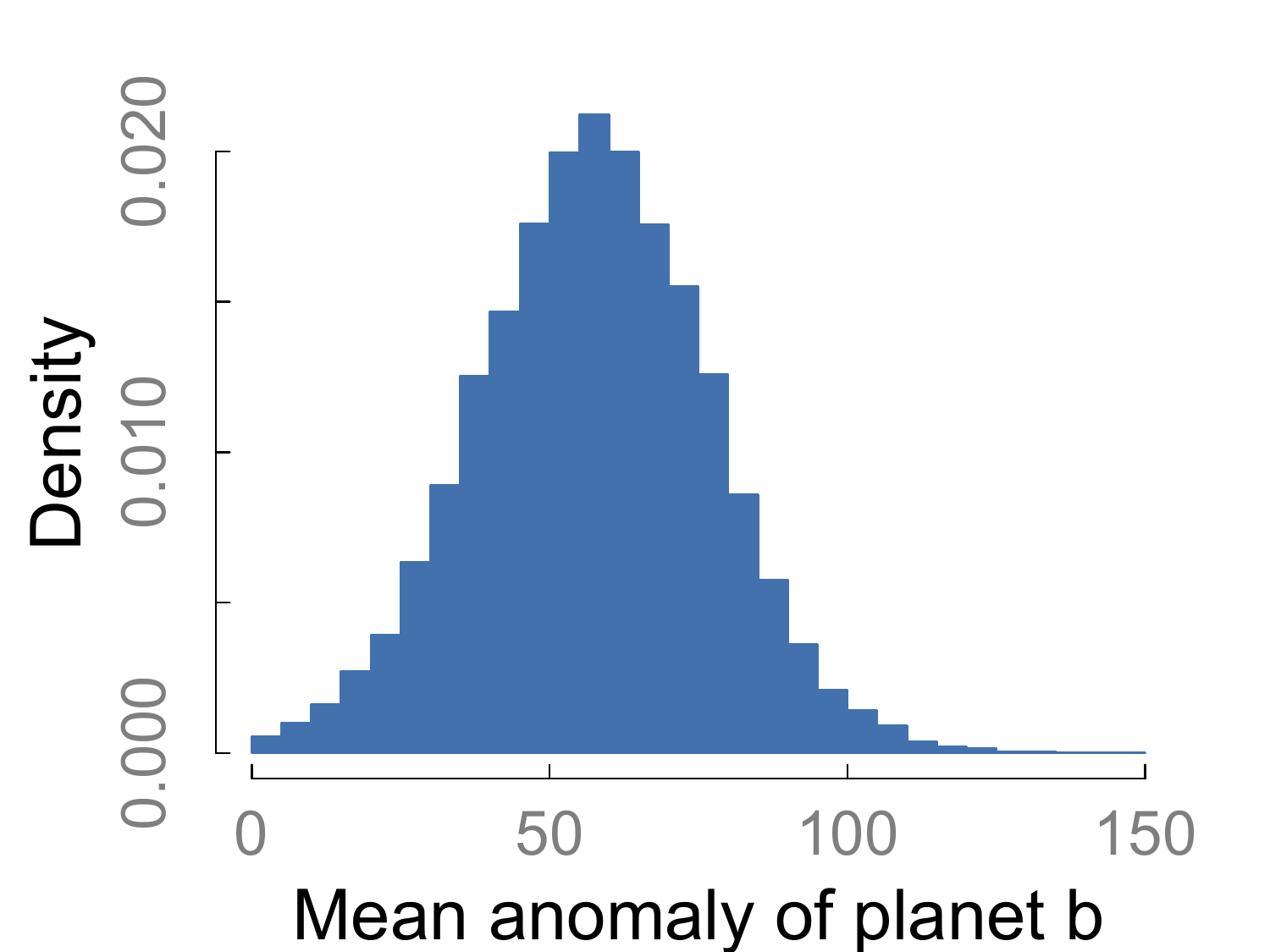}
\plotone{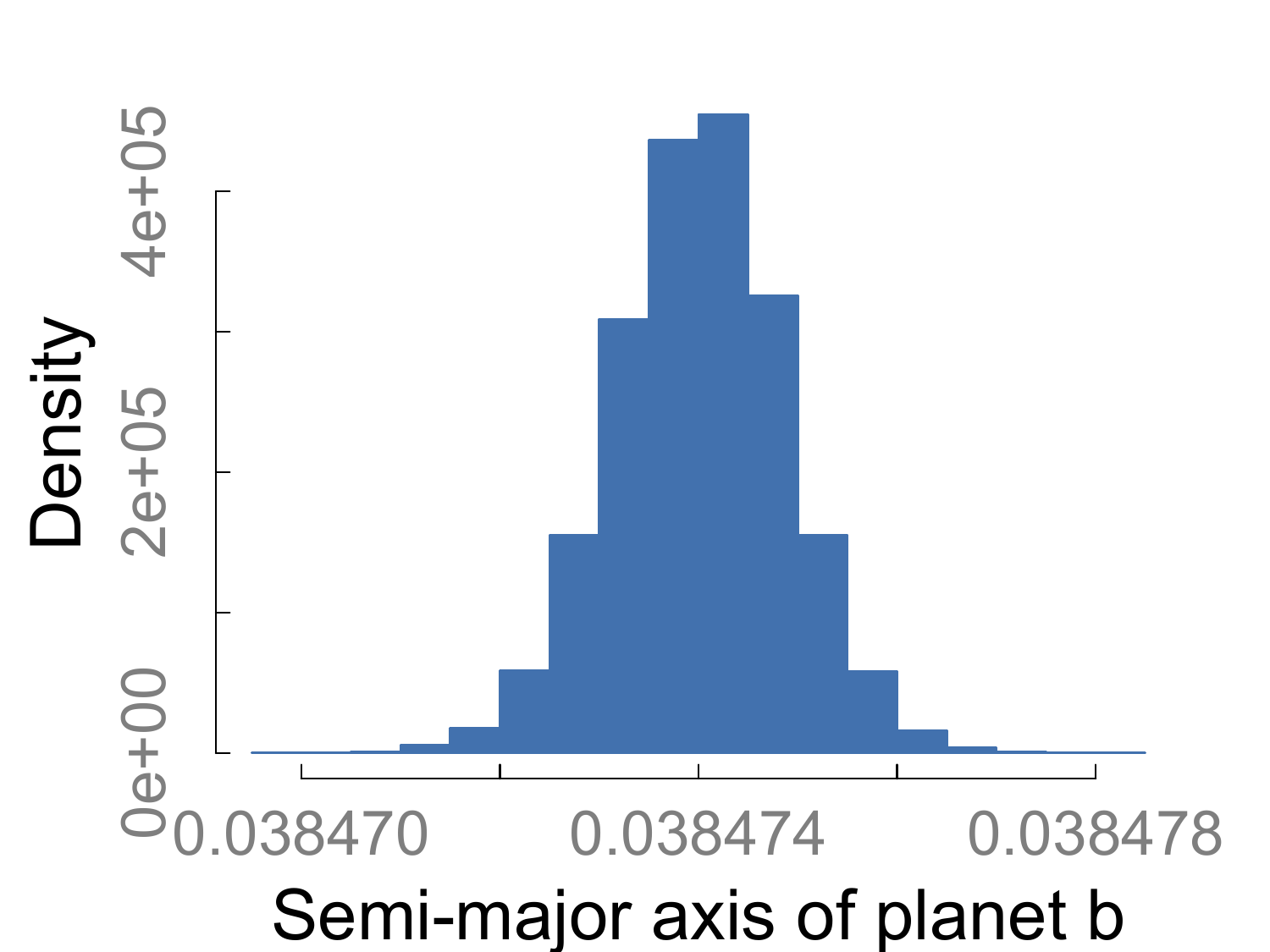}
\plotone{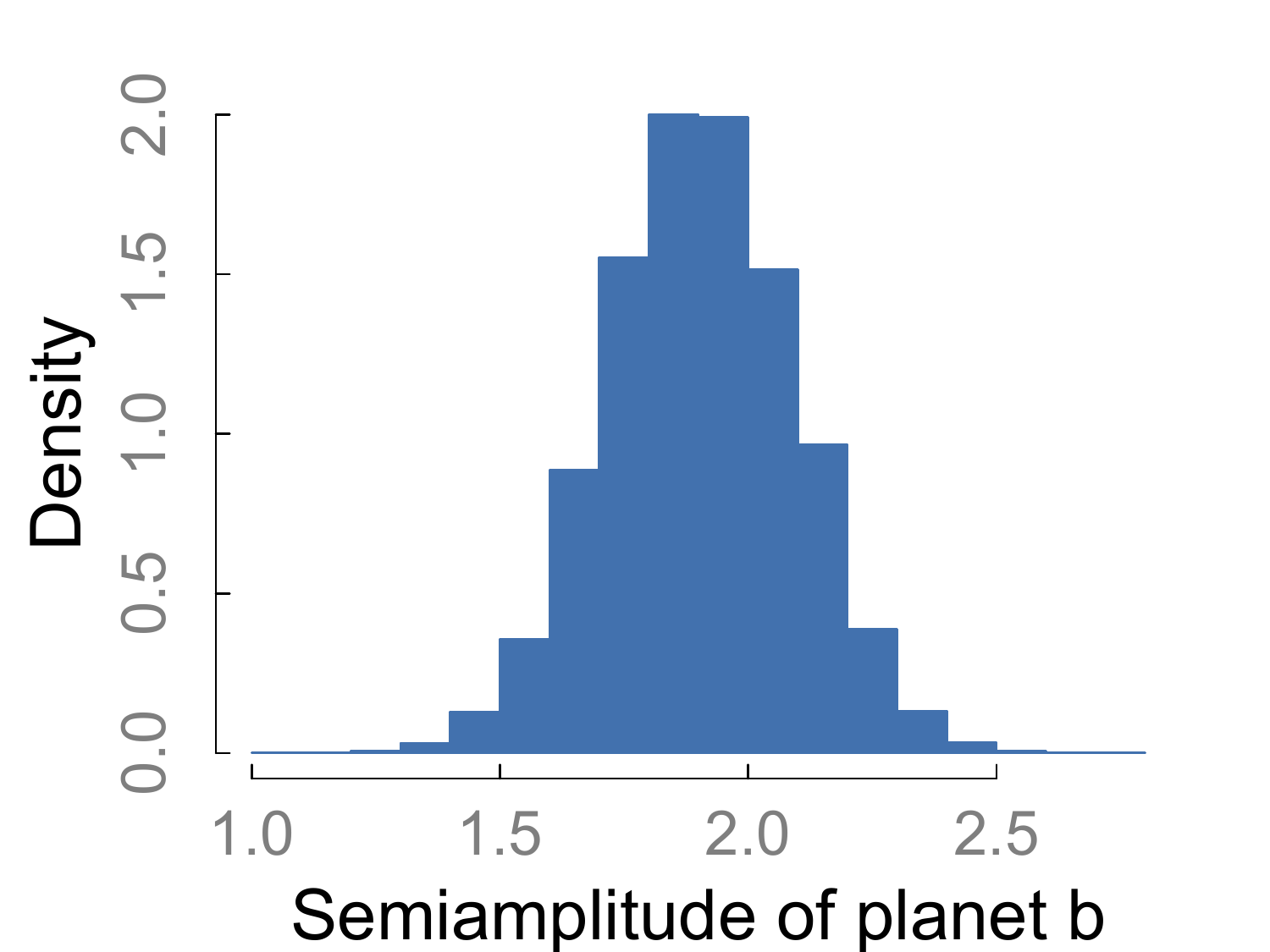}
\plotone{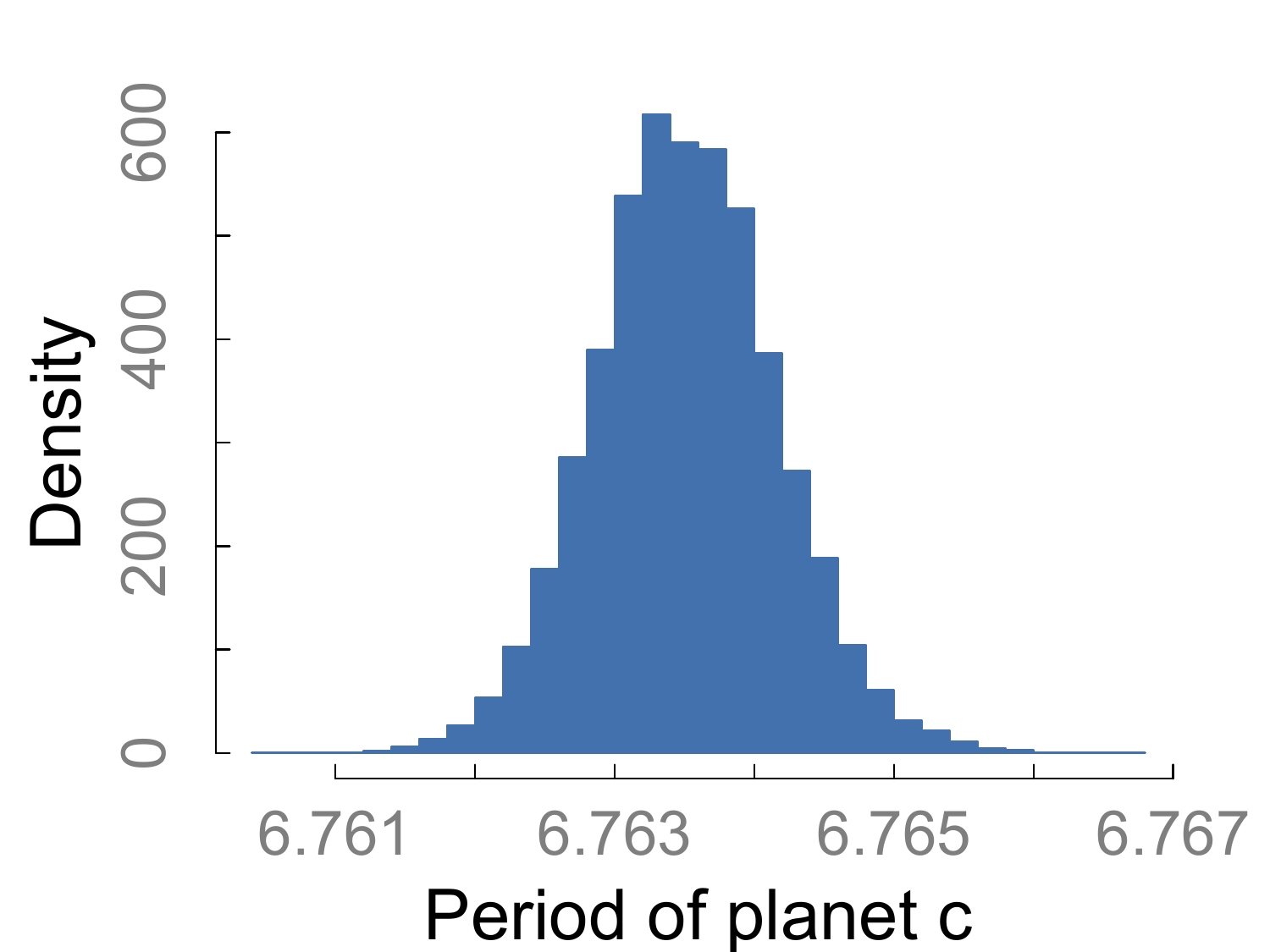}
\plotone{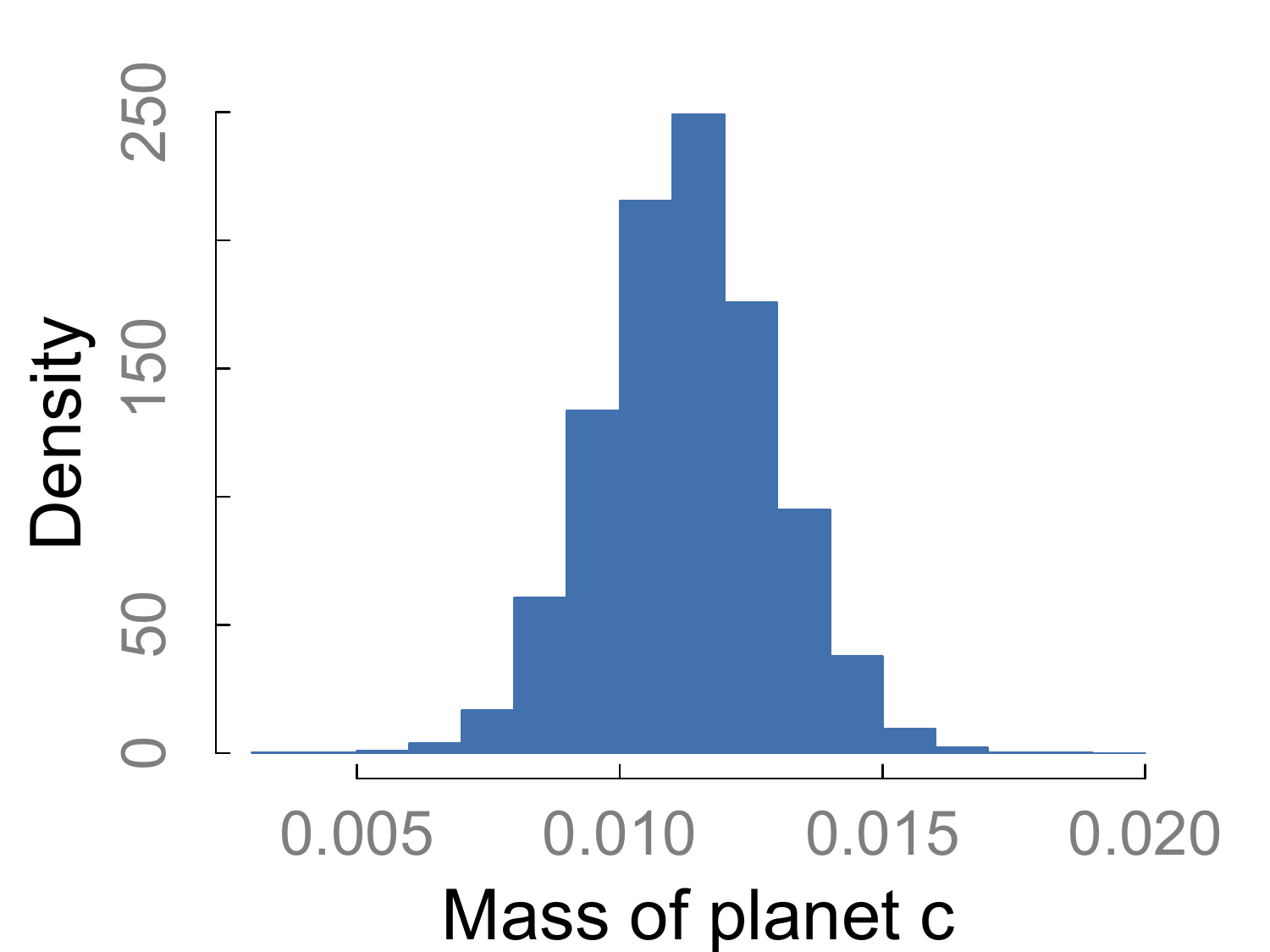}
\plotone{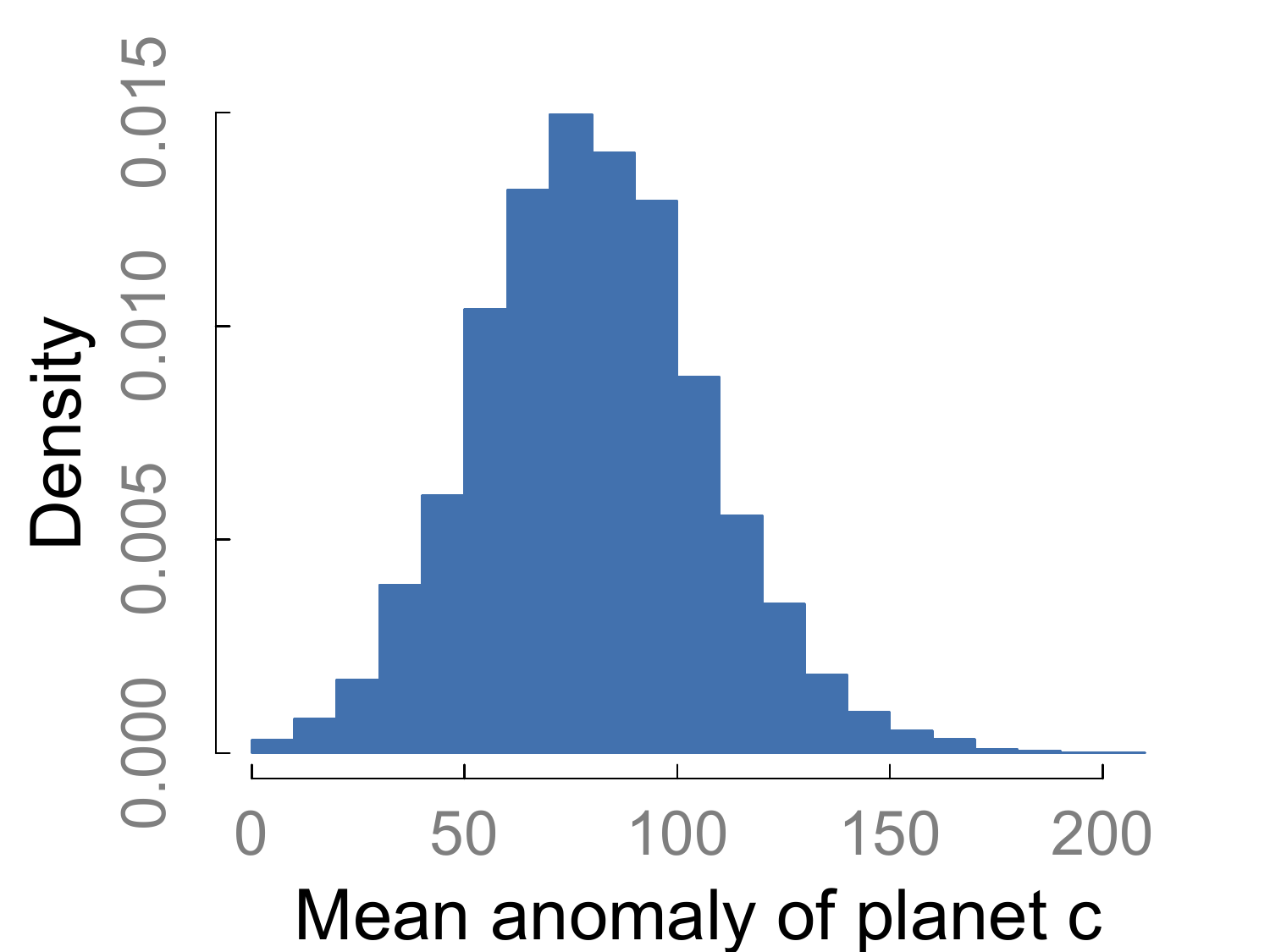}
\plotone{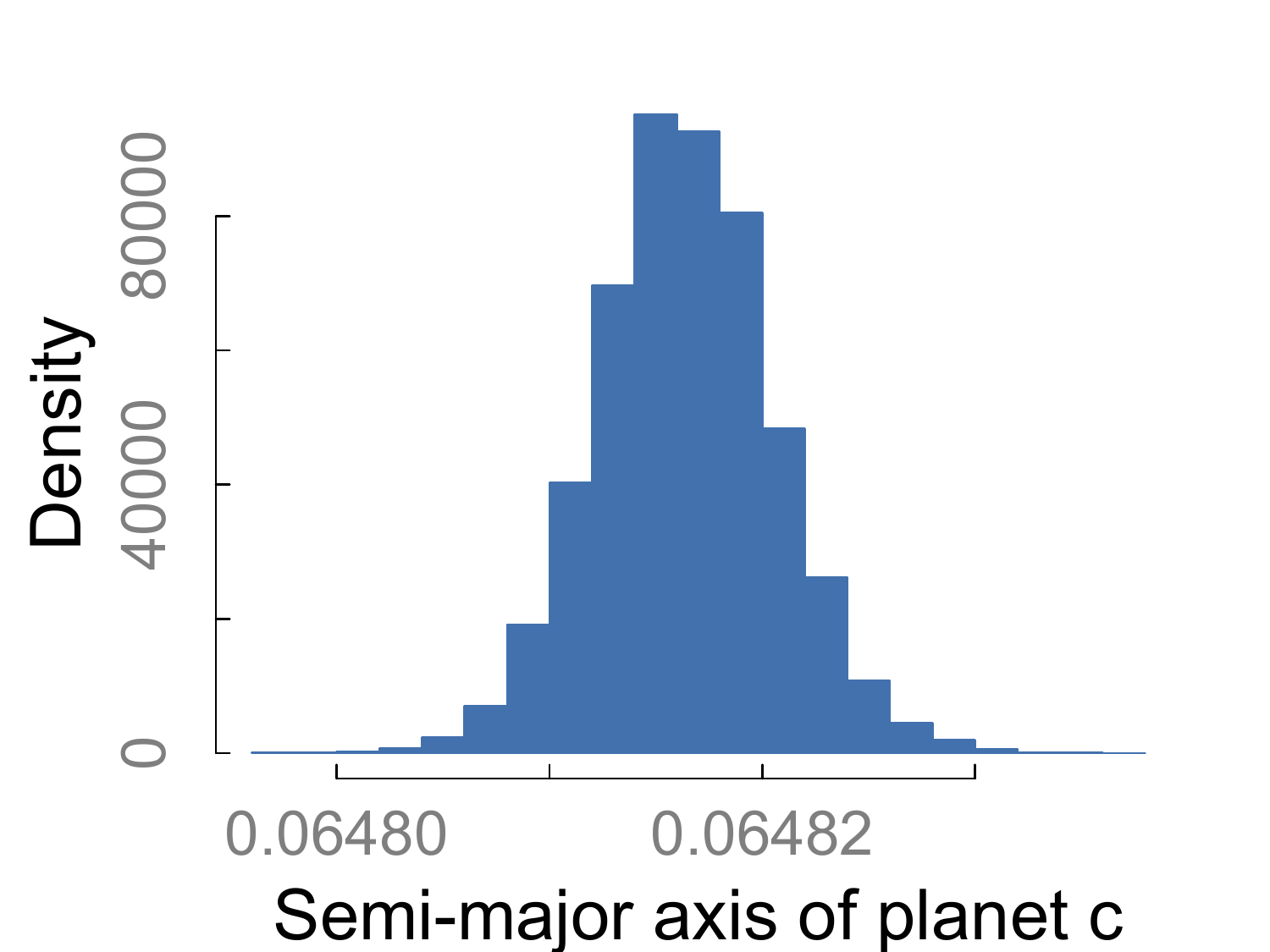}
\plotone{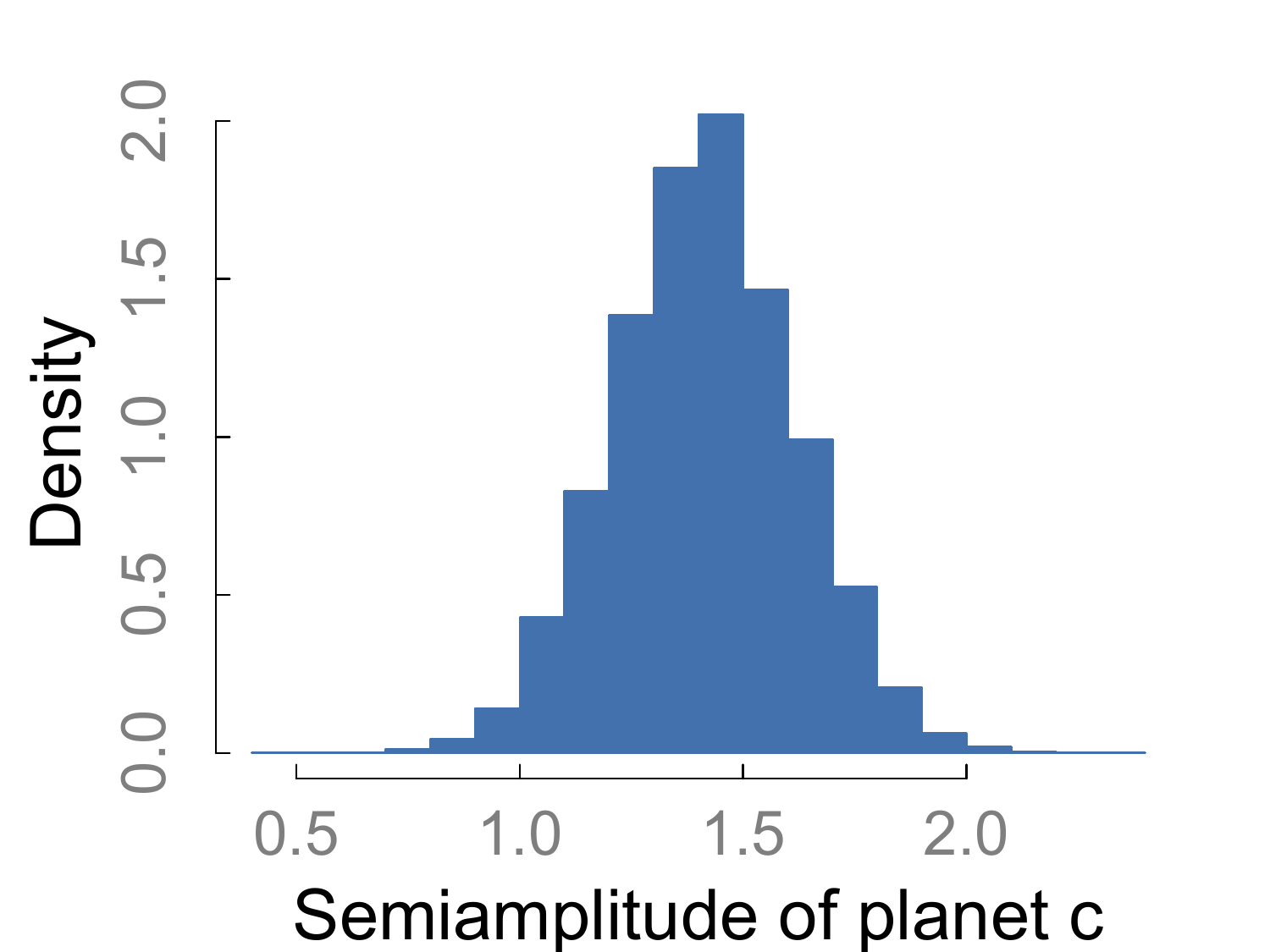}
\plotone{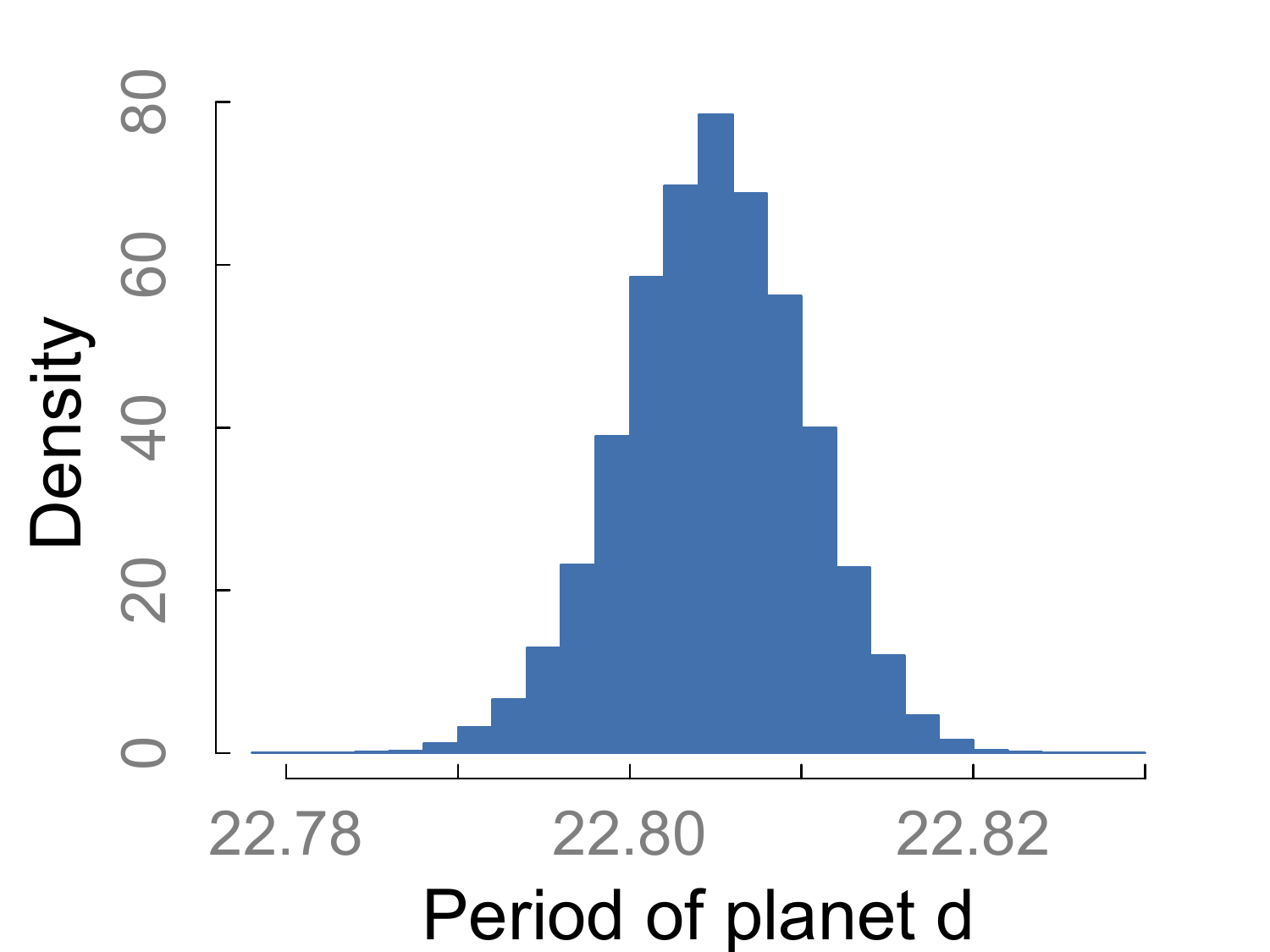}
\plotone{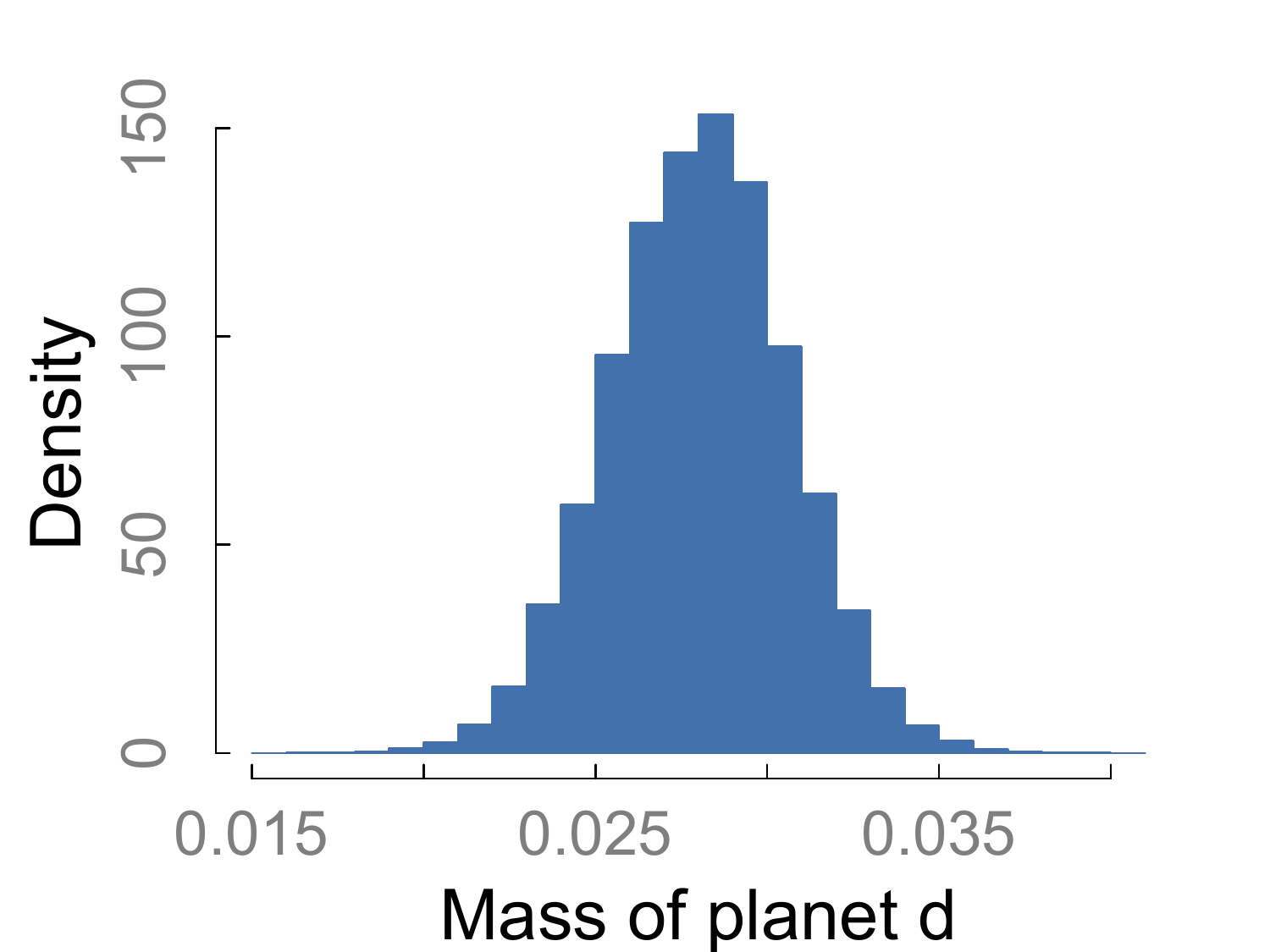}
\plotone{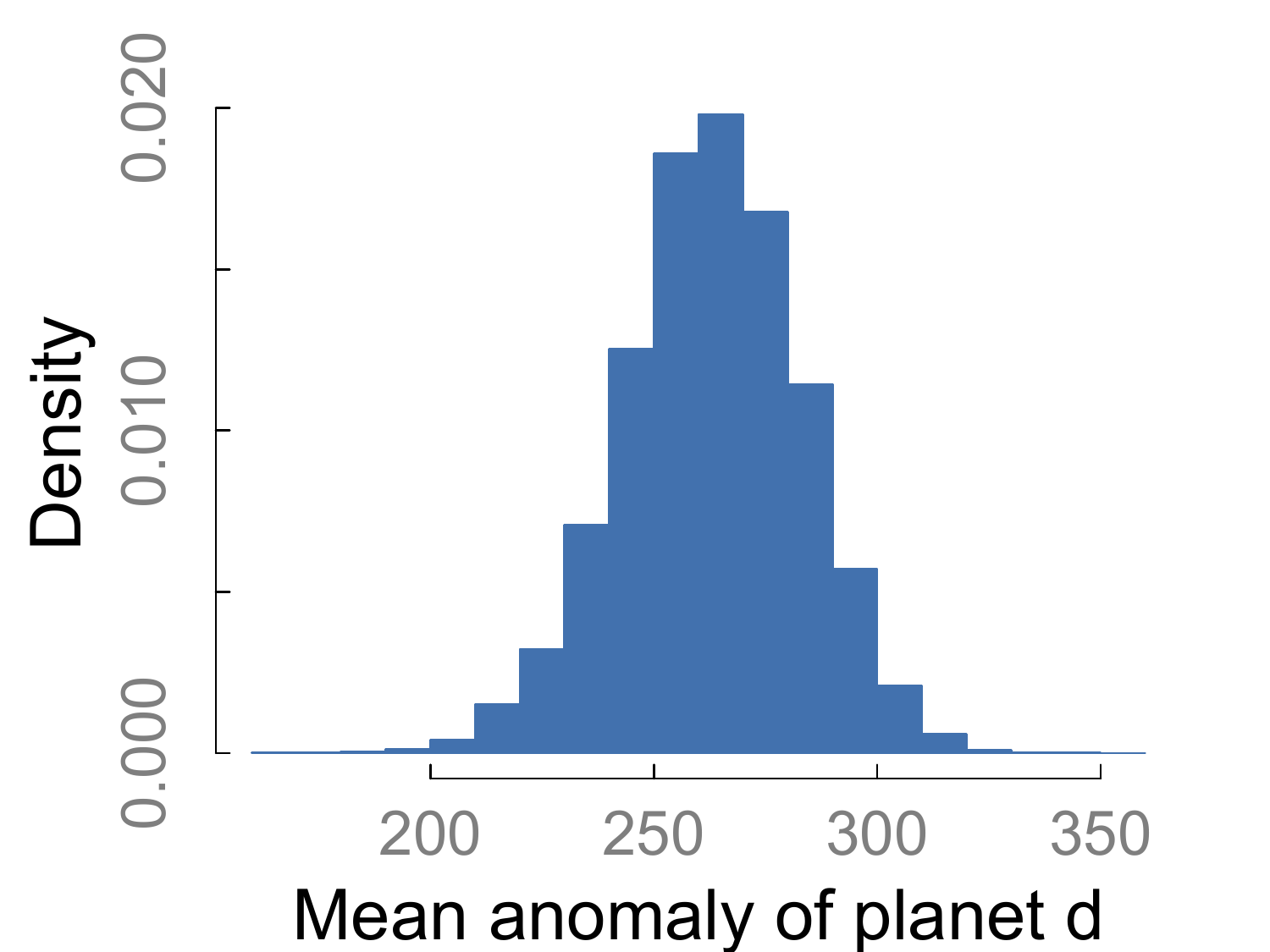}
\plotone{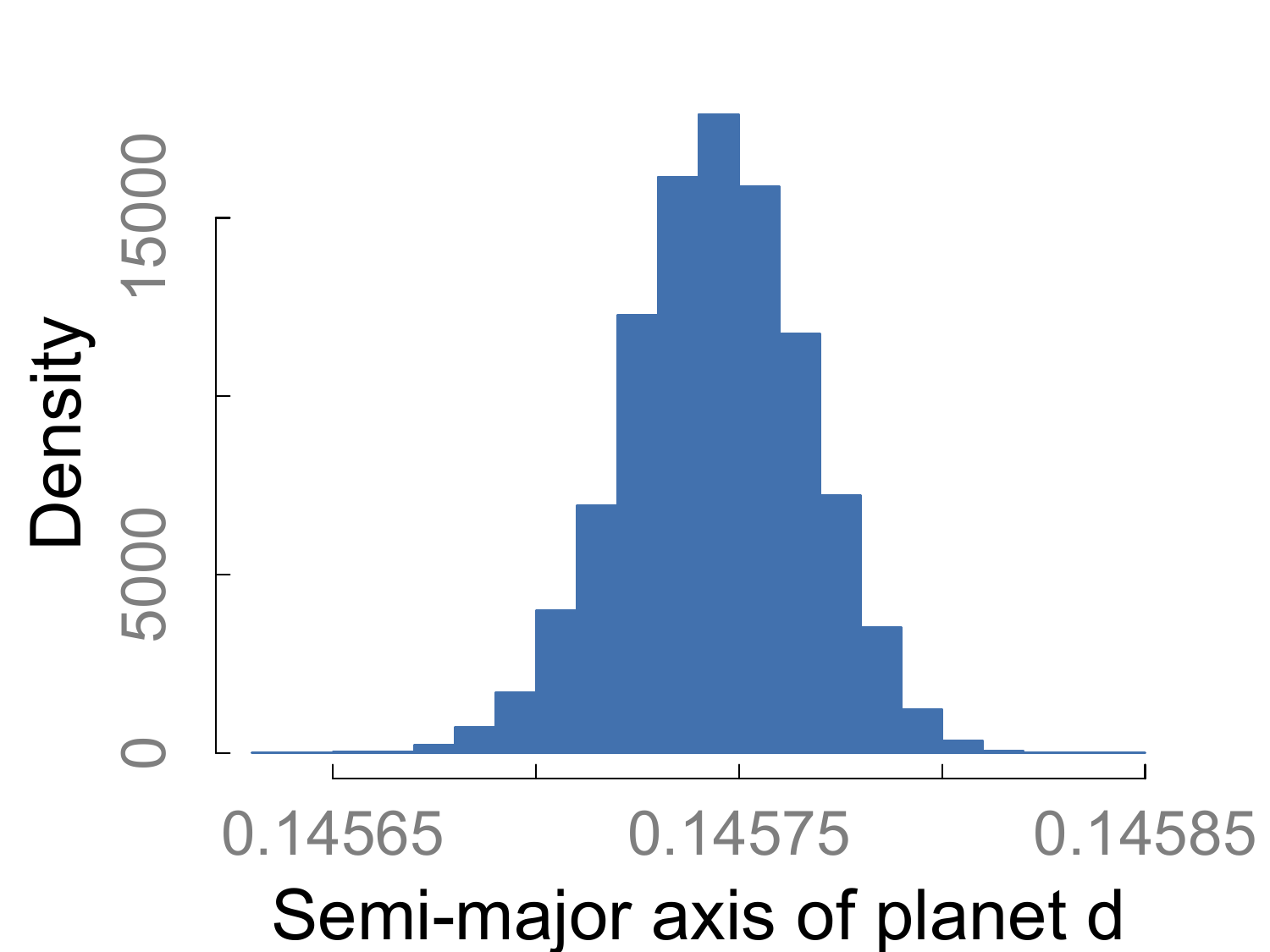}
\plotone{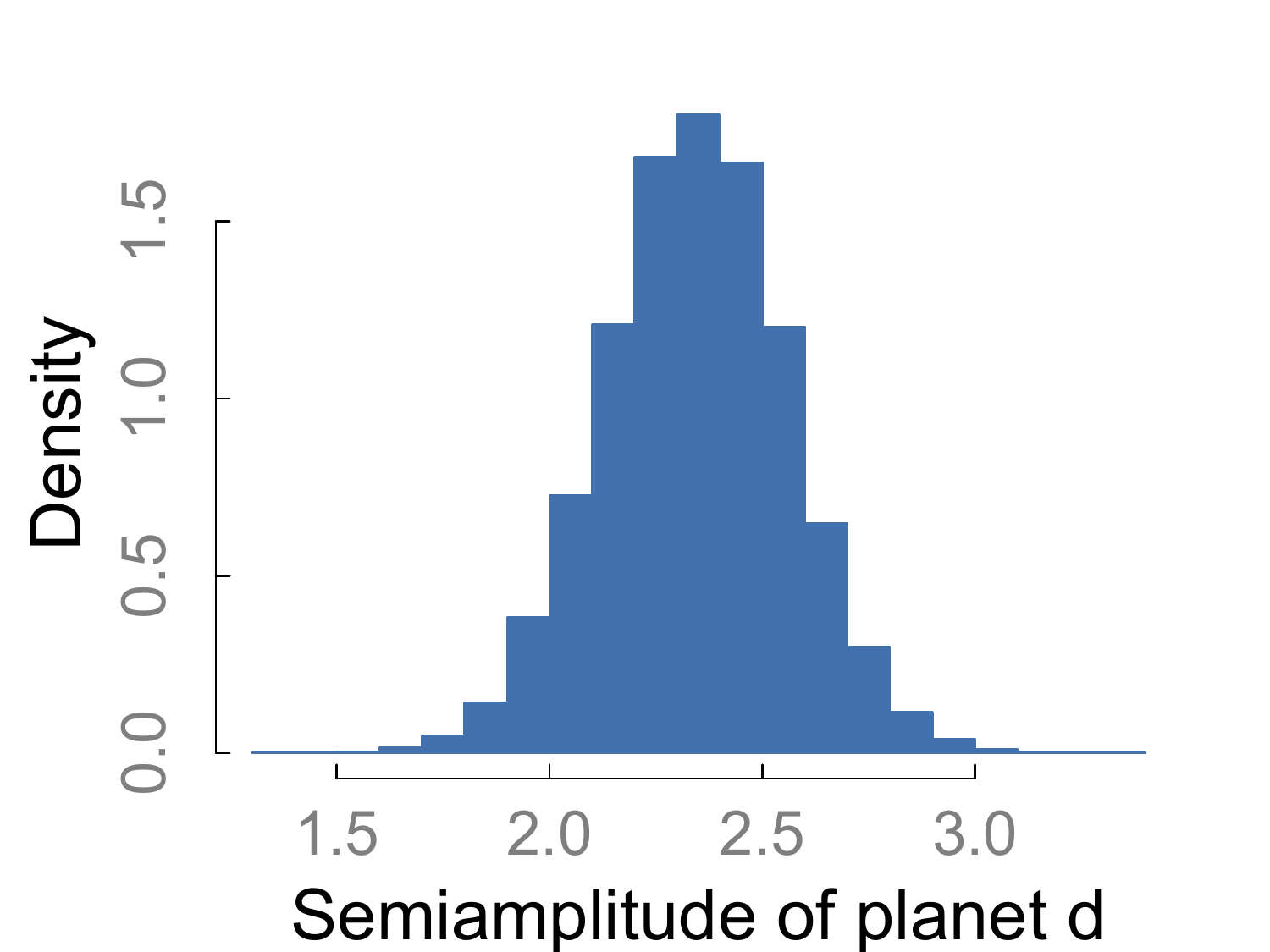}
\plotone{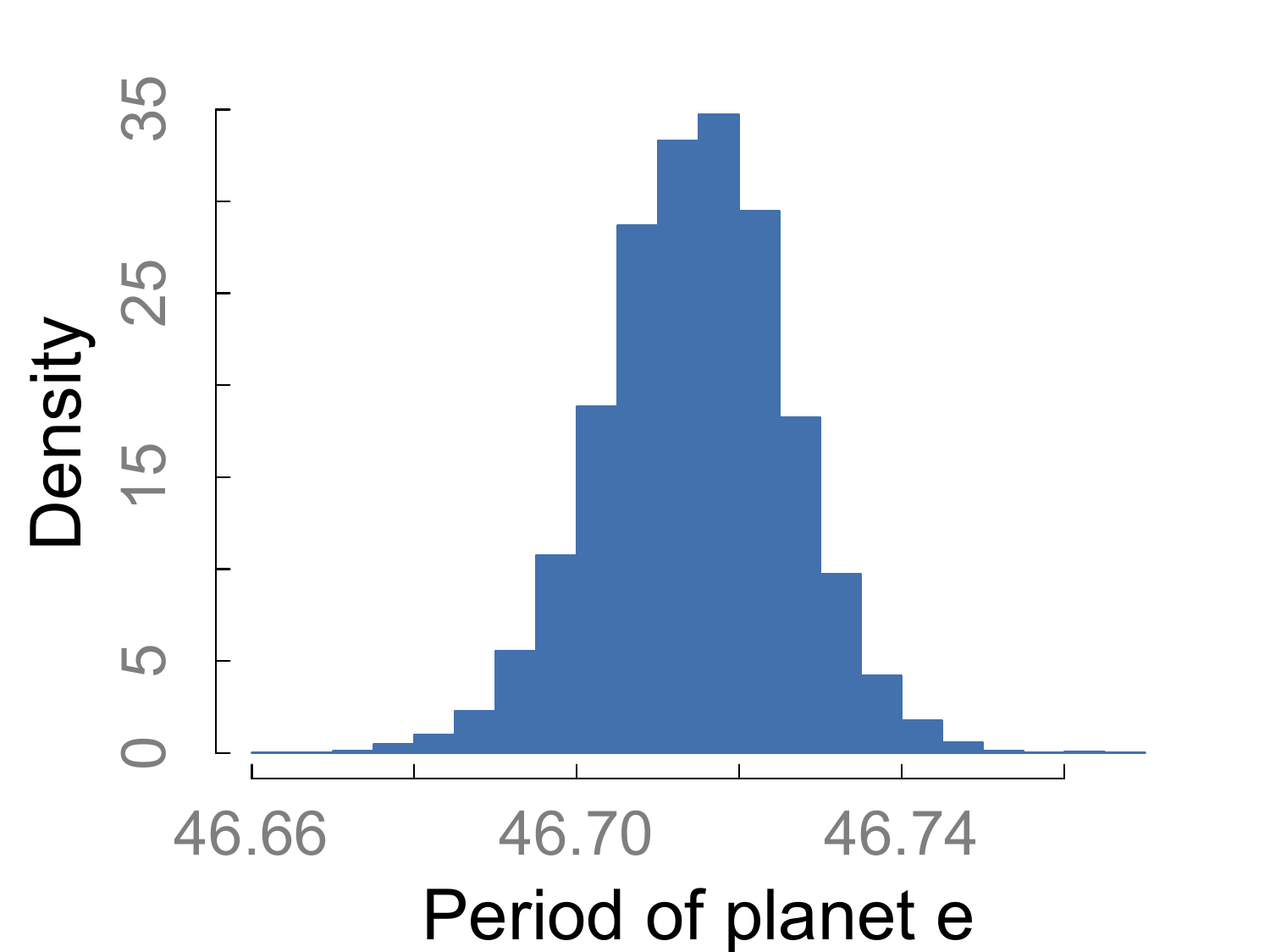}
\plotone{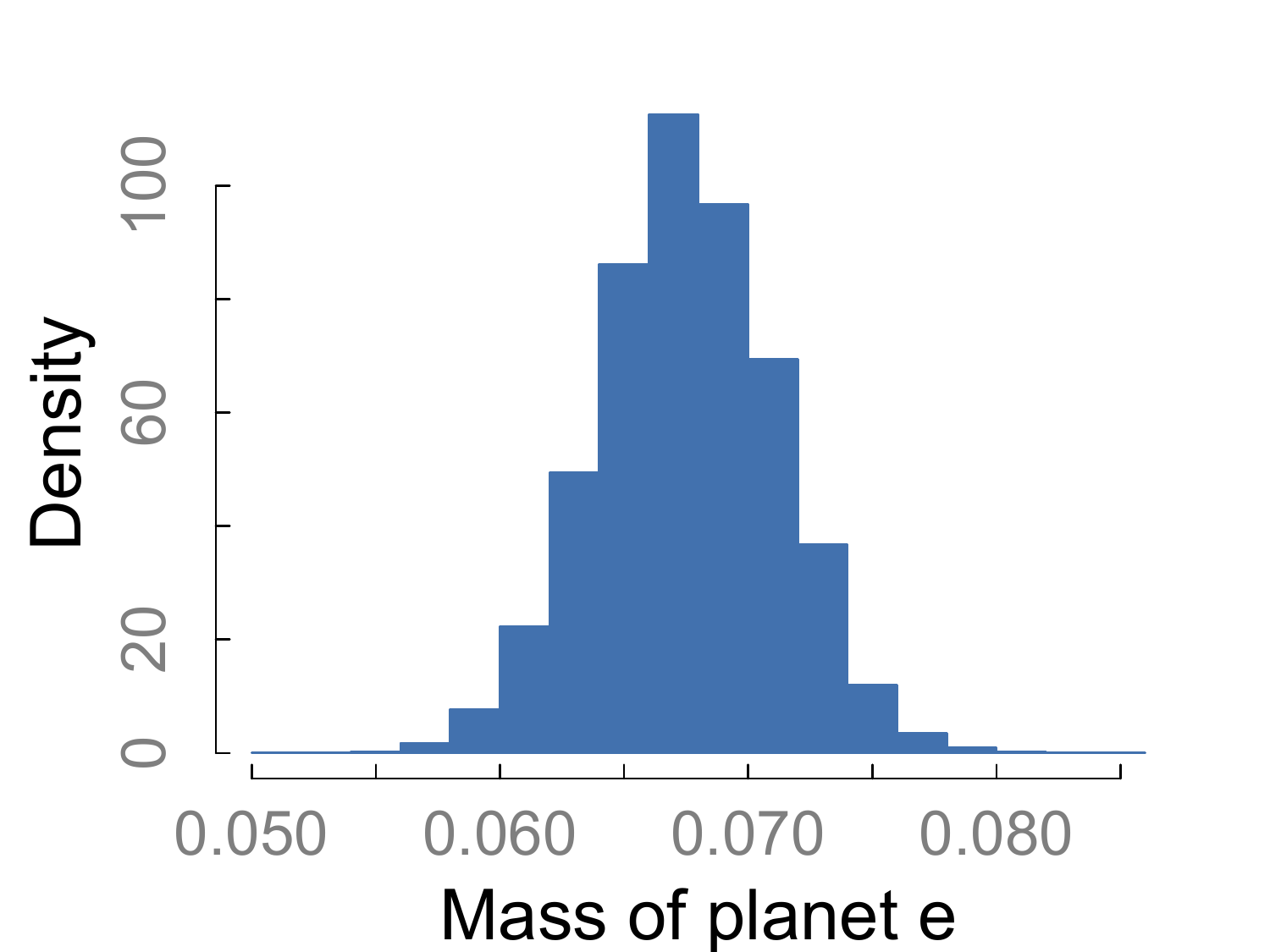}
\plotone{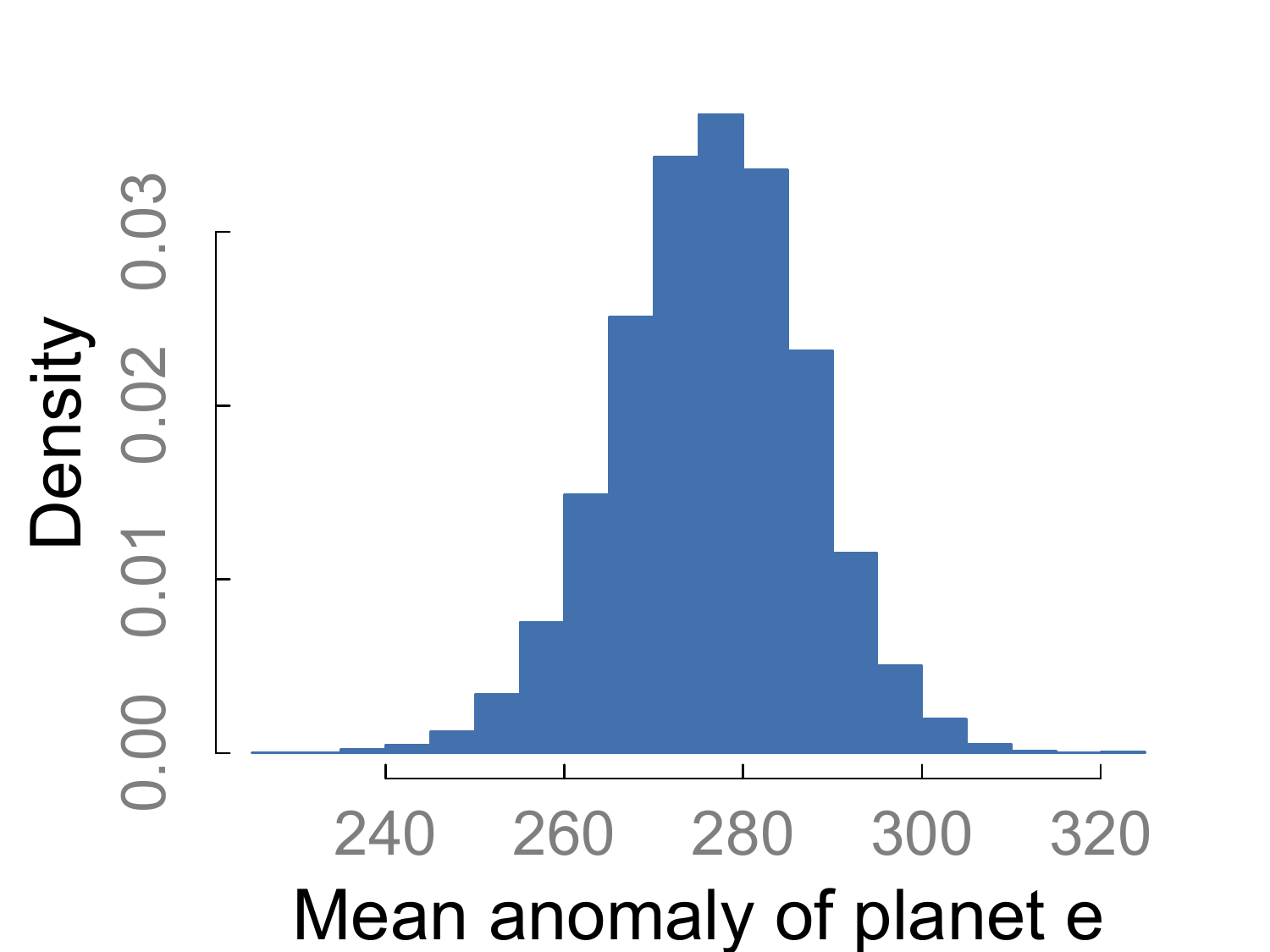}
\plotone{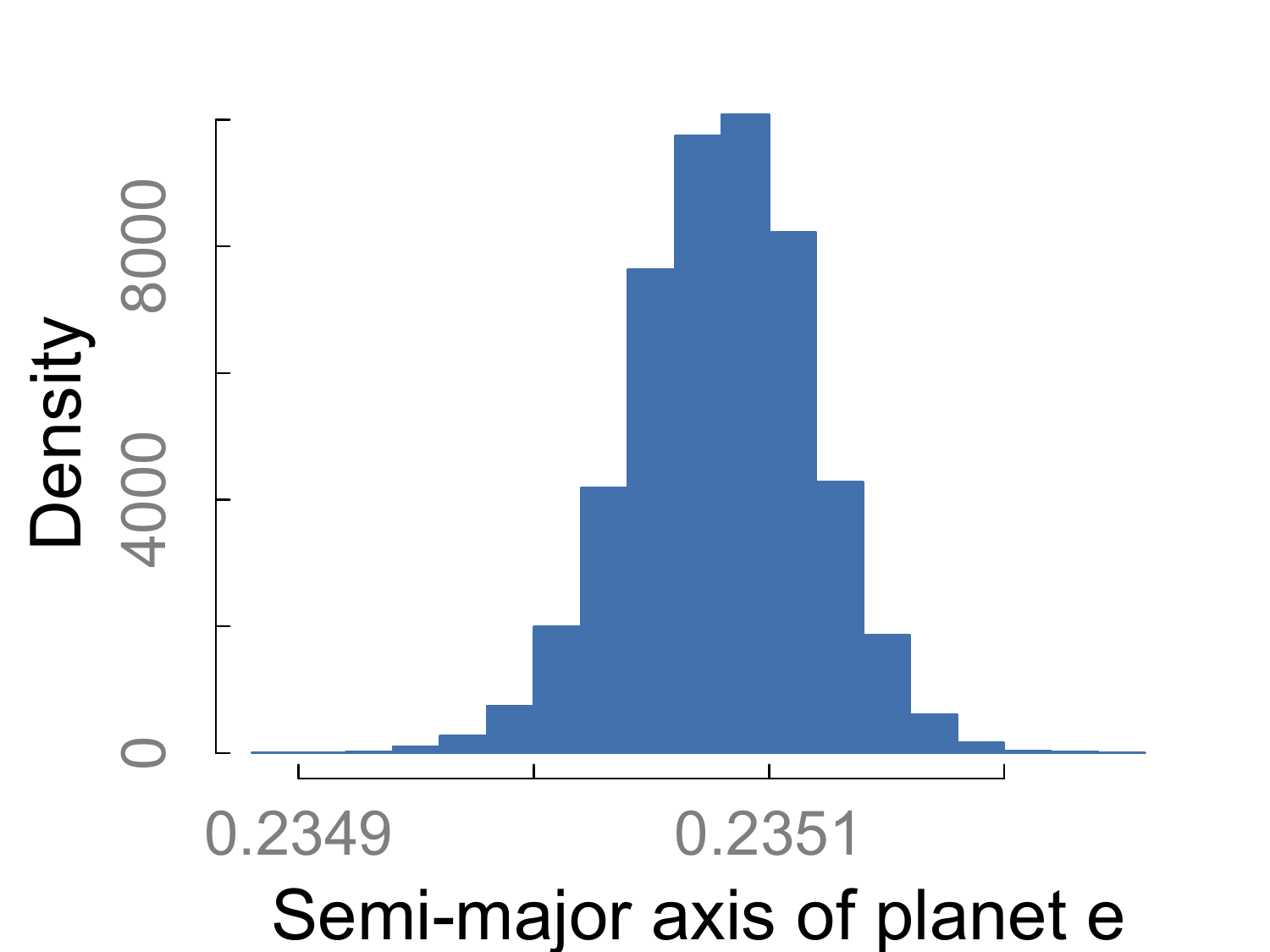}
\plotone{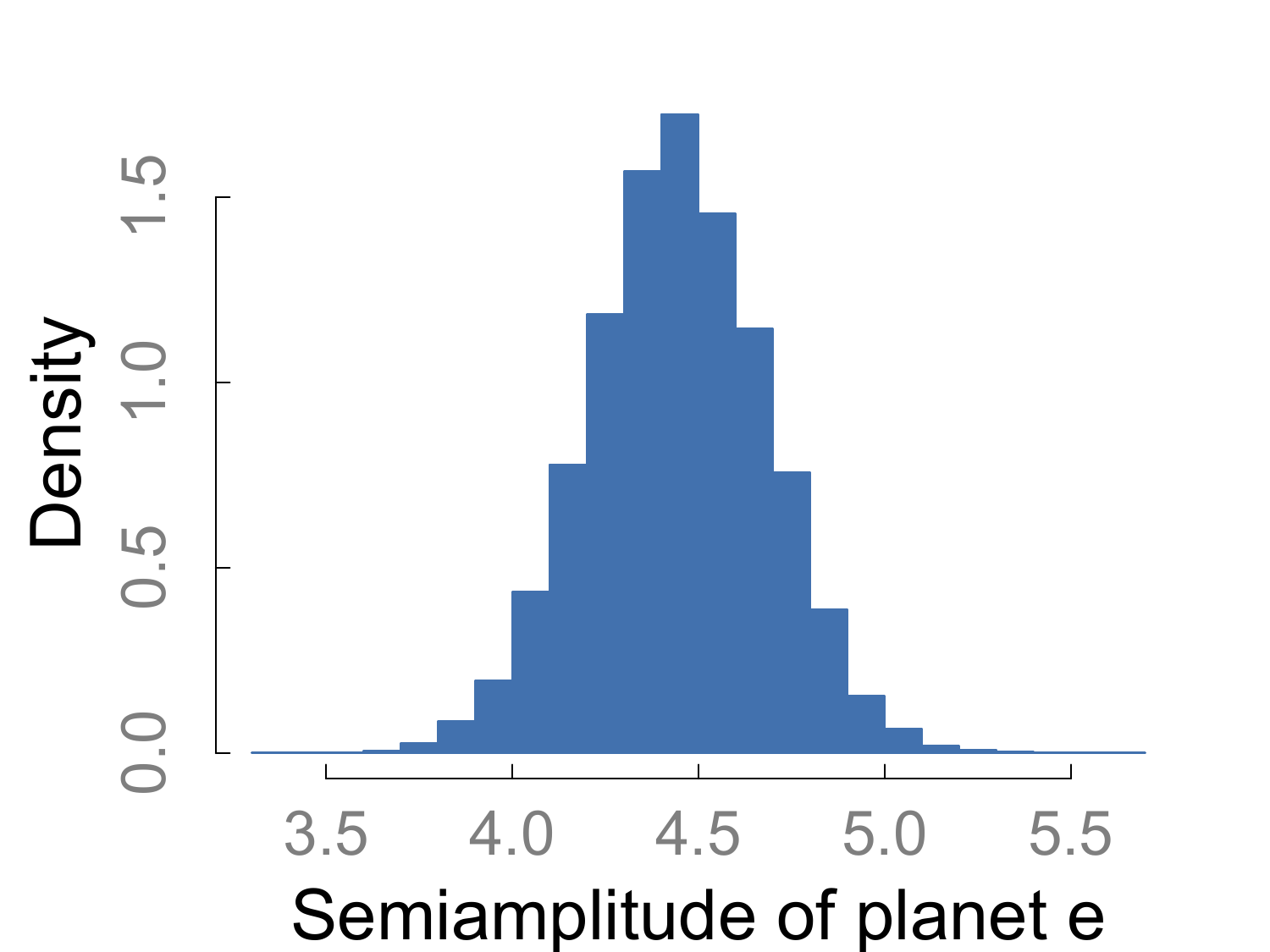}
\plotone{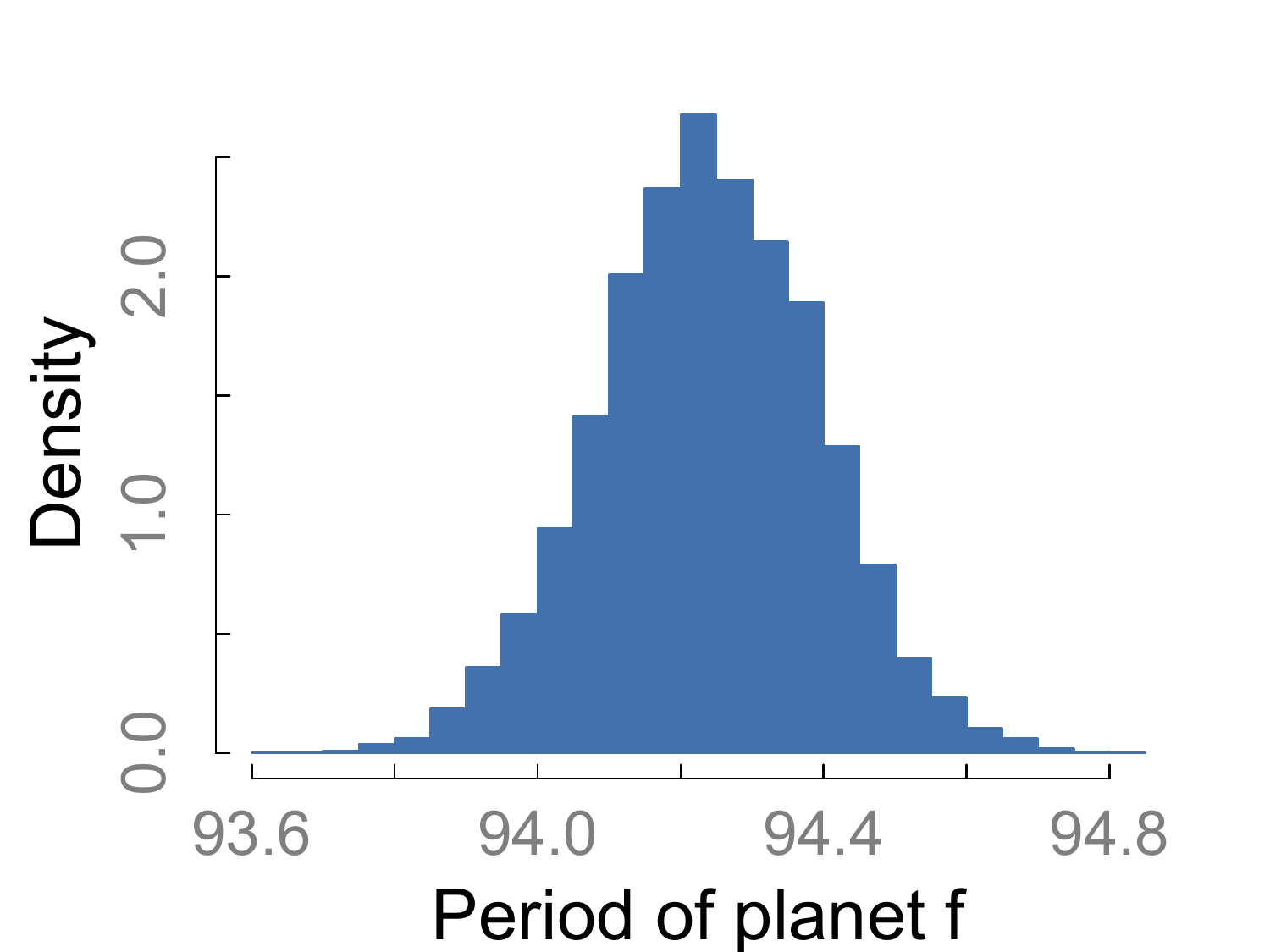}
\plotone{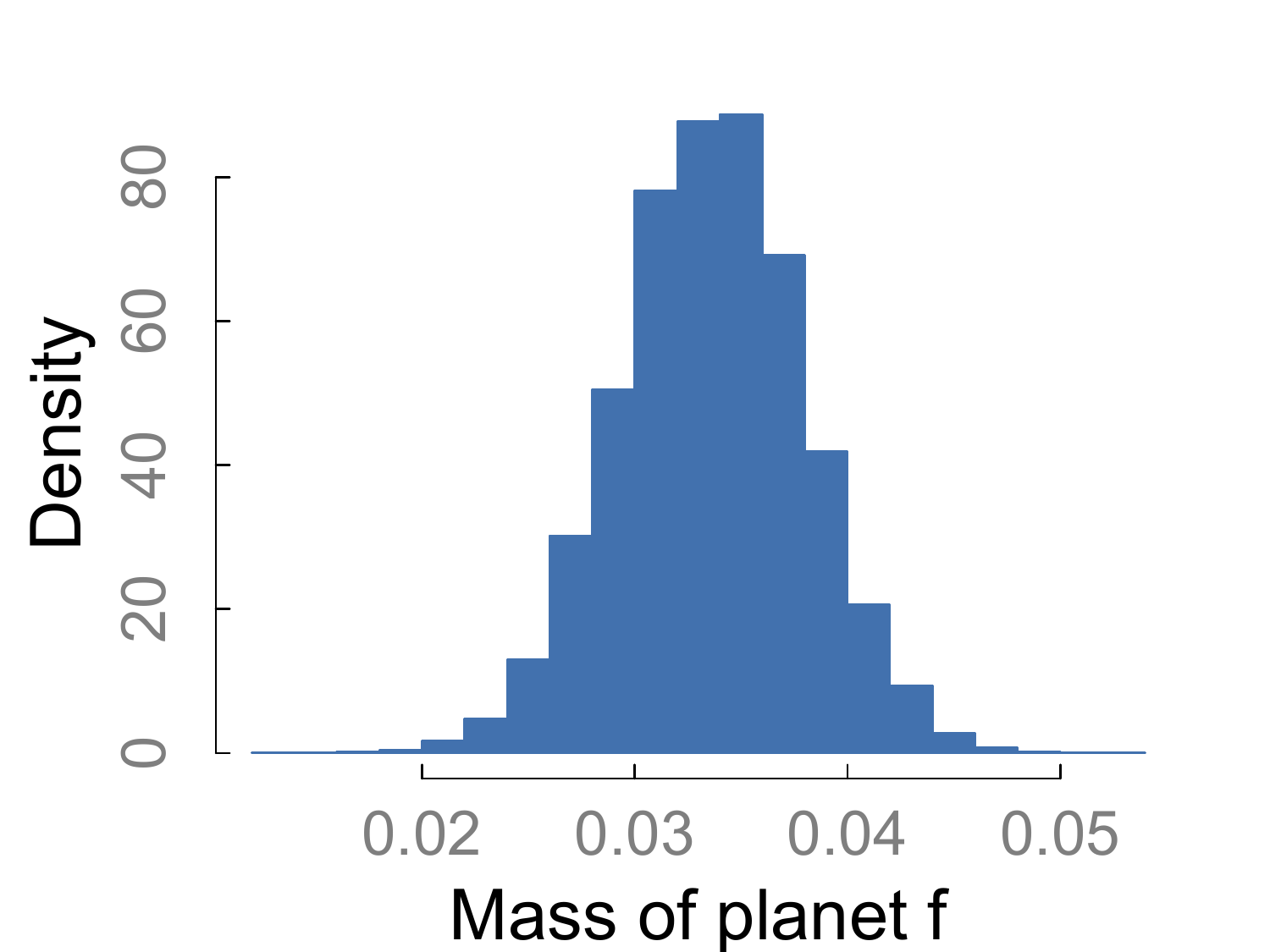}
\plotone{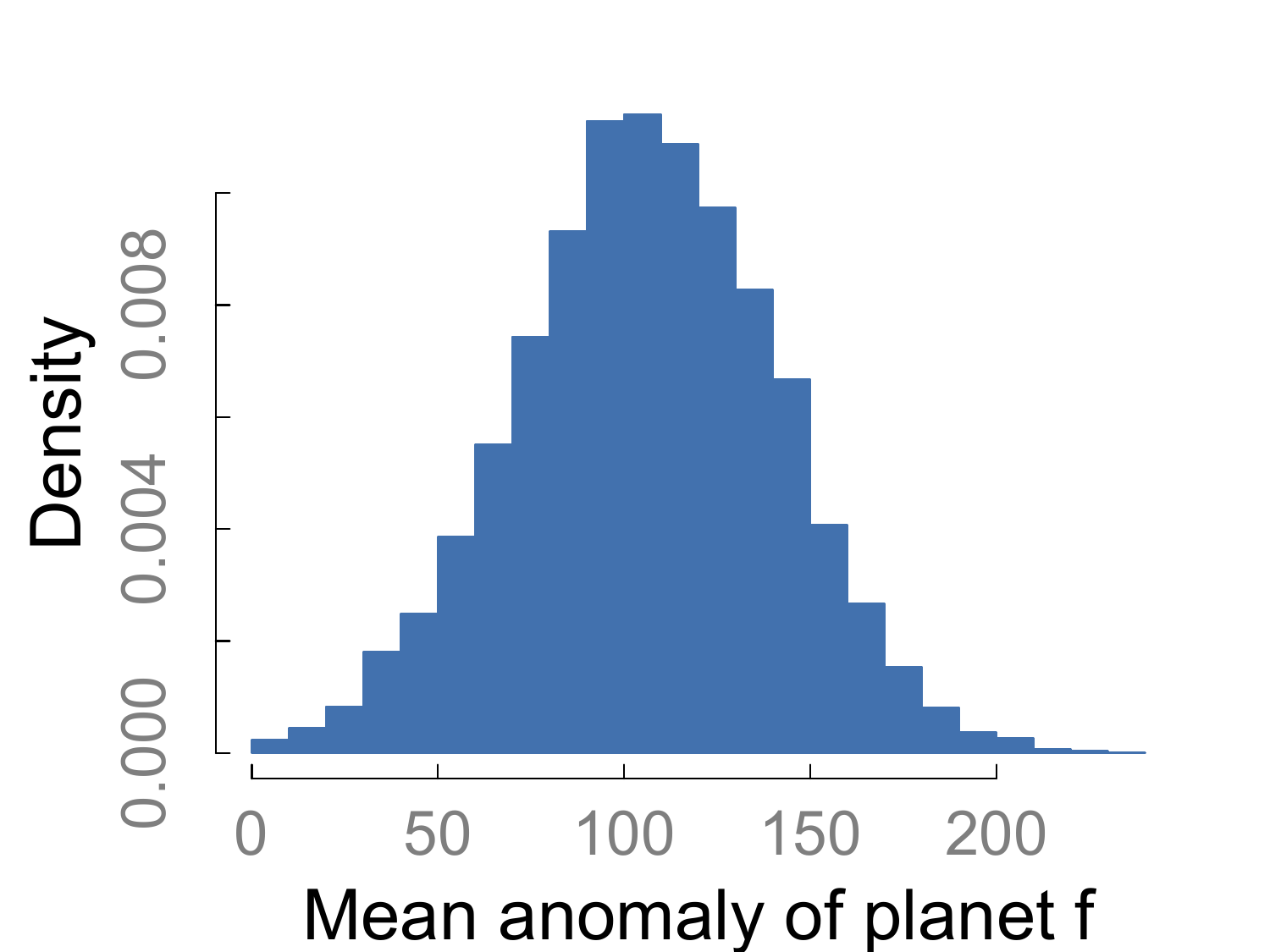}
\plotone{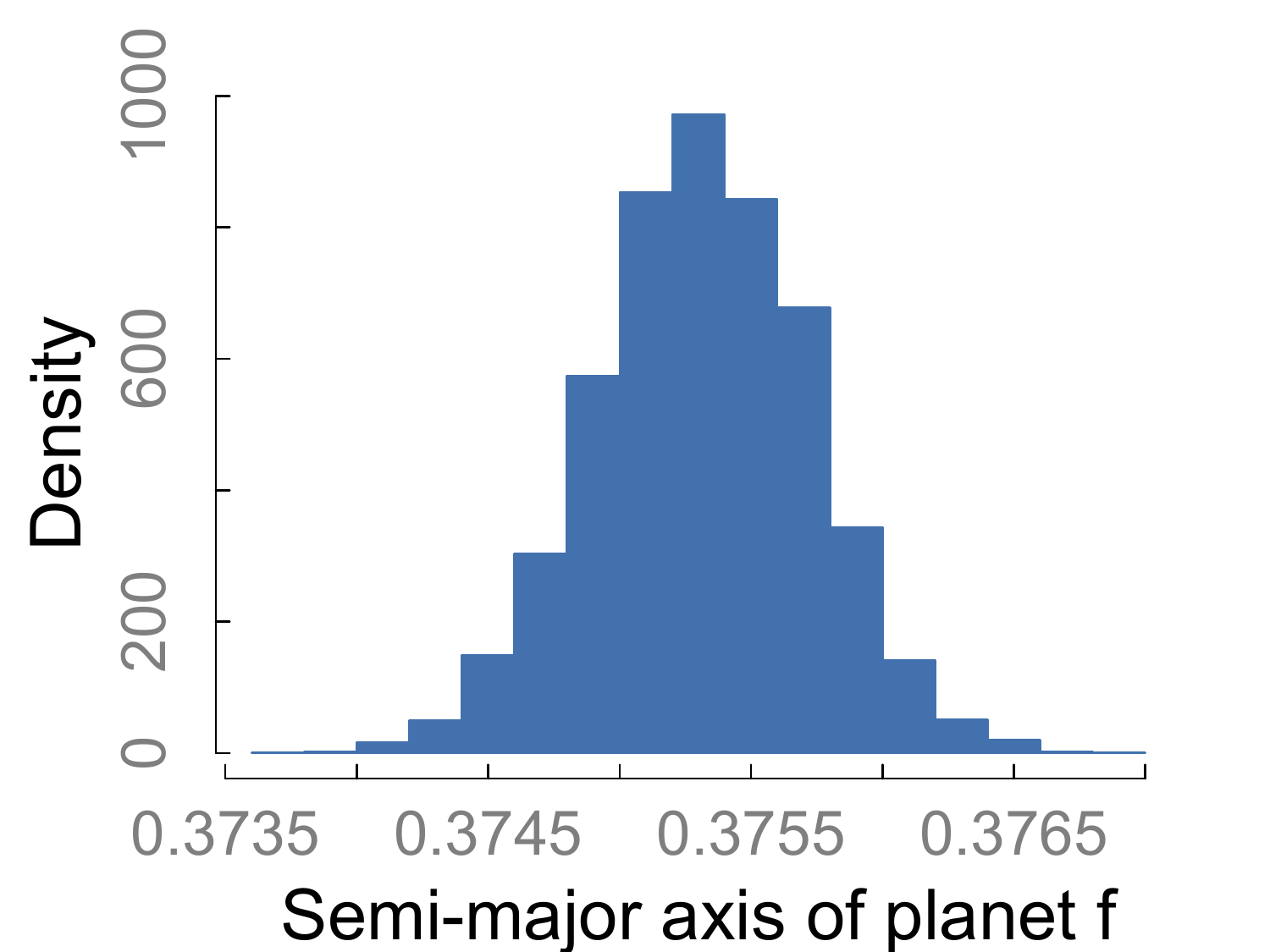}
\plotone{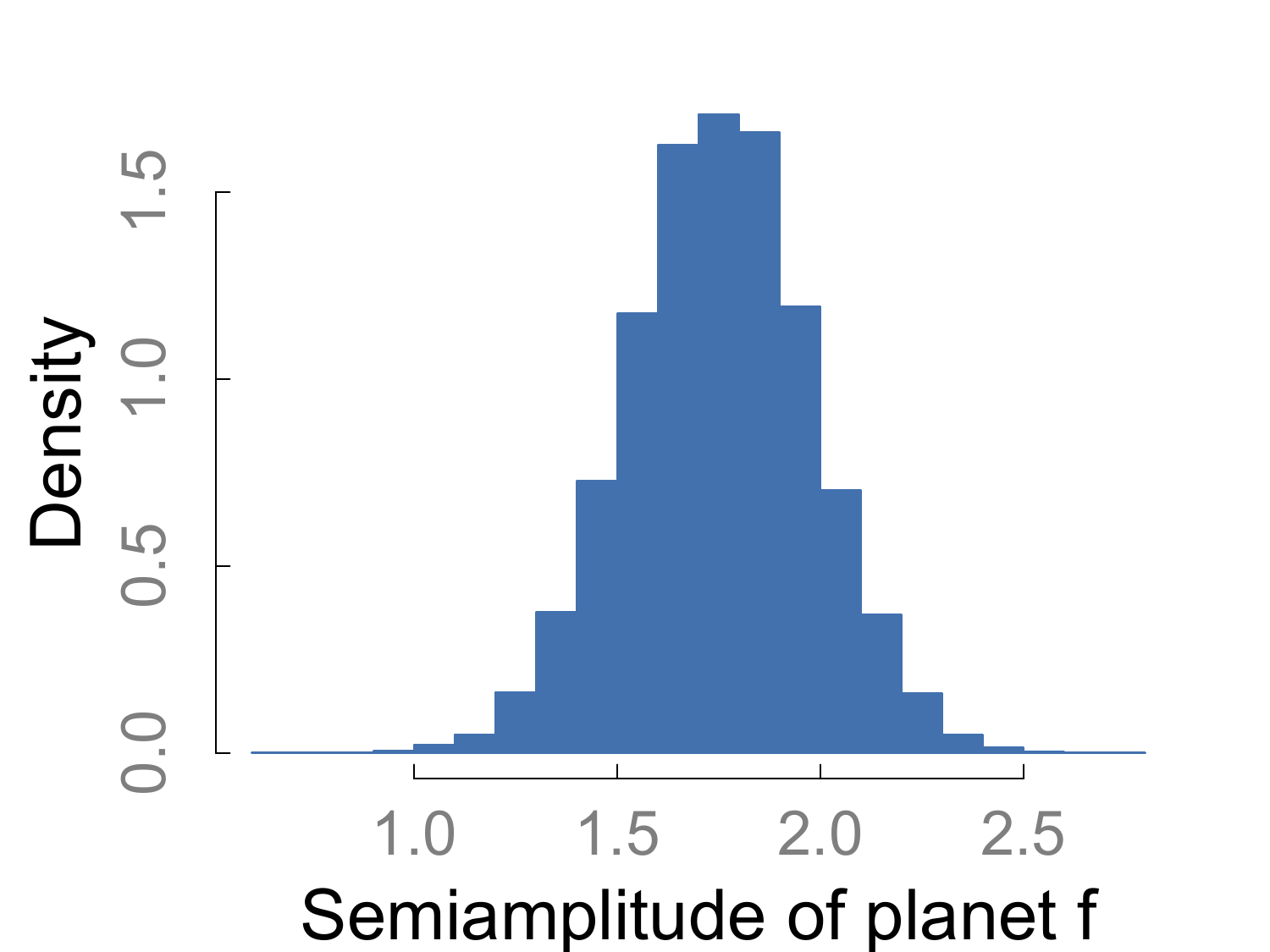}
\plotone{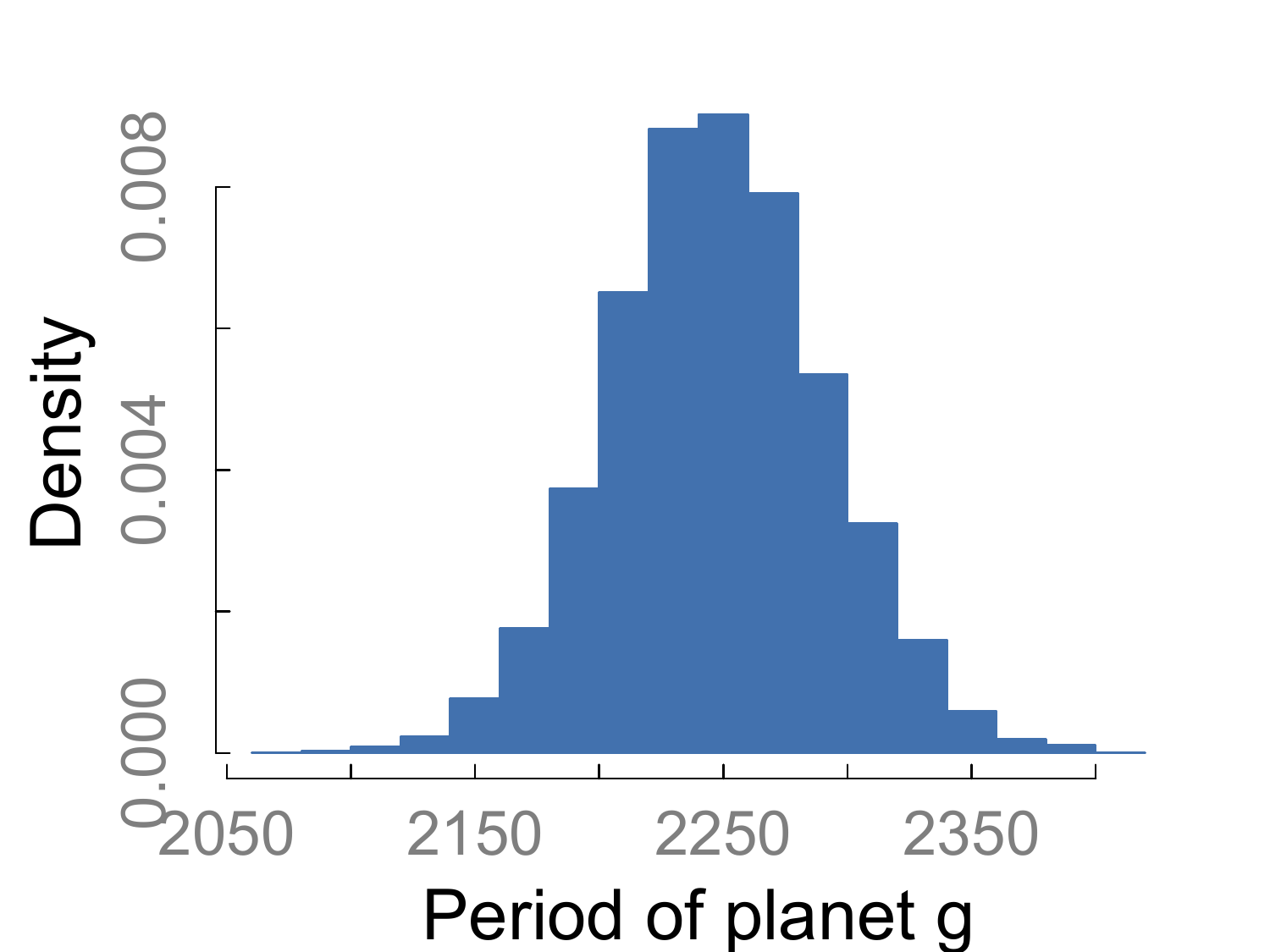}
\plotone{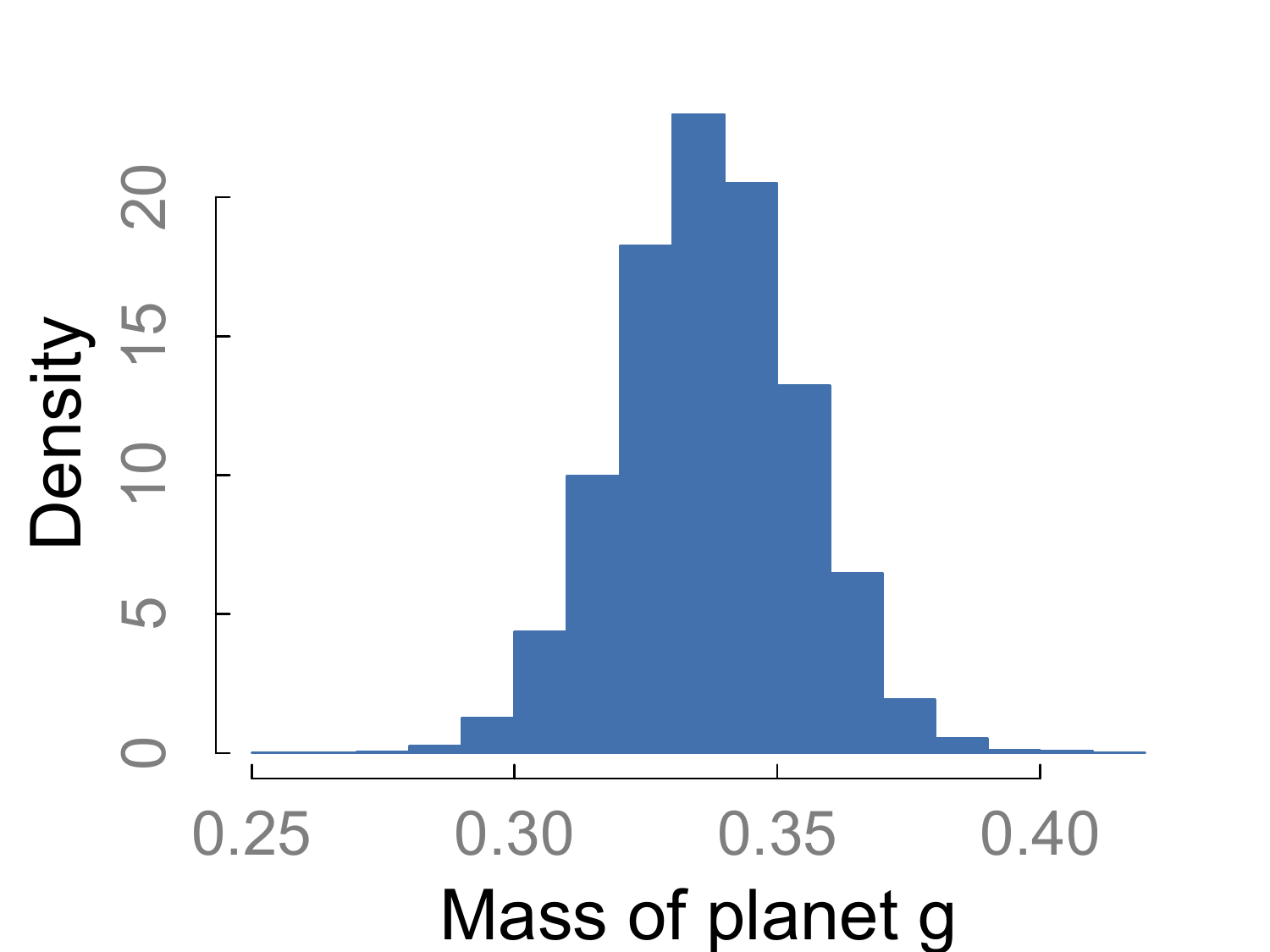}
\plotone{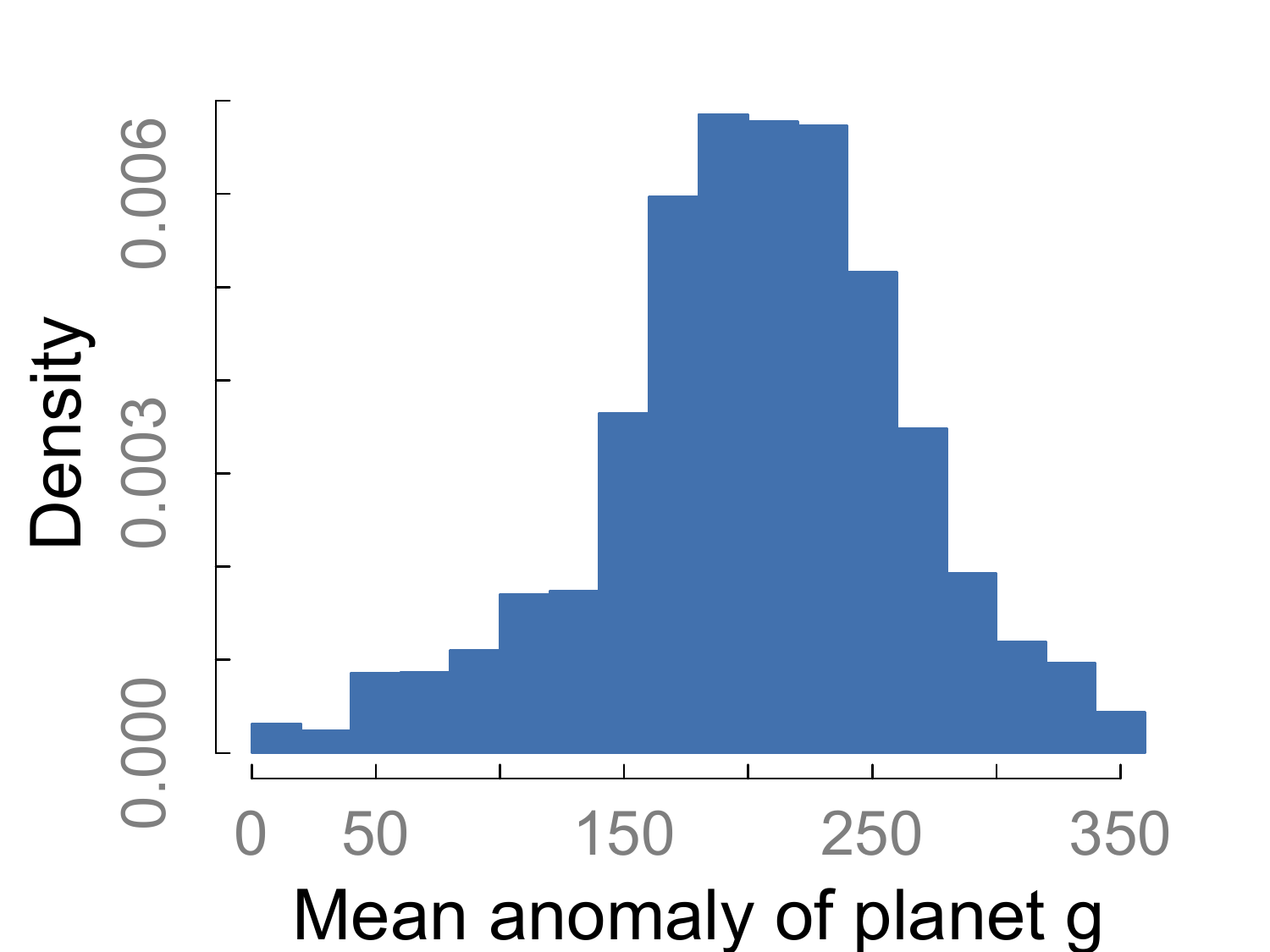}
\plotone{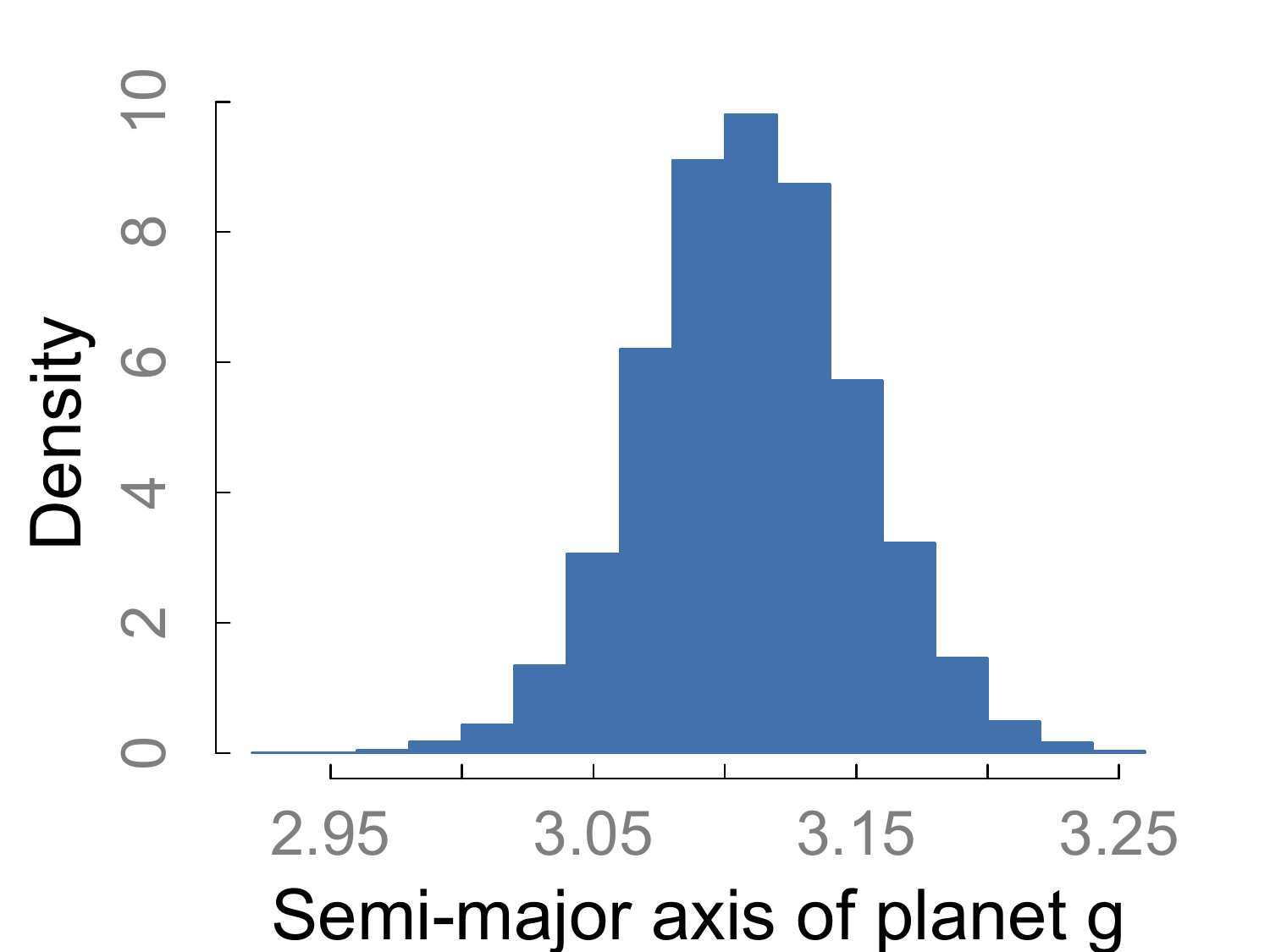}
\plotone{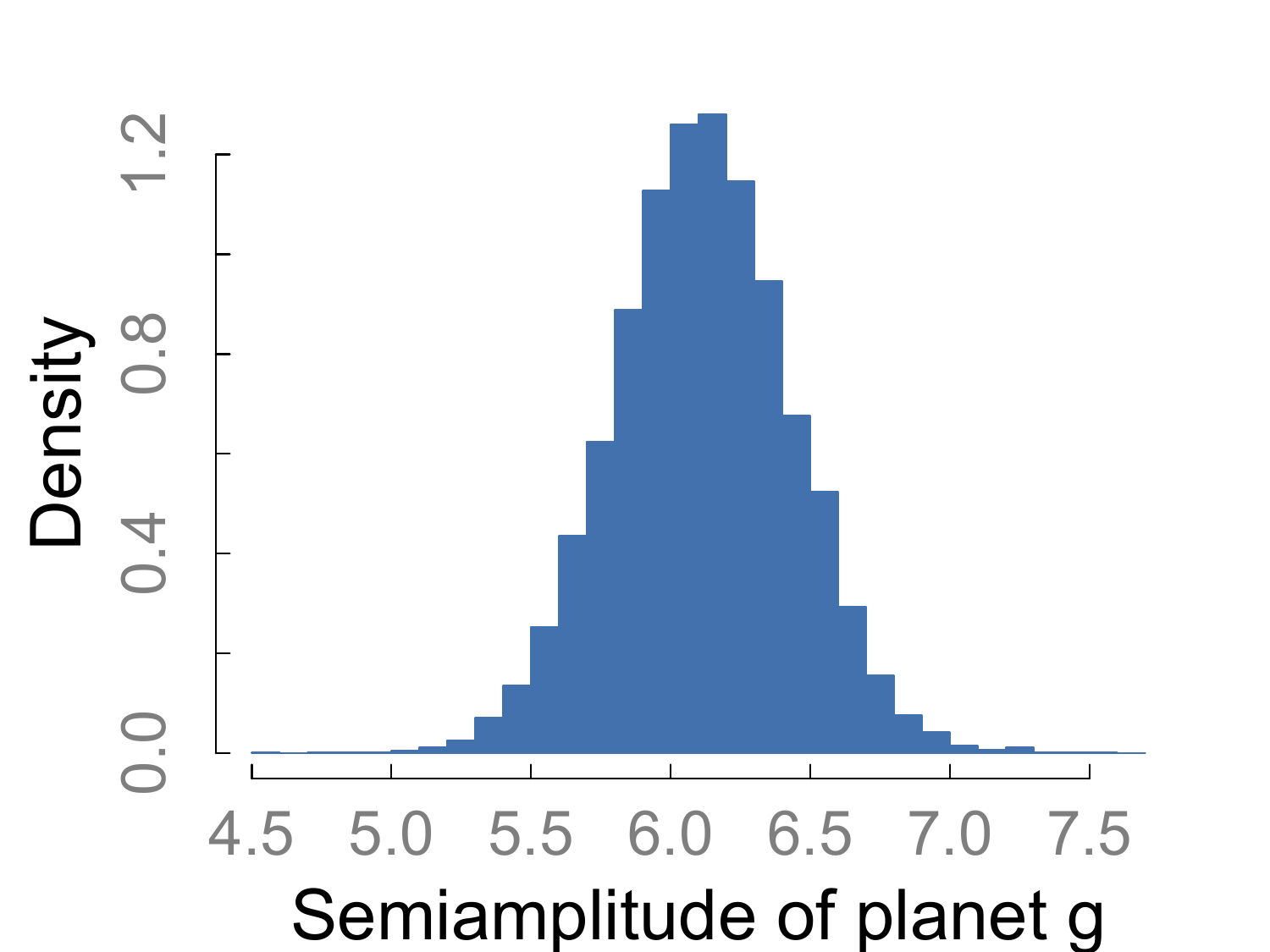}
\plotone{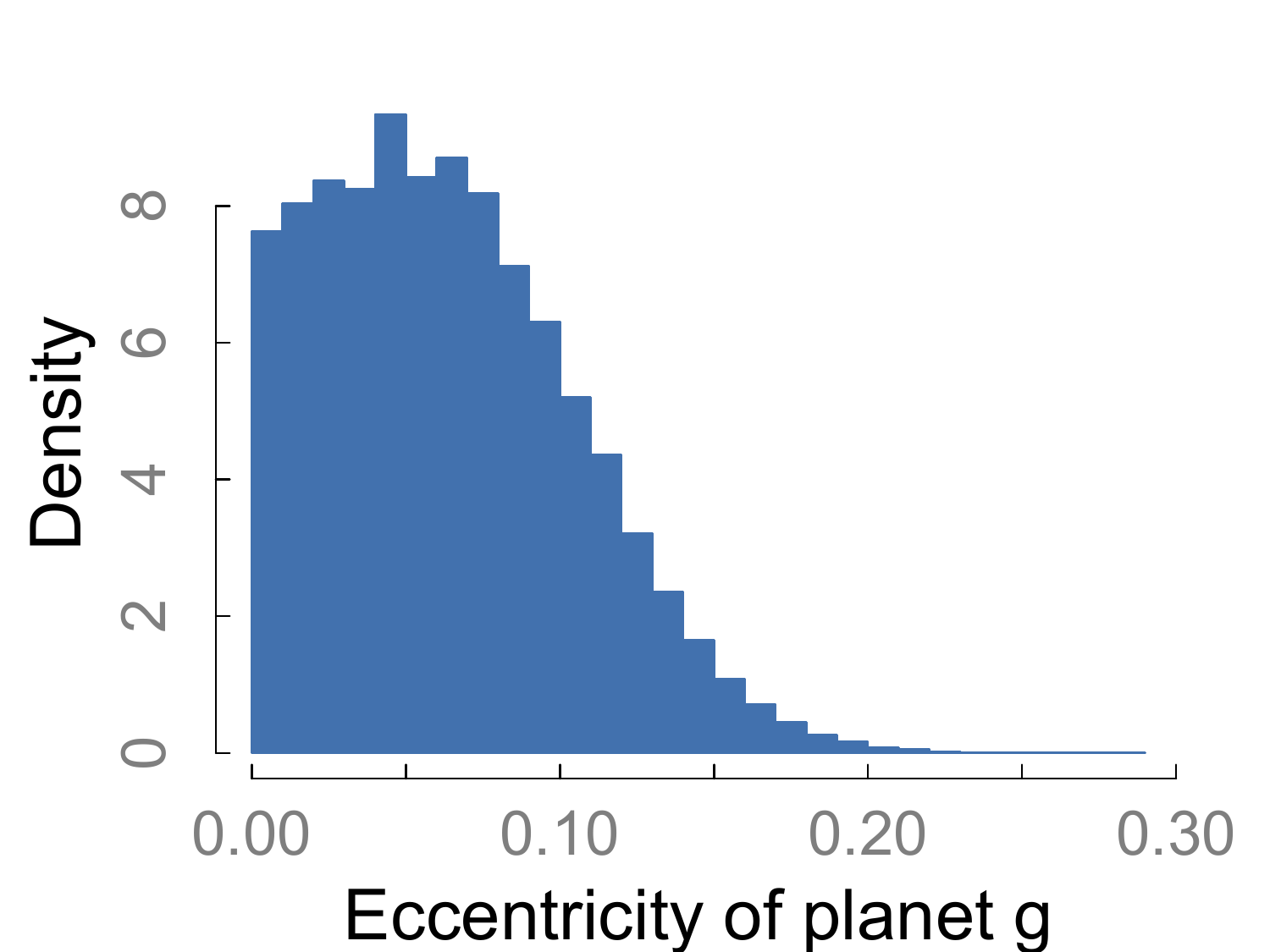}
\plotone{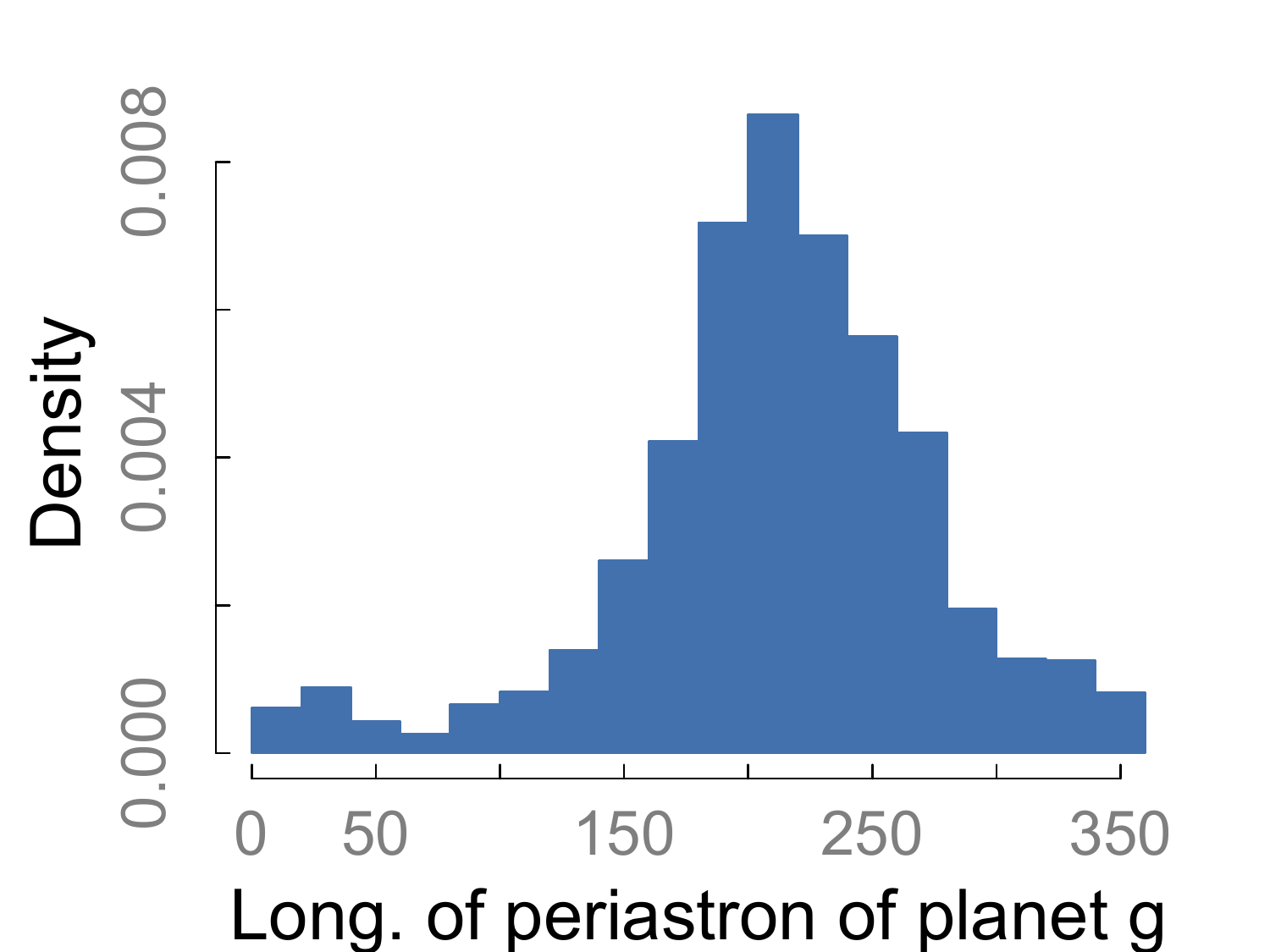}\\
\plotone{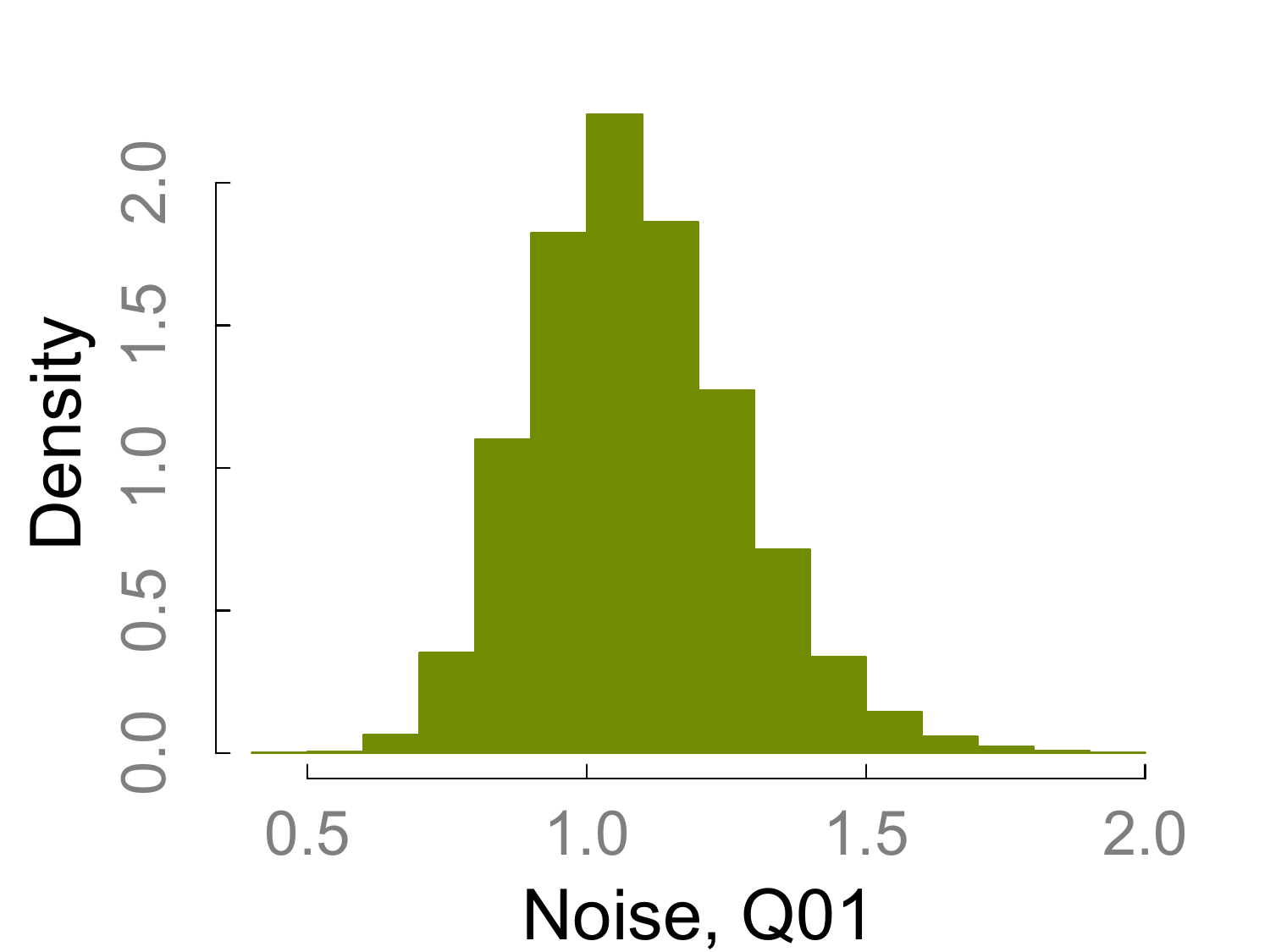}
\plotone{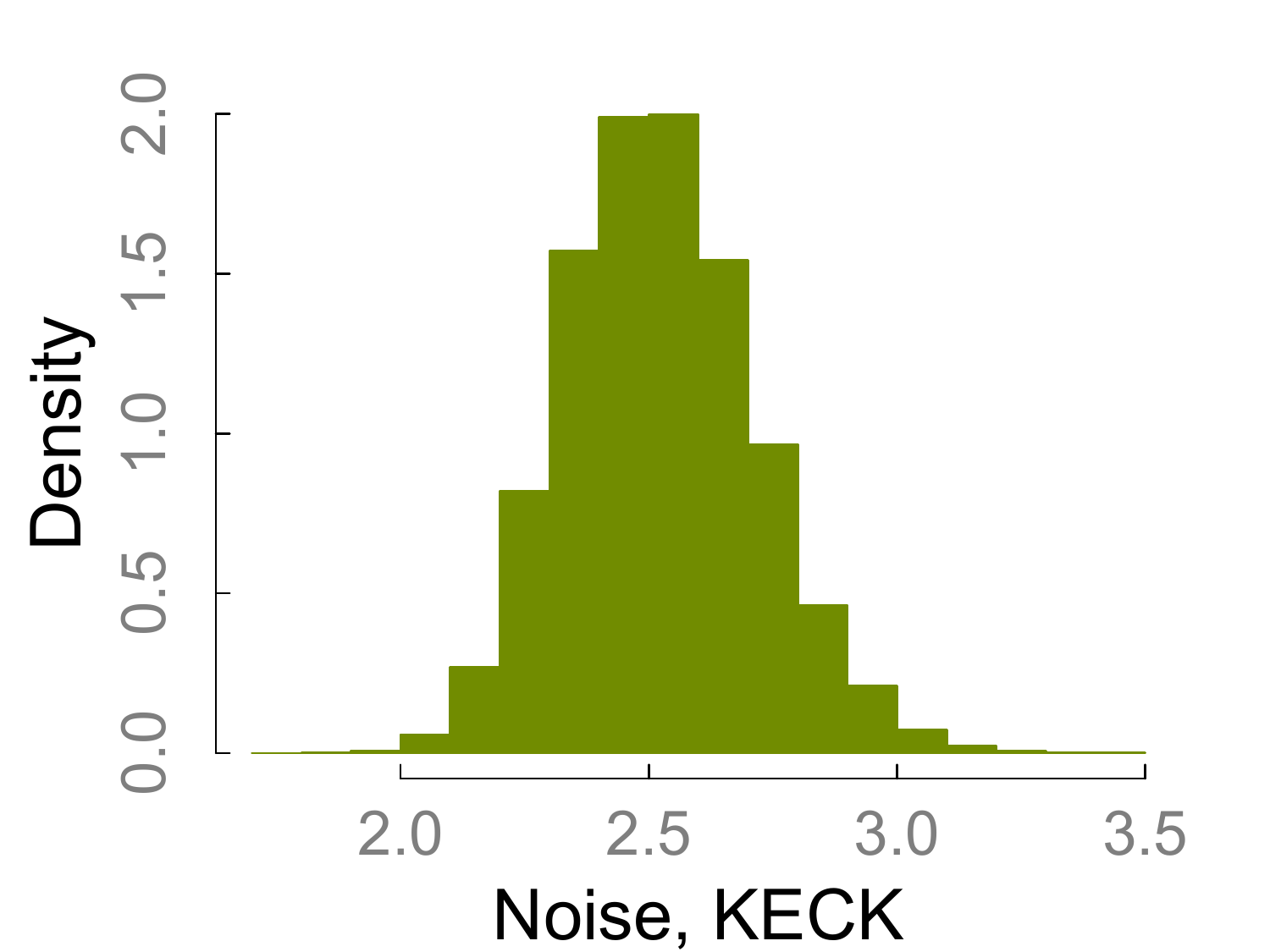}
\plotone{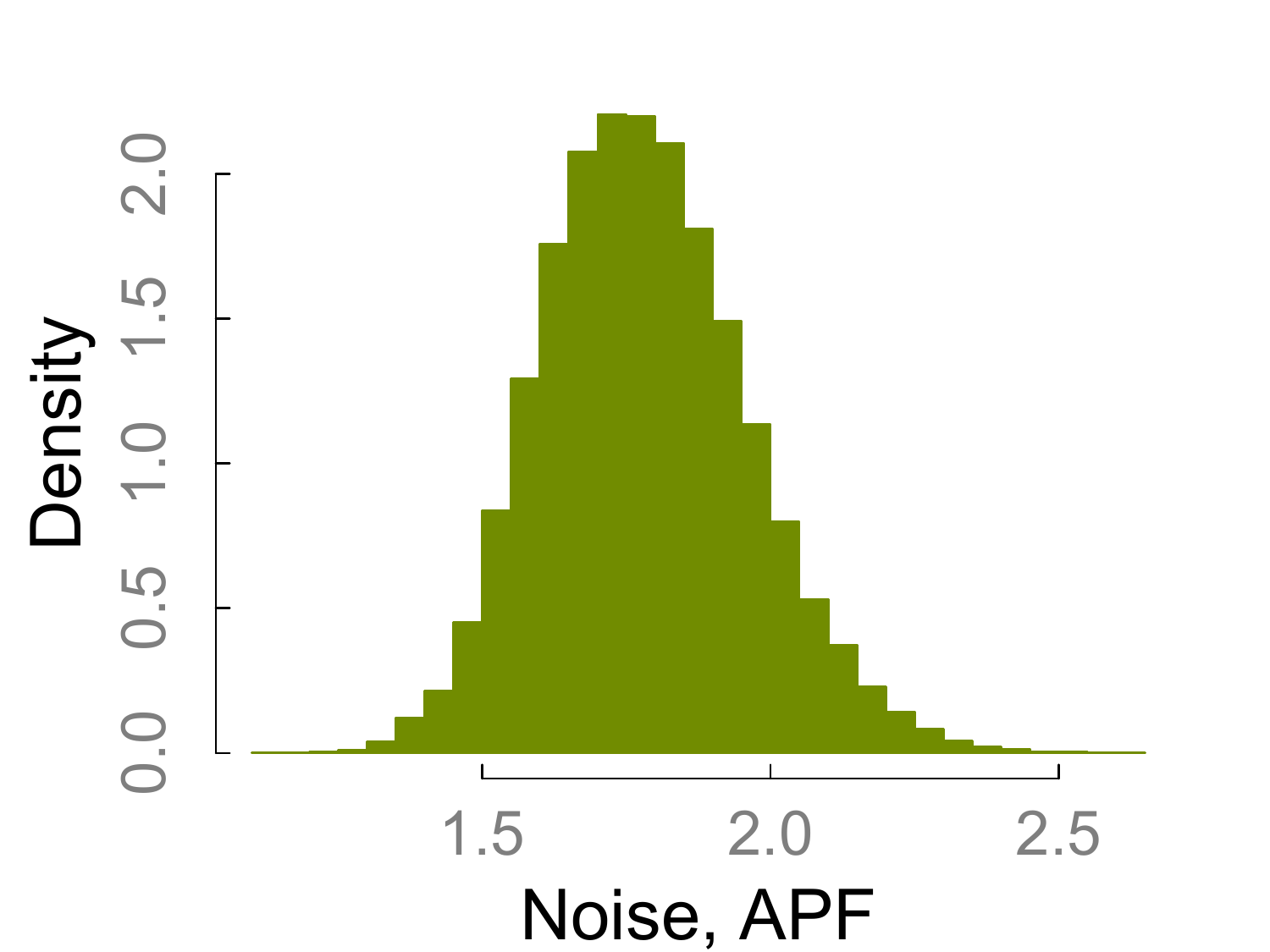}\\
\end{framed}
\caption{\label{fig:dist} Marginal distributions of the orbital elements, drawn from $5\times 10^5$ Markov-Chain Monte Carlo samples.}
\end{figure*}

A simple Markov-Chain Monte Carlo algorithm \citep[MCMC; e.g.][]{Ford05, Ford06, Gregory11}, in conjunction with Equations 1-3 and flat priors on log $P$, log $\mass$, and the other orbital parameters, was used to characterize the distribution of the parameters of the model, using the best fit the starting point. For the noise parameters, $s_\s{j}$, the corresponding prior is a modified Jeffrey function $p(s_\s{j}) = [(s_\s{j}+s_0)\ln(1+s_\s{max}/s_0)]$; for $s_\s{j} \ll s_0$ (which we take equal to 0.3$\, \ms$), the function is a uniform prior (which includes 0), while for $s_\s{j} \gg s_0$ the function is a regular Jeffrey prior \citep{Gregory11}. The MCMC routine is run until sufficient convergence is achieved. The uncertainties in each parameter are reported in Table \ref{tab:elements} within square brackets. The marginal distribution of each parameter, based on $5\times 10^5$ samples from the MCMC routine, is shown in Figure \ref{fig:dist}. 

\begin{figure*}
\centering
\plottwo{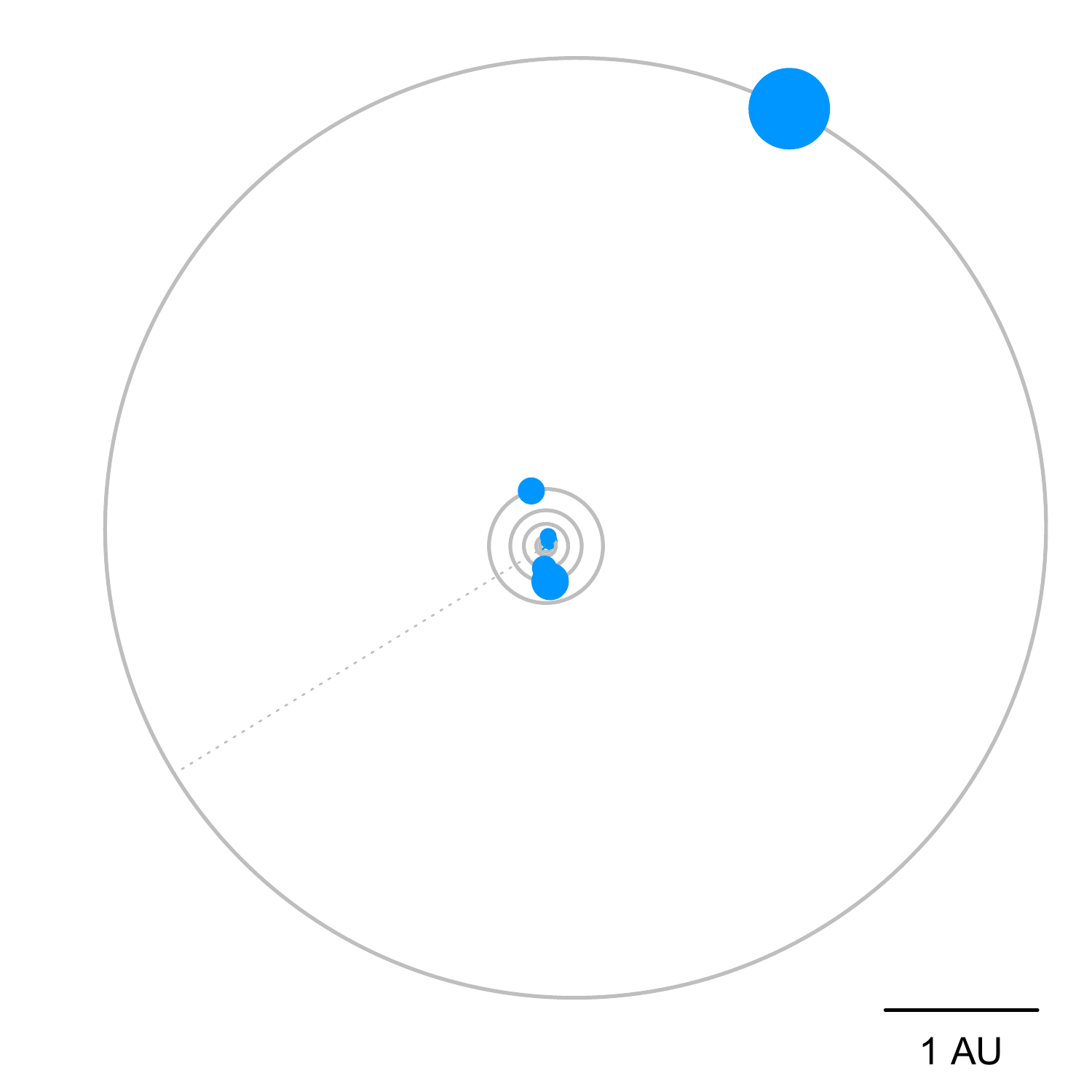}{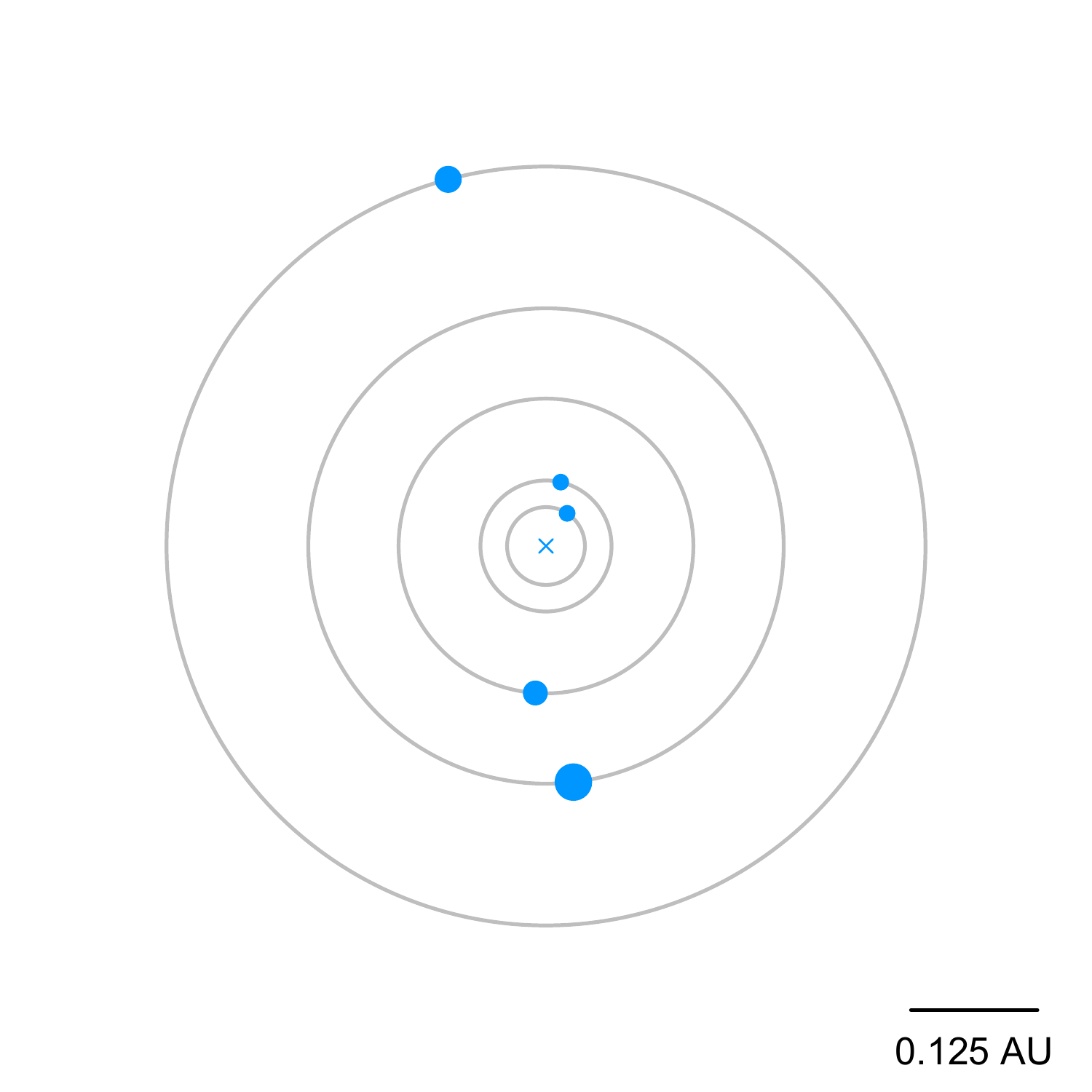}
\caption{\label{fig:orbits} \text{Left panel:} Orbital plot of the 6-planet model. \textit{Right panel:} Orbital plot of the inner 5 planets. The radius of each point is proportional to the square root of its minimum mass.}
\end{figure*}

The final best-fit model is shown in Figure \ref{fig:fit}, with the corresponding orbital elements listed in Table \ref{tab:elements}, where we report median and mean absolute deviation for each parameter. The inner 5 planets are fixed on circular orbits, since the best-fit Keplerian model with eccentric orbits has crossing orbits and is therefore unstable. We also note that the gravitational (non-Keplerian) interaction between the inner 5 planets is small, but not completely negligible, since the planet pair b-c and the planet triple d-e-f lie close to mean-motion resonances (2:1 and 4:2:1 respectively).
Figure \ref{fig:orbits} shows an orbital diagram of the system.

\begin{table}[t]
\centering
\begin{tabular}{rl}
  \hline
 & $\logl_\s{cv}$ \\ 
  \hline
No planets & -804.34 \\ 
  1 planet (g) & -754.59 \\ 
  2 planets (g, e) & -699.17 \\ 
  3 planets (g, e, d) & -672.69 \\ 
  4 planets (g, e, d, b) & -640.45 \\ 
  5 planets (g, e, d, b, f) & -614.71 \\ 
  6 planets (g, e, d, b, f, c) & -594.11 \\ 
   \hline
\end{tabular}
\caption{Cross-validation results} 
\label{tab:logl}
\end{table}

As a final test to assess the quality of the 6-planet model, we use a cross-validation algorithm on the data, comparing an $N$-planet model with an $(N-1)$-planet model. In the ``leave-one-out'' flavor used for the present paper, we divide the full dataset of $N_{\rm o}$ observations into a \textit{training set} of $N_{\rm o}-1$ observations and a \textit{testing set} of a single observation, rotated among all observations; each training set is used to derive a new fit. The likelihood of the prediction made combined from each testing set ($\logl_\s{cv}$; higher is better) measures the predictive power of the model, and is sensitive to both underfitting and overfitting; for instance, a lower (worse) $\logl_\s{cv}$ for a model with a larger set of parameters is indicative of overfitting. Table \ref{tab:logl} reports the values of $\logl_\s{cv}$. Each fit is strictly better (higher likelihood) than the previous, suggesting (but not conclusively proving) that the 6-planet Keplerian model is not overfitting the data.

\begin{figure}
\centering
\plotone{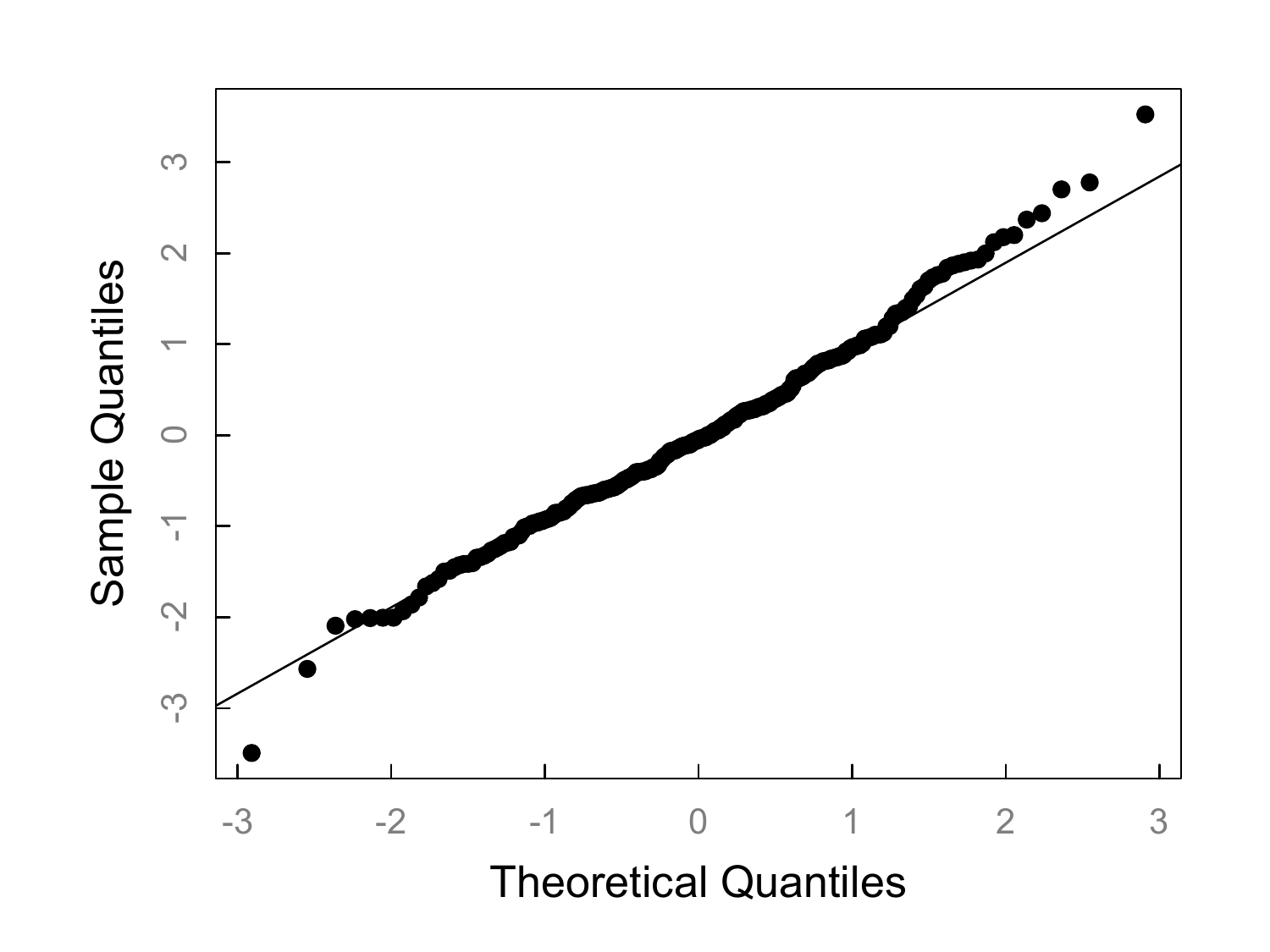}
\plotone{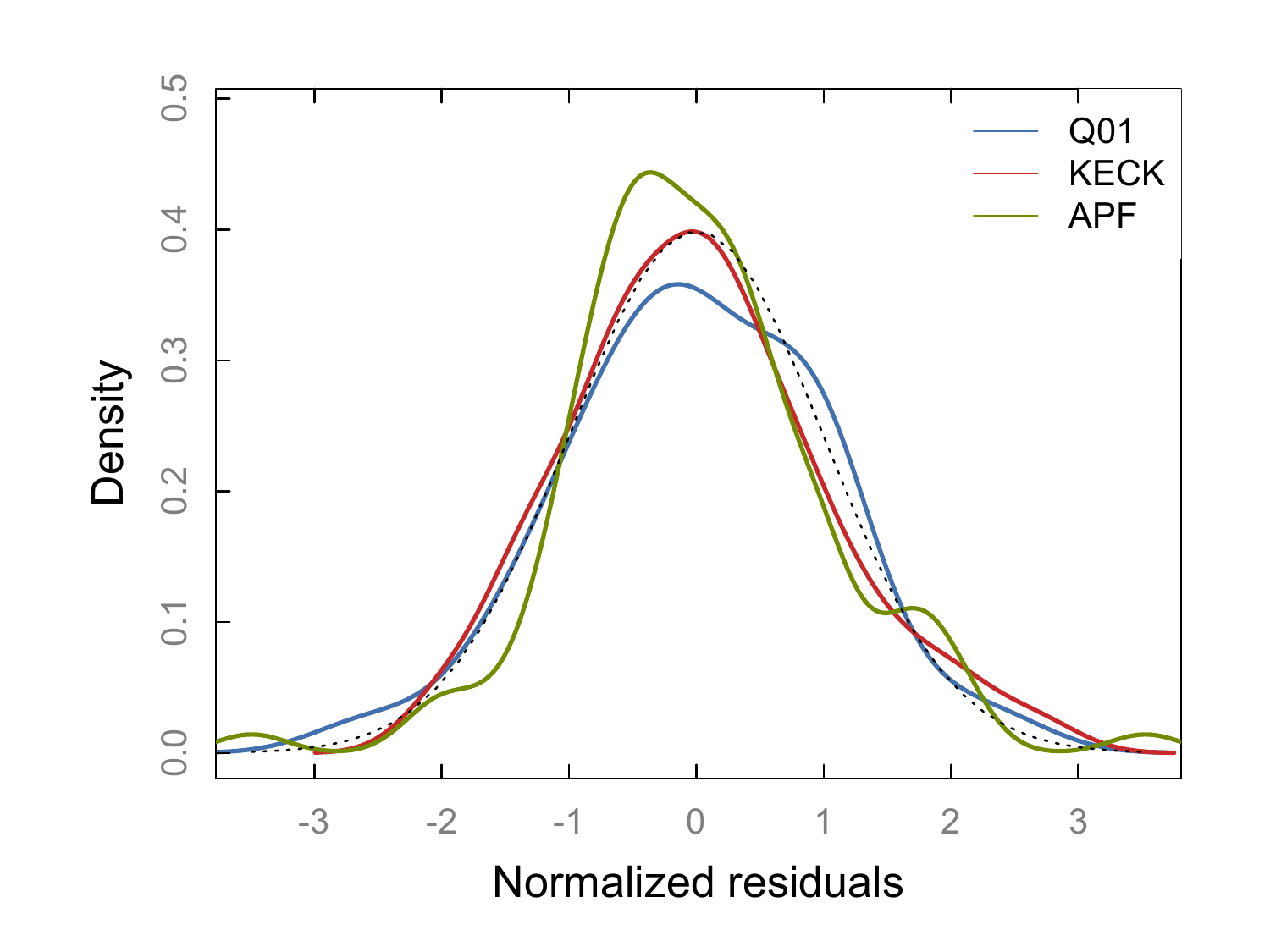}
\caption{\label{fig:qq} \textit{Top panel:} Quantile-quantile plot of the residuals from the 6-planets model. Perfectly normally distributed residuals would fall on the solid line. \textit{Bottom panel:} Kernel density for the normalized residuals (including noise) for each dataset, compared to a Gaussian density distribution (dotted line).}
\end{figure}

Figure \ref{fig:qq} illustrates that the distribution of the residuals is very nearly normal, which suggests that we are not overfitting the data (assuming the real distribution of the residuals is itself normal). Finally, we note that the orbital elements (for signals with $P<100\,$d) derived with the APF data alone are comparable to the orbital elements of the fit derived with the combined datasets, aside for the 94-day candidate planet, which is detected at the lowest significance among the six signals.

\subsection{Risk Assessment}

\begin{figure}
\centering
\plotone{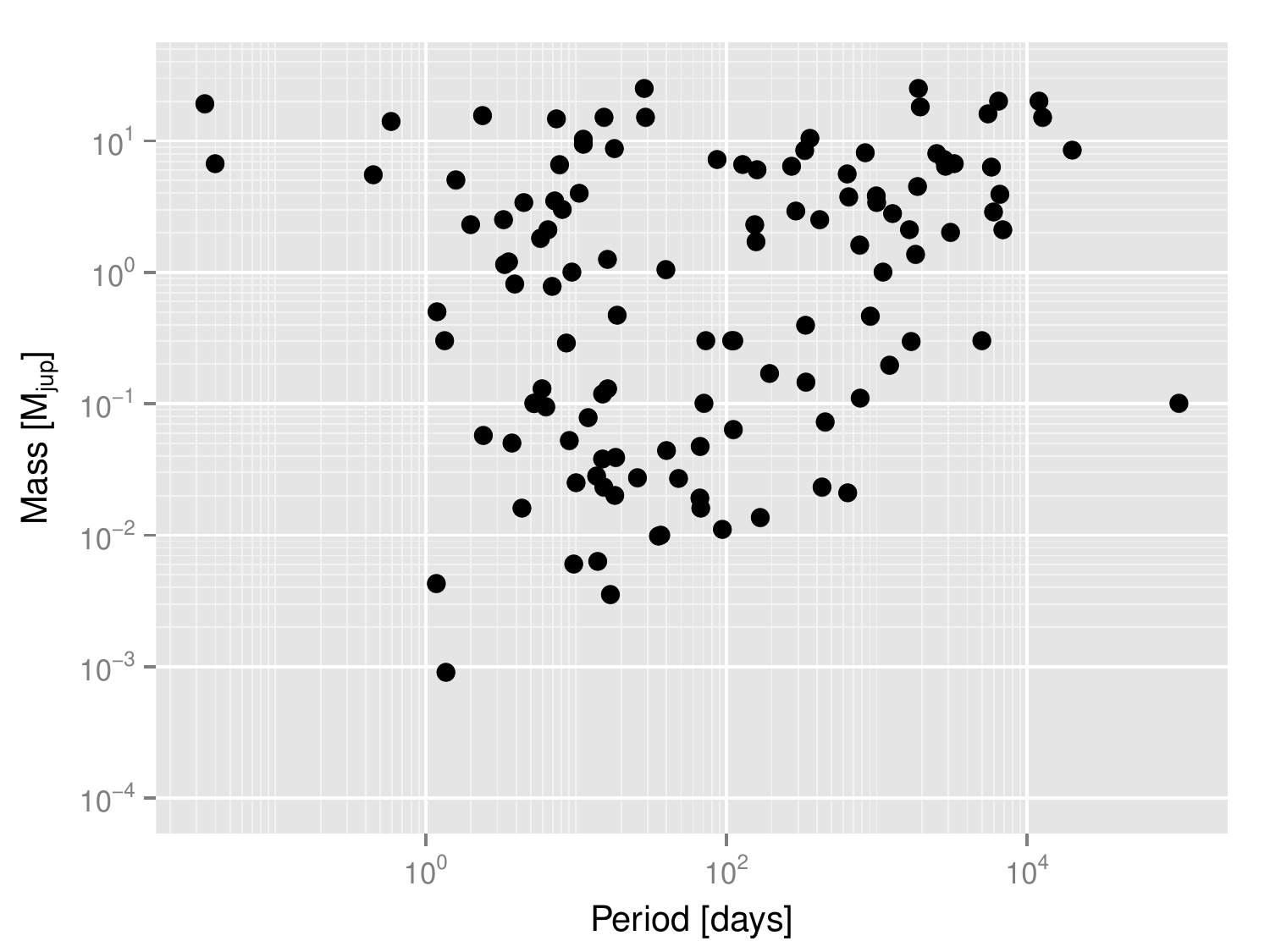}
\caption{\label{fig:controversial} Planets listed in the exoplanet.eu catalog as ``controversial''.}
\end{figure}

Figure \ref{fig:controversial} is a mass-period diagram for planets whose detections have been publicly announced but whose status is currently listed as ``controversial'' in the exoplanet.eu database. Nearly 200 objects are plotted; the concentration with the region of small $K$ (that is, low mass), and $P<100$ days is both noticeable and sobering. The use of precision Doppler velocity measurements to detect the class of planetary systems that dominate the Kepler census is fraught with potential pitfalls.

The time series of radial velocity measurements for HD 219134 is no exception, and there are several specific concerns. (1) With half-amplitudes $K\lesssim2\,{\rm m\,s^{-1}}$ for three of the inner planets (b,c, and d), the signals are weak. (2) For planets with low amplitudes, the presence of aliases can plague correct interpretation of the periodicities in the data. (3) The proximity to mean motion resonances for b-c and for d-e-f leads to inconsistencies between $N$-body and Keplerian models of the system. (4) The eccentricities of all the inner candidates must be very low in order to ensure orbital stability over the long term.

\begin{figure}
\centering
\plotone{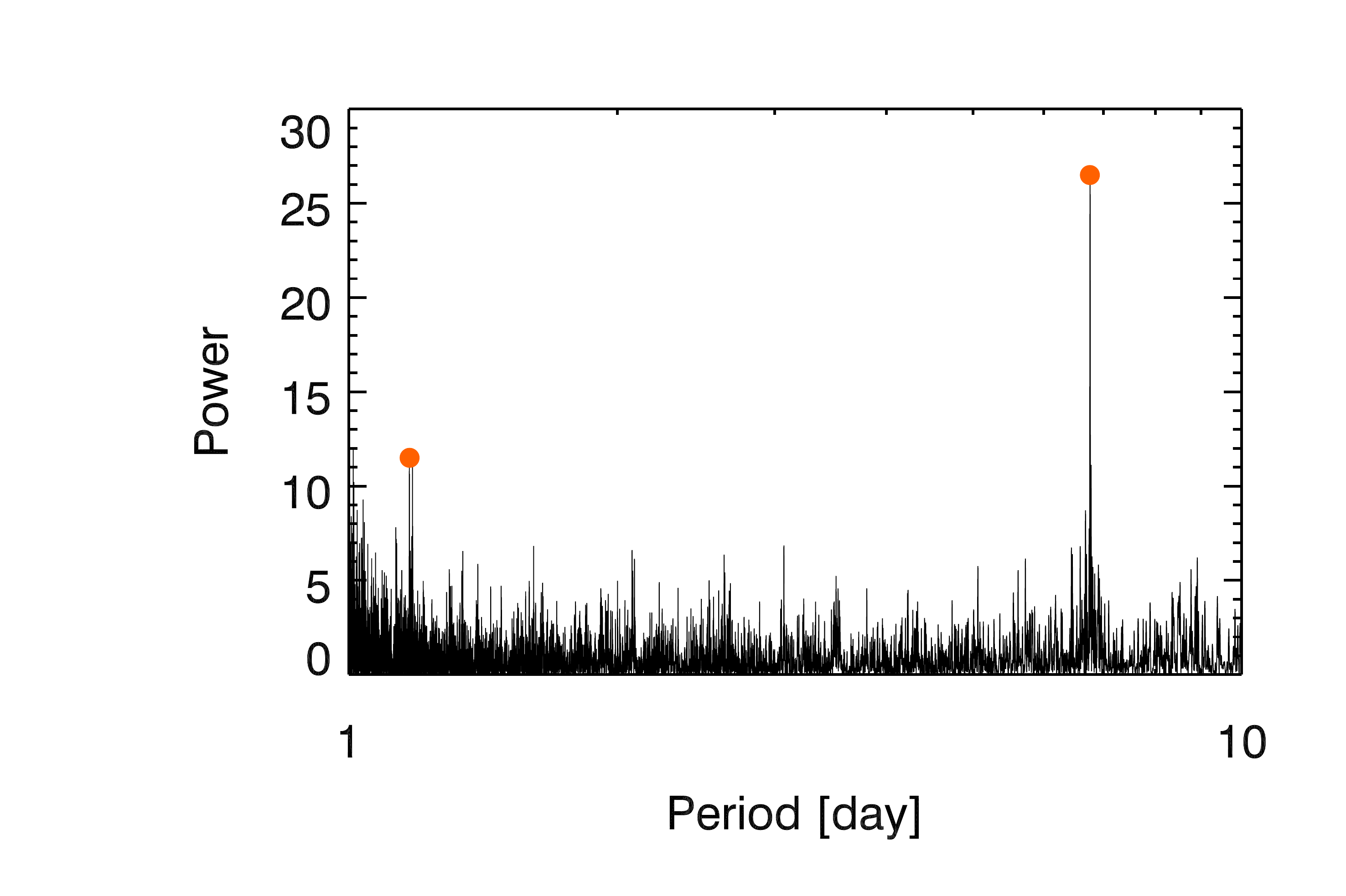}
\caption{\label{fig:aliasFig1} Periodogram for the HD~219134 residuals after subtraction of a five-planet solution.     
There are two power-excess peaks around $P\sim 1.17\,\rm{days}$ and $P\sim 6.76\,\rm{days}$ (indicated by the red points), one of which is likely the daily alias of the other.
}
\end{figure}

The residuals periodogram for a 5-Keplerian solution containing the 3.1, 22.8, 46.7, 94, and 2247 day candidate periods reveals remaining peaks at $P \approx 6.76\, \rm{days}$ and  $P \approx 1.17\, \rm{days}$  (Figure~\ref{fig:aliasFig1}). In our analysis, we have so far assumed that the 1.17-day period is an alias of the 6.76 day signal. Concerns regarding aliases are heightened for the 6.76 day signal as a consequence of the $K=1.4\pm0.2\,\mathrm {m\,s}^{-1}$ RV half-amplitude.  To improve confidence that the 6.76 day signal represents the true physical period, we proceed with the following steps.

\begin{figure}
\centering
\plotone{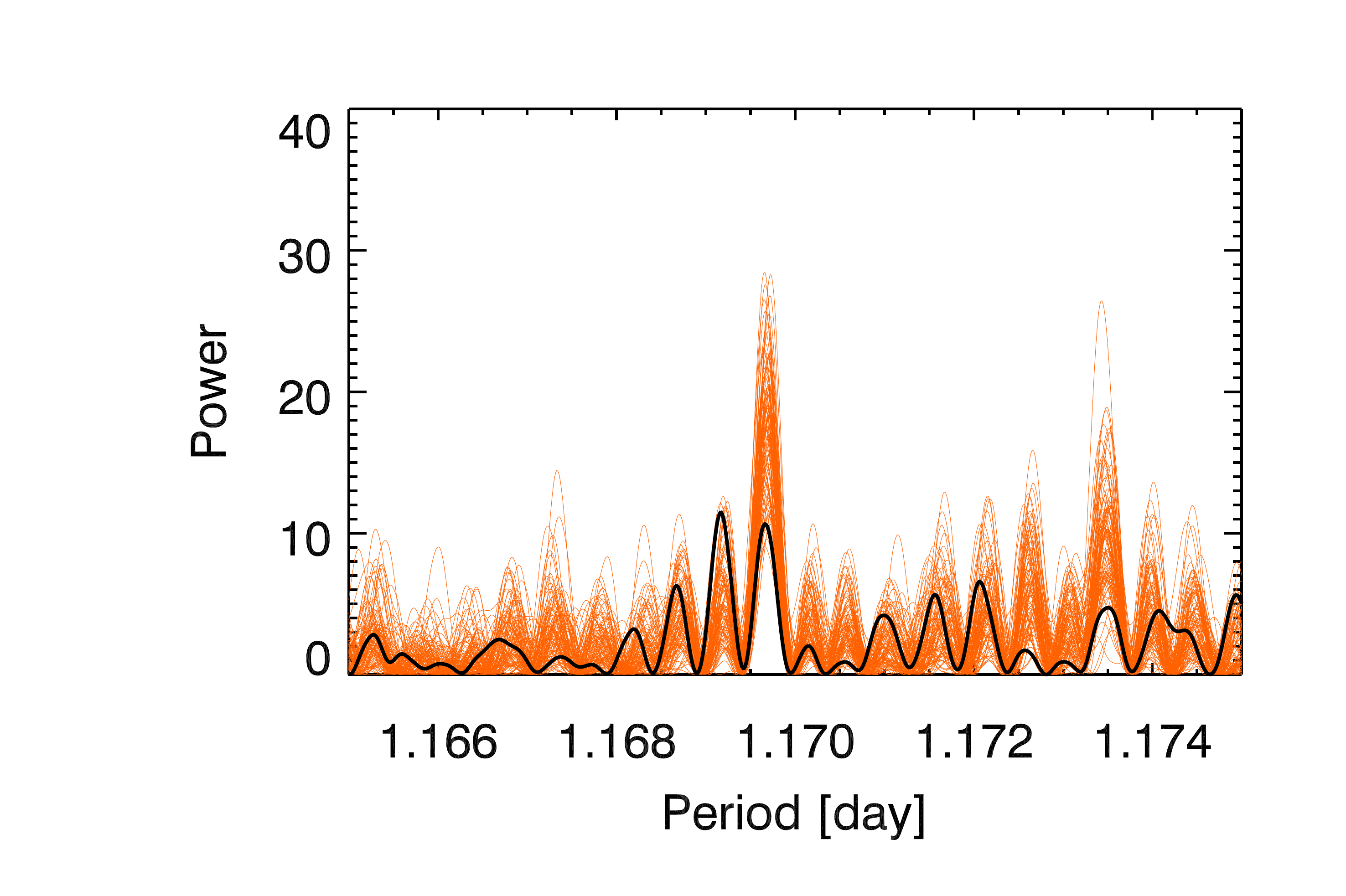}
\caption{ \label{fig:aliasFig2} Periodogram of HD~219134 for the 6.76-day planet c (best five-planet solution have been subtracted). 
The black line represents the periodogram of real data sets.
The red line represents the periodogram of an artificial 6.76-d signal + bootstrapped residuals to six-planet solution.
With artificial signals injected at $P=6.76\,\rm{days}$, the 1.17-d peak in the observed data sets is fully recovered.}
\end{figure}

\begin{figure}
\centering
\plotone{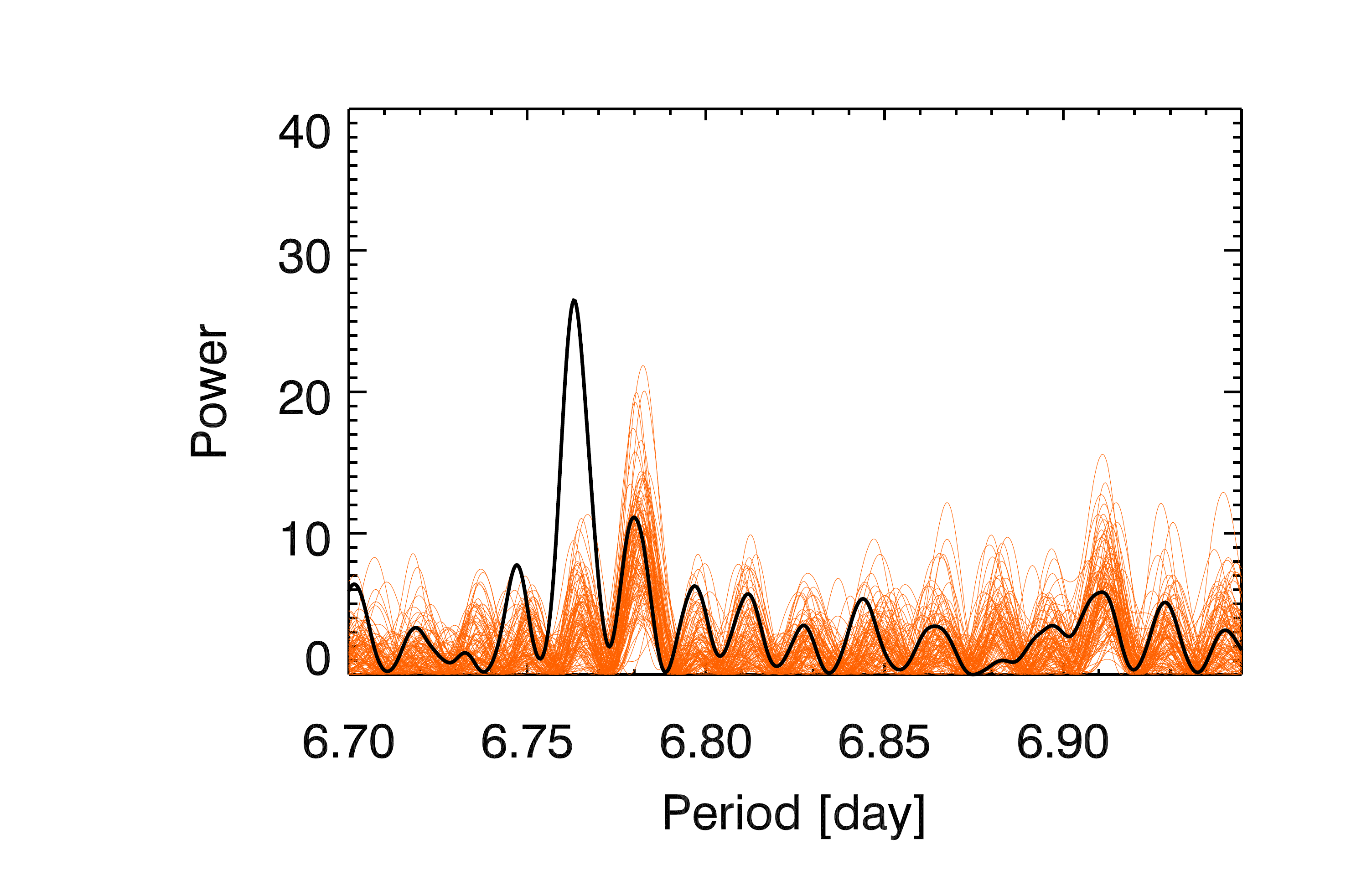}
\caption{ \label{fig:aliasFig3} Periodogram of HD~219134 for the 6.76-day planet c (best five-planet solution have been subtracted). 
The black line represents the periodogram of real data sets.
The red line represents the periodogram of an artificial 1.17-d signal + bootstrapped residuals to six-planet solution.
When injecting an artificial 1.17-d signal, the 6.76-d peak in the observed data sets cannot be correctly reproduced.}
\end{figure}

We perform bootstrap simulations, outlined by \citet{Dawson2010} and adopted in previous analyses of aliases in systems containing multiple low-amplitude planets  \citep[e.g.][]{Lovis11}.  We generate 500 bootstrapped Doppler time series by shuffling the residuals from the six-planet solution (with either $P_{6}=6.76\,\rm{d}$ or  $P_{6}=1.17\,\rm{d}$). For each trial, we inject an artificial noiseless sinusoid at the corresponding period, $P_{6}$, and compare the resulting periodogram to the observed one. Figure~\ref{fig:aliasFig2} and Figure~\ref{fig:aliasFig3} show the results of this exercise, and indicate that the observed periodogram of the residuals to the five-planet solution is much more likely to be correctly reproduced by a $P_6=6.76\,{\rm d}$ signal than a $P_{6}=1.17\,{\rm d}$ signal.

Additionally, we fit six-Keplerian models to the radial velocities using sixth-planet periodicities of $P_{6}=6.76\,\rm{d}$ and  $P_{6}=1.17\,\rm{d}$.   
The 6.76-d signal provides  a better fit with a reduced $\chi^2=9.53$, and in this case, the 1.17-d peak does not remain in the residuals periodogram. The six-planet solution with $P=1.17\,\rm{d}$ yields a worse reduced $\chi^2=10.96$. Moreover, the 6.76-d signal is still clear in its residuals periodogram. We therefore conclude that the peak at $P=6.76\,\rm{d}$ is a true signal, whereas the peak at $P=1.17\,\rm{d}$ is its one-day alias.

Another point of concern is the possibility that one (or more) of the candidate planet signals are an artifact of the stellar rotation period. In \S 2, we have noted that HD 219134's radius and $v\sin(i)$ suggest a rotation period of order $P\sim20$ days, which is close to the period of the 22.8 day planet candidate, raising the possibility that planet ``d'' is a rotationally modulated spot signal. To check whether this might be true, we have looked at different epochs of data to determine that the amplitude and the period of the signal is constant (a spot signal would likely attenuate over a few rotation periods, but could then reappear, and would be expected to show significant period and frequency drift). We have taken the three datasets (Keck, Q01, and APF) and (1) applied the velocity offsets from our best fit, (2) quadrature-augmented the noise by the fitted values of $\sigma_{S}=1.10\,\mathrm {m\,s}^{-1}$ m/s for Q01, $\sigma_{K}=2.50\,\mathrm {m\,s}^{-1}$ for Keck, and $\sigma_{A}=1.80\,\mathrm {m\,s}^{-1}$ for APF, and (3) sorted the joint data by ascending time stamp. We then divided the joint time-sorted data set into three segments, each with equal total signal-to-noise in its component points. The first data segment thus contains Q01 and Keck data, the second segment contains Keck and APF data, and the third contains only APF data. We subtract out the radial velocity signals from the 3.1, 6.7, 46.7, 94.2, and 2247 day candidate planets using the parameters listed in Table 3. We then compute the residuals periodogram for each of the three data segments. In each case, the $P=$22.8-day periodicity is the location of the strongest peak in the power spectrum of the velocity residuals (between $P=10$ days and $P=30$ days). This suggests that the $P=22.8$-day signal is not the product of spot modulation, and that it has been present and stable throughout the full time span of our observations.

\section{Photometric observations}

\begin{table}[]
\centering
\caption{Photometric Amplitudes at HD~219134 Candidate Periods}
\label{tab:photometricPeriods}
\begin{tabular}{lll}
Period & Semi-Amplitude      & RMS Error \\
\hline
3.093  & 0.00025$\pm$0.00016 & 0.00204   \\
6.763  & 0.00010$\pm$0.00016 & 0.00205   \\
22.81  & 0.00040$\pm$0.00016 & 0.00201   \\
46.72  & 0.00025$\pm$0.00017 & 0.00204   \\
94.19  & 0.00014$\pm$0.00017 & 0.00202 \\   
\hline
\end{tabular}
\end{table}
\vskip0.17in

High-precision long-baseline photometric data have been acquired for HD~219134 with the T10 0.8m APT at Fairborn Observatory \citep{Henry1999} in the Stromgren $b$ \& $y$ pass bands.  A total of 313 observations were obtained from the 2010 through 2014 observing seasons.  The two-color observations have been combined to produce a $\Delta (b+y)/2$ joint-filter time series, which improves measurement precision. The time-series is obtained using  the standard quartet observing sequence that compares the target star with a set of three comparison stars \citep{Henry1999}.  For the observations reported here, the three comparison stars (denoted a, b, \& c) are: {\bf  a} --  HD 223421,  (V = 6.36, B-V=0.408, F2 IV), {\bf  b} -- HD 217071  (V=7.45, B-V=0.368, F1 III), {\bf  c} --  HD 215588  (V=6.45, B-V=0.430, F5 V), whereas the target star HD 219134 (5.57, B-V=1.000, K3 V) is denoted star {\bf d}.

 \begin{figure}
 \plotone{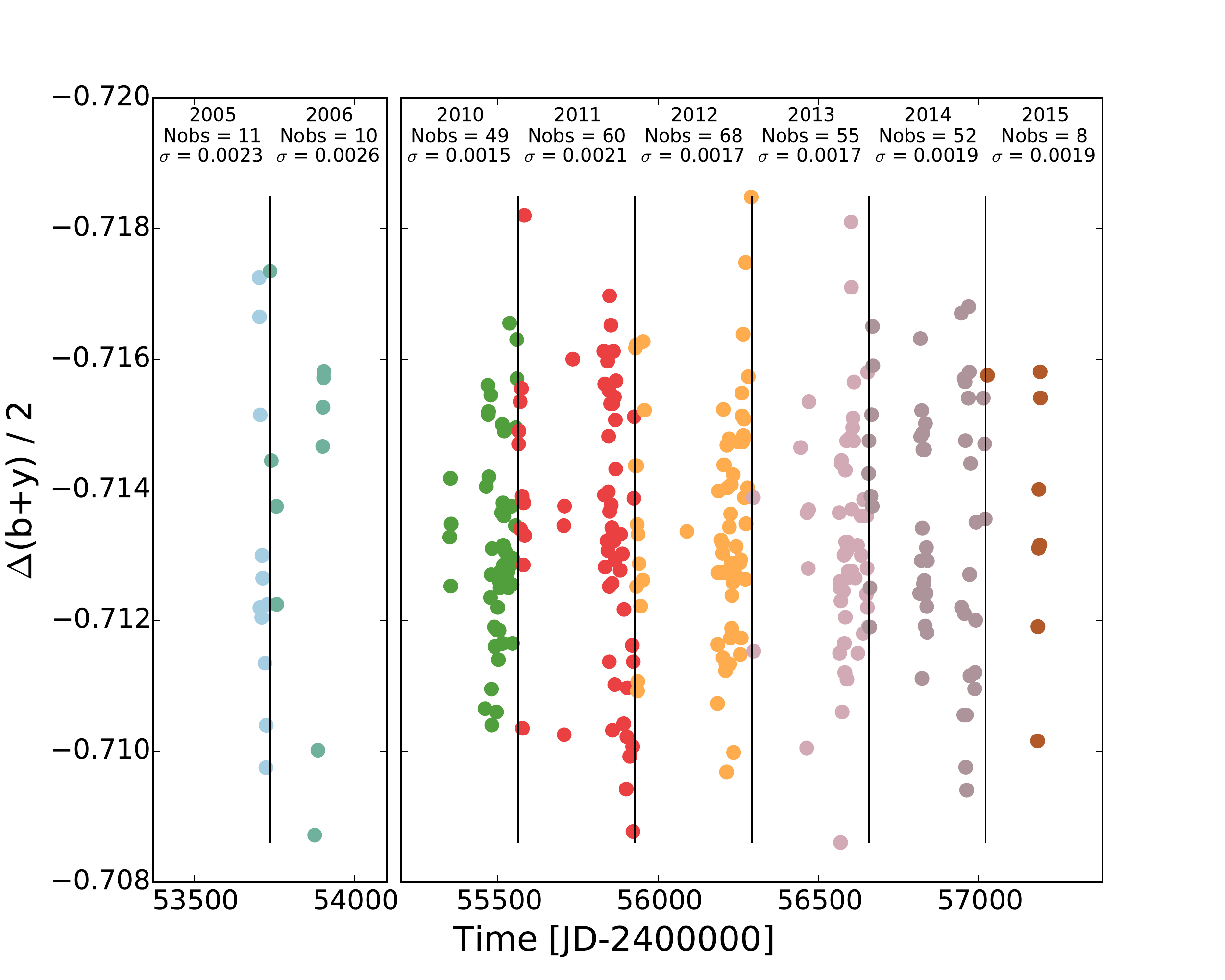}
 \caption{Differential photometric measurements of HD~219134 using the T10 0.8m APT at Fairborn Observatory \citep{Henry1999}. }
 \label{fig:Henry219134_all_photometry}
 \end{figure}
 
  \begin{figure}
 \plotone{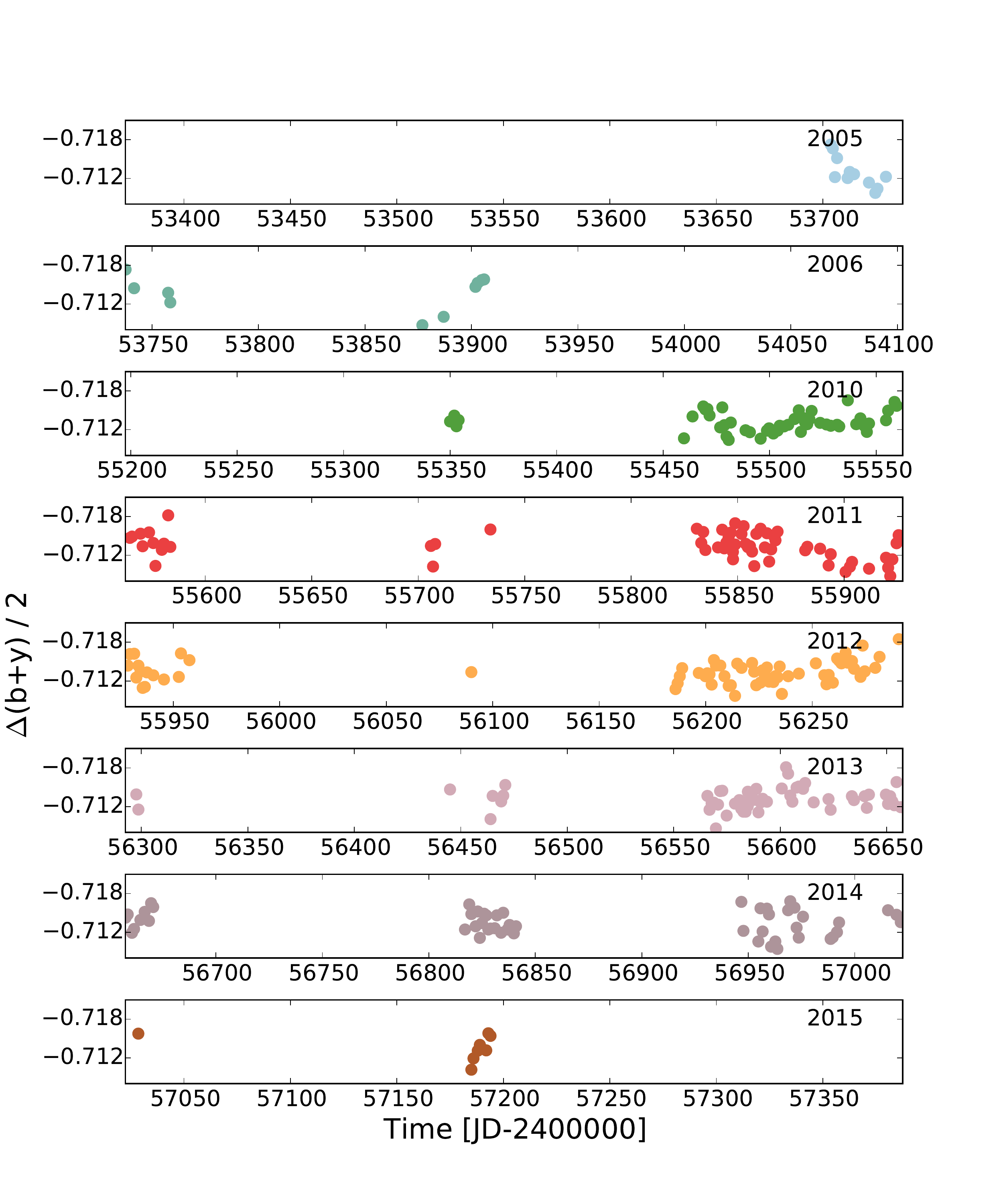}
 \caption{Differential photometric measurements of HD~219134 using the T10 0.8m APT at Fairborn Observatory \citep{Henry1999}. }
 \label{fig:Henry219134_yearly_photometry}
 \end{figure}

In reducing the data, all six permutations of differential magnitudes of the four stars are evaluated.  As described in \citet{Henry1999}, only data that survive a cloud filter are retained, and the photometry is normalized so that all observing seasons have the same mean as the first season.  We note, however, that there is little, if any, observable change in the mean magnitude from year to year.  A 3-$\sigma$ filter was applied, which removed six outlying photometric points. In keeping with standard procedure, however, the outliers were not removed until after the data had been phased to each of the planetary periods to be sure they were not transit points. 

The full photometric time series is shown in Figure \ref{fig:Henry219134_all_photometry} and a year-by-year breakdown is shown in Figure \ref{fig:Henry219134_yearly_photometry}. The observations within each year are constant from night to night with a mean standard deviation of ~0.0014 mag.  This is the approximate limit of precision for a single nightly observation with the T10 APT.  The yearly means are also constant to a limit of only 0.00055 mag.  No significant photometric period is found in any year or across the entire data set.   In addition, we have normalized the data so all yearly means are identical and fit least-squares sine curves on the candidate planetary periods with the results shown in Table 5.  None of the candidate planetary periods exhibit significant photometric periodicities. Phase plots on the planetary periods show no sign of transits.

The five inner planets of the HD~219134 system have a-priori geometric probabilities of transit of 9.2\%, 5.4\%, 2.4\%, 1.5\% and 0.93\% for the $P=\,$3.093, 6.763, 22.81, 46.72, and 94.19 day periods, respectively. As reviewed by authors such as \citet{Wolfgang2015}, super-Earth mass planets that are members of systems containing multiple planets with $P<100\,{\mathrm d}$ display a very large range in radii at given mass, and in expectation, often contain $\sim1$\% of the total planetary mass in hydrogen and helium. The presence of these light gasses, in turn, generates a substantial contribution to the overall planetary radius. If we use the solar system mass-radius relation, $M_{\rm p}/M_{\oplus}=(R_{\rm p}/R_{\oplus})^{2.06}$ \citep{Lissauer2011}, we expect that the transit depths for HD~219134~b and c will be of order $\delta_\mathrm{t}<0.1$\%, which is too small a signal for phase-folded long-term ground-based photometry of the type reported here, but is within reach if platforms such as MOST, Warm Spitzer, or JWST are employed for a targeted transit check. Indeed, systems such as HD~219134 form a strong basis for the scientific case of the forthcoming CHEOPS Mission, scheduled for launch in 2017.

\section{Discussion}

\begin{figure}
\centering
\plotone{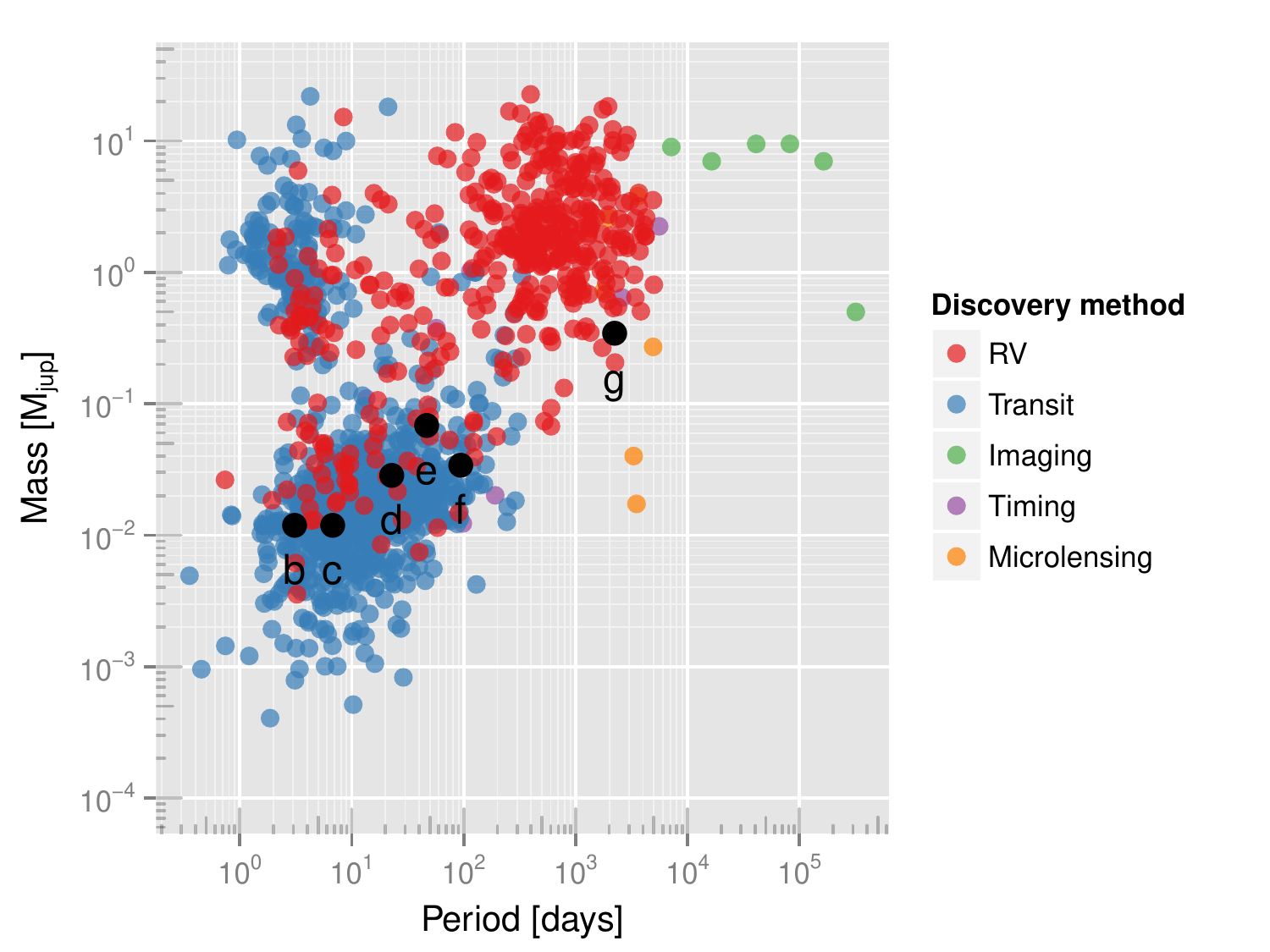}
\caption{\label{fig:population} Mass-Period diagram showing planets logged by the Exoplanet Data Explorer \citep{Wright11} (as of June 2014), and color-coded according to discovery method. The planetary candidates associated with HD~219134 (including the long-period signal that may be a signature of stellar activity) are labeled and plotted in black.}
\end{figure}

In comparison to our own Solar System, HD 219134 has an exotic architecture, with at least five super-Earth mass planets orbiting with periods of less than 100 days. Discoveries in recent years, however, have indicated that such systems are surprisingly common \citep{Mayor2009}. This result has received strong, and indeed dramatic confirmation from the Kepler Mission \citep{Batalha2013}, which revealed hundreds of candidate multiple-transiting multiple-planet systems that (at least in broad-brush strokes) call to mind HD 219134 b-f. This resemblance is underscored by Figure \ref{fig:population}, in which the HD 219134 planets are shown in conjunction with Kepler's transiting planet candidates.

A question of substantial interest is whether planetary systems such as HD 219134 are assembled \textit{in situ} \citep{Montgomery2009, Hansen2012, Chiang2013}, or whether the planets form at large distances and then migrate inward to their final locations. At present, it is not fully clear how to realistically distinguish between the two scenarios. Recent work by \citet{Batygin2015} has emphasized the role of outer giant planets in triggering collisional cascades among planetesimals that can potentially destroy systems of super-Earths such as those described in this paper. It will therefore be of substantial interest to understand the nature of the long-term ($P\gtrsim2000$ day) periodicity in the radial velocity time series. 

Among the thousands of planetary systems that are now known, HD 219134 stands out. The bright primary star has demonstrated excellent radial velocity stability over two decades of measurement, and, given more data, there is a tantalizing possibility of finding additional low-mass planets in the system. With the parent star luminosity estimated at $0.31\,L_{\odot}$, a planet orbiting HD 219134 at 0.56 AU would receive the same energy flux that the Earth receives from the Sun. Such a planet would have an orbital period of 167 days. If it had a mass equal to that of Earth, its radial velocity half amplitude would be $K=14\,\mathrm{cm\,s^{-1}}$. Such a signal would be challenging, but given current projections for the Doppler velocity technique, almost certainly not impossible to detect. Going forward, HD~219134 looks to be an ideal target for platforms such as APF, HARPS-N, the APTs, and other high-precision Doppler and photometric facilities with access to the far northern sky.

\vskip0.2in
\section{Acknowledgments}

GL acknowledges support from the NASA Astrobiology Institute through a cooperative agreement between NASA Ames Research Center and the University of California at Santa Cruz, and from the NASA TESS Mission through a cooperative agreement between M.I.T. and UCSC. SSV gratefully acknowledges support from NSF grants AST-0307493 and AST-0908870. RPB gratefully acknowledges support from NASA OSS Grant NNX07AR40G, the NASA Keck PI program, and from the Carnegie Institution of Washington. SM acknowledges support from the W. J. McDonald Postdoctoral Fellowship and the Longhorn Innovation Fund for Technology grant. G.W.H. acknowledges support from Tennessee State University and the State of Tennessee through its Centers of Excellence program. The work herein is based on observations obtained at the W. M. Keck Observatory, which is operated jointly by the University of California and the California Institute of Technology, and we thank the UC-Keck, UH, and NASA-Keck Time Assignment Committees for their support. This research has additionally made use of the Keck Observatory Archive (KOA), which is operated by the W. M. Keck Observatory and the NASA Exoplanet Science Institute (NExScI), under contract with the National Aeronautics and Space Administration. We also wish to extend our special thanks to those of Hawaiian ancestry on whose sacred mountain of Mauna Kea we are privileged to be guests. Without their generous hospitality, the Keck observations presented herein would not have been possible. This research has made use of the SIMBAD database, operated at CDS, Strasbourg, France, and it has made use of the Keck Observatory Archive (KOA), which is operated by the W. M. Keck Observatory and the NASA Exoplanet Science Institute (NExScI) under contract with the National Aeronautics and Space Administration. This paper was produced using $^BA^M$.
 \vskip0.1in
 \textit{Final Note}: As this paper was being prepared for submission, we learned of the independent detection and submission for publication of the HD 219134 system by a team employing the HARPS-North Telescope (Segransan \& Udry 2015, Personal Communication).  \vskip0.1in

{\it Facilities:} \facility{Keck (HIRES)} \facility{APF (Levy Spectrometer)}
\bibliographystyle{apj}
\bibliography{biblio}

\begin{thebibliography}{41}
\expandafter\ifx\csname natexlab\endcsname\relax\def\natexlab#1{#1}\fi

\bibitem[{REV(????)}]{REVTEX41Control}
 ????

\bibitem[{08(1)}]{apsrev41Control}
08. 1

\bibitem[{{Allende Prieto} {et~al.}(2004){Allende Prieto}, {Barklem},
  {Lambert}, \& {Cunha}}]{AllendePrieto2004}
{Allende Prieto}, C., {Barklem}, P.~S., {Lambert}, D.~L., \& {Cunha}, K. 2004,
  \aap, 420, 183

\bibitem[{{Batalha} {et~al.}(2013){Batalha}, {Rowe}, {Bryson}, {Barclay},
  {Burke}, {Caldwell}, {Christiansen}, {Mullally}, {Thompson}, {Brown},
  {Dupree}, {Fabrycky}, {Ford}, {Fortney}, {Gilliland}, {Isaacson}, {Latham},
  {Marcy}, {Quinn}, {Ragozzine}, {Shporer}, {Borucki}, {Ciardi}, {Gautier},
  {Haas}, {Jenkins}, {Koch}, {Lissauer}, {Rapin}, {Basri}, {Boss}, {Buchhave},
  {Carter}, {Charbonneau}, {Christensen-Dalsgaard}, {Clarke}, {Cochran},
  {Demory}, {Desert}, {Devore}, {Doyle}, {Esquerdo}, {Everett}, {Fressin},
  {Geary}, {Girouard}, {Gould}, {Hall}, {Holman}, {Howard}, {Howell},
  {Ibrahim}, {Kinemuchi}, {Kjeldsen}, {Klaus}, {Li}, {Lucas}, {Meibom},
  {Morris}, {Pr{\v s}a}, {Quintana}, {Sanderfer}, {Sasselov}, {Seader},
  {Smith}, {Steffen}, {Still}, {Stumpe}, {Tarter}, {Tenenbaum}, {Torres},
  {Twicken}, {Uddin}, {Van Cleve}, {Walkowicz}, \& {Welsh}}]{Batalha2013}
{Batalha}, N.~M., {Rowe}, J.~F., {Bryson}, S.~T., {Barclay}, T., {Burke},
  C.~J., {Caldwell}, D.~A., {Christiansen}, J.~L., {Mullally}, F., {Thompson},
  S.~E., {Brown}, T.~M., {Dupree}, A.~K., {Fabrycky}, D.~C., {Ford}, E.~B.,
  {Fortney}, J.~J., {Gilliland}, R.~L., {Isaacson}, H., {Latham}, D.~W.,
  {Marcy}, G.~W., {Quinn}, S.~N., {Ragozzine}, D., {Shporer}, A., {Borucki},
  W.~J., {Ciardi}, D.~R., {Gautier}, III, T.~N., {Haas}, M.~R., {Jenkins},
  J.~M., {Koch}, D.~G., {Lissauer}, J.~J., {Rapin}, W., {Basri}, G.~S., {Boss},
  A.~P., {Buchhave}, L.~A., {Carter}, J.~A., {Charbonneau}, D.,
  {Christensen-Dalsgaard}, J., {Clarke}, B.~D., {Cochran}, W.~D., {Demory},
  B.-O., {Desert}, J.-M., {Devore}, E., {Doyle}, L.~R., {Esquerdo}, G.~A.,
  {Everett}, M., {Fressin}, F., {Geary}, J.~C., {Girouard}, F.~R., {Gould}, A.,
  {Hall}, J.~R., {Holman}, M.~J., {Howard}, A.~W., {Howell}, S.~B., {Ibrahim},
  K.~A., {Kinemuchi}, K., {Kjeldsen}, H., {Klaus}, T.~C., {Li}, J., {Lucas},
  P.~W., {Meibom}, S., {Morris}, R.~L., {Pr{\v s}a}, A., {Quintana}, E.,
  {Sanderfer}, D.~T., {Sasselov}, D., {Seader}, S.~E., {Smith}, J.~C.,
  {Steffen}, J.~H., {Still}, M., {Stumpe}, M.~C., {Tarter}, J.~C., {Tenenbaum},
  P., {Torres}, G., {Twicken}, J.~D., {Uddin}, K., {Van Cleve}, J.,
  {Walkowicz}, L., \& {Welsh}, W.~F. 2013, \apjs, 204, 24

\bibitem[{{Batygin} \& {Laughlin}(2015)}]{Batygin2015}
{Batygin}, K., \& {Laughlin}, G. 2015, Proceedings of the National Academy of
  Science, 112, 4214

\bibitem[{{Burt} {et~al.}(2014){Burt}, {Vogt}, {Butler}, {Hanson}, {Meschiari},
  {Rivera}, {Henry}, \& {Laughlin}}]{Burt2014}
{Burt}, J., {Vogt}, S.~S., {Butler}, R.~P., {Hanson}, R., {Meschiari}, S.,
  {Rivera}, E.~J., {Henry}, G.~W., \& {Laughlin}, G. 2014, \apj, 789, 114

\bibitem[{{Butler} {et~al.}(1996){Butler}, {Marcy}, {Williams}, {McCarthy},
  {Dosanjh}, \& {Vogt}}]{Butler96}
{Butler}, R.~P., {Marcy}, G.~W., {Williams}, E., {McCarthy}, C., {Dosanjh}, P.,
  \& {Vogt}, S.~S. 1996, \pasp, 108, 500

\bibitem[{{Chiang} \& {Laughlin}(2013)}]{Chiang2013}
{Chiang}, E., \& {Laughlin}, G. 2013, \mnras, 431, 3444

\bibitem[{{Dawson} \& {Fabrycky}(2010)}]{Dawson2010}
{Dawson}, R.~I., \& {Fabrycky}, D.~C. 2010, \apj, 722, 937

\bibitem[{{Eggleton} \& {Tokovinin}(2008)}]{Eggleton2008}
{Eggleton}, P.~P., \& {Tokovinin}, A.~A. 2008, \mnras, 389, 869

\bibitem[{{Ford}(2005)}]{Ford05}
{Ford}, E.~B. 2005, \aj, 129, 1706

\bibitem[{{Ford}(2006)}]{Ford06}
---. 2006, \apj, 642, 505

\bibitem[{{Fulton} {et~al.}(2015){Fulton}, {Weiss}, {Sinukoff}, {Isaacson},
  {Howard}, {Marcy}, {Henry}, {Holden}, \& {Kibrick}}]{Fulton2015}
{Fulton}, B.~J., {Weiss}, L.~M., {Sinukoff}, E., {Isaacson}, H., {Howard},
  A.~W., {Marcy}, G.~W., {Henry}, G.~W., {Holden}, B.~P., \& {Kibrick}, R.~I.
  2015, \apj, 805, 175

\bibitem[{{Gregory}(2005)}]{Gregory05}
{Gregory}, P.~C. 2005, \apj, 631, 1198

\bibitem[{{Gregory}(2011)}]{Gregory11}
---. 2011, \mnras, 415, 2523

\bibitem[{{Hansen} \& {Murray}(2012)}]{Hansen2012}
{Hansen}, B.~M.~S., \& {Murray}, N. 2012, \apj, 751, 158

\bibitem[{{Henry}(1999)}]{Henry1999}
{Henry}, G.~W. 1999, \pasp, 111, 845

\bibitem[{{Howard} {et~al.}(2010){Howard}, {Johnson}, {Marcy}, {Fischer},
  {Wright}, {Bernat}, {Henry}, {Peek}, {Isaacson}, {Apps}, {Endl}, {Cochran},
  {Valenti}, {Anderson}, \& {Piskunov}}]{Howard2010}
{Howard}, A.~W., {Johnson}, J.~A., {Marcy}, G.~W., {Fischer}, D.~A., {Wright},
  J.~T., {Bernat}, D., {Henry}, G.~W., {Peek}, K.~M.~G., {Isaacson}, H.,
  {Apps}, K., {Endl}, M., {Cochran}, W.~D., {Valenti}, J.~A., {Anderson}, J.,
  \& {Piskunov}, N.~E. 2010, \apj, 721, 1467

\bibitem[{{Isaacson} \& {Fischer}(2010)}]{Isaacson2010}
{Isaacson}, H., \& {Fischer}, D. 2010, \apj, 725, 875

\bibitem[{{Kirkwood}(1866)}]{Kirkwood1866}
{Kirkwood}, D. 1866, Proceedings of the American Association for the
  Advancement of Science for 1866, 8

\bibitem[{{Lissauer} {et~al.}(2011){Lissauer}, {Ragozzine}, {Fabrycky},
  {Steffen}, {Ford}, {Jenkins}, {Shporer}, {Holman}, {Rowe}, {Quintana},
  {Batalha}, {Borucki}, {Bryson}, {Caldwell}, {Carter}, {Ciardi}, {Dunham},
  {Fortney}, {Gautier}, {Howell}, {Koch}, {Latham}, {Marcy}, {Morehead}, \&
  {Sasselov}}]{Lissauer2011}
{Lissauer}, J.~J., {Ragozzine}, D., {Fabrycky}, D.~C., {Steffen}, J.~H.,
  {Ford}, E.~B., {Jenkins}, J.~M., {Shporer}, A., {Holman}, M.~J., {Rowe},
  J.~F., {Quintana}, E.~V., {Batalha}, N.~M., {Borucki}, W.~J., {Bryson},
  S.~T., {Caldwell}, D.~A., {Carter}, J.~A., {Ciardi}, D., {Dunham}, E.~W.,
  {Fortney}, J.~J., {Gautier}, III, T.~N., {Howell}, S.~B., {Koch}, D.~G.,
  {Latham}, D.~W., {Marcy}, G.~W., {Morehead}, R.~C., \& {Sasselov}, D. 2011,
  \apjs, 197, 8

\bibitem[{{Lovis} {et~al.}(2011){Lovis}, {S{\'e}gransan}, {Mayor}, {Udry},
  {Benz}, {Bertaux}, {Bouchy}, {Correia}, {Laskar}, {Lo Curto}, {Mordasini},
  {Pepe}, {Queloz}, \& {Santos}}]{Lovis11}
{Lovis}, C., {S{\'e}gransan}, D., {Mayor}, M., {Udry}, S., {Benz}, W.,
  {Bertaux}, J.-L., {Bouchy}, F., {Correia}, A.~C.~M., {Laskar}, J., {Lo
  Curto}, G., {Mordasini}, C., {Pepe}, F., {Queloz}, D., \& {Santos}, N.~C.
  2011, \aap, 528, A112

\bibitem[{{Marcy} {et~al.}(2005){Marcy}, {Butler}, {Vogt}, {Fischer}, {Henry},
  {Laughlin}, {Wright}, \& {Johnson}}]{Marcy05}
{Marcy}, G.~W., {Butler}, R.~P., {Vogt}, S.~S., {Fischer}, D.~A., {Henry},
  G.~W., {Laughlin}, G., {Wright}, J.~T., \& {Johnson}, J.~A. 2005, \apj, 619,
  570

\bibitem[{{Mayor} {et~al.}(2009){Mayor}, {Udry}, {Lovis}, {Pepe}, {Queloz},
  {Benz}, {Bertaux}, {Bouchy}, {Mordasini}, \& {Segransan}}]{Mayor2009}
{Mayor}, M., {Udry}, S., {Lovis}, C., {Pepe}, F., {Queloz}, D., {Benz}, W.,
  {Bertaux}, J.-L., {Bouchy}, F., {Mordasini}, C., \& {Segransan}, D. 2009,
  \aap, 493, 639

\bibitem[{{Montgomery} \& {Laughlin}(2009)}]{Montgomery2009}
{Montgomery}, R., \& {Laughlin}, G. 2009, Icarus, 202, 1

\bibitem[{{Oja}(1993)}]{Oja1993}
{Oja}, T. 1993, \aaps, 100, 591

\bibitem[{{Ram{\'{\i}}rez} {et~al.}(2013){Ram{\'{\i}}rez}, {Allende Prieto}, \&
  {Lambert}}]{Ramirez2013}
{Ram{\'{\i}}rez}, I., {Allende Prieto}, C., \& {Lambert}, D.~L. 2013, \apj,
  764, 78

\bibitem[{{Robinson} {et~al.}(2006){Robinson}, {Laughlin}, {Bodenheimer}, \&
  {Fischer}}]{Robinson2006}
{Robinson}, S.~E., {Laughlin}, G., {Bodenheimer}, P., \& {Fischer}, D. 2006,
  \apj, 643, 484

\bibitem[{{Soubiran} {et~al.}(2008){Soubiran}, {Bienayme}, {Mishenina}, \&
  {Kovtyukh}}]{Soubiran2008}
{Soubiran}, C., {Bienayme}, O., {Mishenina}, T.~V., \& {Kovtyukh}, V.~V. 2008,
  VizieR Online Data Catalog, 348, 91

\bibitem[{{Takeda} {et~al.}(2007){Takeda}, {Ford}, {Sills}, {Rasio}, {Fischer},
  \& {Valenti}}]{Takeda2007}
{Takeda}, G., {Ford}, E.~B., {Sills}, A., {Rasio}, F.~A., {Fischer}, D.~A., \&
  {Valenti}, J.~A. 2007, \apjs, 168, 297

\bibitem[{{Tanner} {et~al.}(2009){Tanner}, {Beichman}, {Bryden}, {Lisse}, \&
  {Lawler}}]{Tanner09}
{Tanner}, A., {Beichman}, C., {Bryden}, G., {Lisse}, C., \& {Lawler}, S. 2009,
  \apj, 704, 109

\bibitem[{{Tanner} {et~al.}(2010){Tanner}, {Gelino}, \& {Law}}]{Tanner2010}
{Tanner}, A.~M., {Gelino}, C.~R., \& {Law}, N.~M. 2010, \pasp, 122, 1195

\bibitem[{{Valenti} \& {Fischer}(2005)}]{FischerValenti05}
{Valenti}, J.~A., \& {Fischer}, D.~A. 2005, \apjs, 159, 141

\bibitem[{Vogt {et~al.}(1994)Vogt, Allen, Bigelow, Bresee, Brown, Cantrall,
  Conrad, Couture, Delaney, Epps, Hilyard, Hilyard, Horn, Jern, Kanto, Keane,
  Kibrick, Lewis, Osborne, Pardeilhan, Pfister, Ricketts, Robinson, Stover,
  Tucker, Ward, \& Wei}]{Vogt94}
Vogt, S.~S., Allen, S.~L., Bigelow, B.~C., Bresee, L., Brown, B., Cantrall, T.,
  Conrad, A., Couture, M., Delaney, C., Epps, H.~W., Hilyard, D., Hilyard,
  D.~F., Horn, E., Jern, N., Kanto, D., Keane, M.~J., Kibrick, R.~I., Lewis,
  J.~W., Osborne, J., Pardeilhan, G.~H., Pfister, T., Ricketts, T., Robinson,
  L.~B., Stover, R.~J., Tucker, D., Ward, J., \& Wei, M.~Z. 1994, in Proc. SPIE
  Instrumentation in Astronomy VIII, ed. D.~L. C. E. R.~C. Eds., Vol. 2198,
  362--+

\bibitem[{{Vogt} {et~al.}(2014{\natexlab{a}}){Vogt}, {Butler}, {Rivera},
  {Kibrick}, {Burt}, {Hanson}, {Meschiari}, {Henry}, \& {Laughlin}}]{Vogt2014b}
{Vogt}, S.~S., {Butler}, R.~P., {Rivera}, E.~J., {Kibrick}, R., {Burt}, J.,
  {Hanson}, R., {Meschiari}, S., {Henry}, G.~W., \& {Laughlin}, G.
  2014{\natexlab{a}}, \apj, 787, 97

\bibitem[{{Vogt} {et~al.}(2014{\natexlab{b}}){Vogt}, {Radovan}, {Kibrick},
  {Butler}, {Alcott}, {Allen}, {Arriagada}, {Bolte}, {Burt}, {Cabak},
  {Chloros}, {Cowley}, {Deich}, {Dupraw}, {Earthman}, {Epps}, {Faber},
  {Fischer}, {Gates}, {Hilyard}, {Holden}, {Johnston}, {Keiser}, {Kanto},
  {Katsuki}, {Laiterman}, {Lanclos}, {Laughlin}, {Lewis}, {Lockwood}, {Lynam},
  {Marcy}, {McLean}, {Miller}, {Misch}, {Peck}, {Pfister}, {Phillips},
  {Rivera}, {Sandford}, {Saylor}, {Stover}, {Thompson}, {Walp}, {Ward},
  {Wareham}, {Wei}, \& {Wright}}]{Vogt2014a}
{Vogt}, S.~S., {Radovan}, M., {Kibrick}, R., {Butler}, R.~P., {Alcott}, B.,
  {Allen}, S., {Arriagada}, P., {Bolte}, M., {Burt}, J., {Cabak}, J.,
  {Chloros}, K., {Cowley}, D., {Deich}, W., {Dupraw}, B., {Earthman}, W.,
  {Epps}, H., {Faber}, S., {Fischer}, D., {Gates}, E., {Hilyard}, D., {Holden},
  B., {Johnston}, K., {Keiser}, S., {Kanto}, D., {Katsuki}, M., {Laiterman},
  L., {Lanclos}, K., {Laughlin}, G., {Lewis}, J., {Lockwood}, C., {Lynam}, P.,
  {Marcy}, G., {McLean}, M., {Miller}, J., {Misch}, T., {Peck}, M., {Pfister},
  T., {Phillips}, A., {Rivera}, E., {Sandford}, D., {Saylor}, M., {Stover}, R.,
  {Thompson}, M., {Walp}, B., {Ward}, J., {Wareham}, J., {Wei}, M., \&
  {Wright}, C. 2014{\natexlab{b}}, \pasp, 126, 359

\bibitem[{{Walker}(2012)}]{Walker2012}
{Walker}, G.~A.~H. 2012, New Astronomy, 56, 9

\bibitem[{{Walker} {et~al.}(1995){Walker}, {Walker}, {Irwin}, {Larson}, {Yang},
  \& {Richardson}}]{Walker1995}
{Walker}, G.~A.~H., {Walker}, A.~R., {Irwin}, A.~W., {Larson}, A.~M., {Yang},
  S.~L.~S., \& {Richardson}, D.~C. 1995, Icarus, 116, 359

\bibitem[{{Wolfgang} \& {Lopez}(2015)}]{Wolfgang2015}
{Wolfgang}, A., \& {Lopez}, E. 2015, \apj, 806, 183

\bibitem[{{Wright} {et~al.}(2011){Wright}, {Fakhouri}, {Marcy}, {Han}, {Feng},
  {Johnson}, {Howard}, {Fischer}, {Valenti}, {Anderson}, \&
  {Piskunov}}]{Wright11}
{Wright}, J.~T., {Fakhouri}, O., {Marcy}, G.~W., {Han}, E., {Feng}, Y.,
  {Johnson}, J.~A., {Howard}, A.~W., {Fischer}, D.~A., {Valenti}, J.~A.,
  {Anderson}, J., \& {Piskunov}, N. 2011, \pasp, 123, 412

\bibitem[{{Zechmeister} \& {K{\"u}rster}(2009)}]{Zechmeister09}
{Zechmeister}, M., \& {K{\"u}rster}, M. 2009, \aap, 496, 577

\end{thebibliography}

\end{document}